\documentclass[12pt]{article}
\usepackage{float}
\usepackage[utf8]{inputenc}
\usepackage{newunicodechar}
\newunicodechar{́}{\'}
\usepackage[paper=a4paper,centering,total={6.5in, 8.8in}]{geometry}
\usepackage{soul}
\usepackage[sorting=none,backend=biber,style=numeric]{biblatex}
\DeclareFieldFormat[article]{volume}{\textbf{#1}}
\renewbibmacro{in:}{}
\DeclareFieldFormat{pages}{#1}

\DefineBibliographyStrings{english}{%
  andothers = {\emph{et\addabbrvspace al\adddot}}
}
\usepackage{xpatch}
\usepackage{etoolbox}

\AtEveryBibitem{%
  \ifentrytype{article}{%
  }{}%
}

\usepackage[absolute,overlay]{textpos}
\textblockorigin{10mm}{10mm} 

\usepackage{graphicx}
\usepackage{amsmath}
\usepackage{amssymb}
\usepackage{setspace}
\usepackage[pdftex,colorlinks,citecolor=blue,bookmarks]{hyperref}
\usepackage{bm}
\usepackage{authblk}
\makeatletter
\renewcommand\AB@affilsepx{\\[1pt]}  
\makeatother

\usepackage{xcolor}
\usepackage[nottoc]{tocbibind}
\usepackage{physics}
\usepackage{multirow}
\usepackage{url}
\usepackage{csquotes}
\usepackage{textcomp}
\usepackage{newunicodechar}
\newunicodechar{−}{\textminus}
\usepackage{xspace}

\usepackage{booktabs}

\usepackage{xspace}  

\usepackage{dcolumn,booktabs}
\newcolumntype{d}[1]{D{.}{.}{#1}}

\newcommand{\trento}{T$\mathrel{\protect\raisebox{-2.1pt}{R}}$ENTo}

\newcommand{\nene}{$^{20}$Ne+$^{20}$Ne}
\newcommand{\oooo}{$^{16}$O+$^{16}$O}
\newcommand{\pbpb}{$^{208}$Pb+$^{208}$Pb}
\newcommand{\xexe}{$^{129}$Xe+$^{129}$Xe}
\newcommand{\auau}{$^{197}$Au+$^{197}$Au}
\newcommand{\uuuu}{$^{238}$U+$^{238}$U}

\newcommand{\ruru}{$^{96}$Ru+$^{96}$Ru}
\newcommand{\zrzr}{$^{96}$Zr+$^{96}$Zr}
\newcommand{\cucu}{$^{62}$Cu+$^{62}$Cu}

\newcommand{\xT}{\mathbf{x_\perp}}

\newcommand{\zT}{\mathbf{z_\perp}}

\usepackage{longtable}

\title{\vspace{-50pt}\textbf{\Large{EMMI Rapid Reaction Task Force}}\\\vspace{10pt}\textbf{Nuclear Physics Confronts Relativistic Collisions Of Isobars}\\\vspace{10pt}}

\author[1,2]{Giuliano Giacalone%
  ~\textcolor{red}{\small $\dagger$}%
  ~\textcolor{green}{\scriptsize $\spadesuit$}%
  \thanks{correspond to: \texttt{giuliano.giacalone@cern.ch}~~~~ \textcolor{red}{\small $\dagger$} = original RRTF member~~~~
          \textcolor{green}{\scriptsize $\spadesuit$} = organizer}%
          }
\affil[1]{Institut f\"ur Theoretische Physik, Universit\"at Heidelberg, Philosophenweg 16, 69120 Heidelberg, Germany}
\affil[2]{Theoretical Physics Department, CERN, CH-1211 Gen\`eve 23, Switzerland}

\author[3,4]{Jiangyong Jia~\textcolor{red}{\small $\dagger$}~\textcolor{green}{\scriptsize $\spadesuit$}}
\affil[3]{Department of Chemistry, Stony Brook University, Stony Brook, NY 11794, USA}
\affil[4]{Physics Department, Brookhaven National Laboratory, Upton, NY 11976, USA}

\author[5]{Vittorio Somà~\textcolor{red}{\small $\dagger$}~\textcolor{green}{\scriptsize $\spadesuit$}}
\affil[5]{IRFU, CEA, Universit\'e Paris-Saclay, 91191 Gif-sur-Yvette, France}

\author[6]{\\You Zhou~\textcolor{red}{\small $\dagger$}~\textcolor{green}{\scriptsize $\spadesuit$}}
\affil[6]{Niels Bohr Institute, University of Copenhagen, Jagtvej 155A, 2200 Copenhagen, Denmark}

\author[7]{Anatoli Afanasjev~\textcolor{red}{\small $\dagger$}}
\affil[7]{Department of Physics and Astronomy, Mississippi State University, Mississippi 39762, USA}

\author[8,9]{Massimiliano Alvioli}
\affil[8]{Istituto di Ricerca per la Protezione Idrogeologica, Consiglio Nazionale delle Ricerche, \hspace{100pt} via Madonna Alta 126, I-06128, Perugia, Italy}
\affil[9]{Istituto Nazionale di Fisica Nucleare, Sezione di Perugia, via Pascoli 23c, I-06123, Perugia, Italy}

\author[10,11]{\\Benjamin Bally~\textcolor{red}{\small $\dagger$}}
\affil[10]{ESNT, IRFU, CEA, Universit\'e Paris-Saclay, 91191 Gif-sur-Yvette, France}
\affil[11]{Technische Universit\"at Darmstadt, Department of Physics, 64289 Darmstadt, Germany}

\author[12,13]{Federica Capellino~\textcolor{red}{\small $\dagger$}}
\affil[12]{GSI Helmholtzzentrum f\"{u}r Schwerionenforschung, D-64291 Darmstadt, Germany}
\affil[13]{Physikalisches Institut Heidelberg, D-69120 Heidelberg, Germany}

\author[14,15]{Jean-Paul Ebran~\textcolor{red}{\small $\dagger$}}
\affil[14]{CEA, DAM, DIF, 91297 Arpajon, France}
\affil[15]{Universit\'e Paris-Saclay, CEA, Laboratoire Mati\`ere en Conditions Extr\^emes, \hspace{200pt} 91680 Bruy\`eres-le-Ch\^atel, France}

\author[12,16,17]{\\Hannah Elfner~\textcolor{red}{\small $\dagger$}}
\affil[16]{Frankfurt Institute for Advanced Studies, Ruth-Moufang-Strasse 1,  60438 Frankfurt am Main, Germany}
\affil[17]{Helmholtz Research Academy Hesse for FAIR (HFHF), GSI Helmholtz Center, \hspace{140pt} Campus Frankfurt,  60438 Frankfurt am Main, Germany}

\author[18]{Fernando G. Gardim}
\affil[18]{Instituto de Ci\^encia e Tecnologia, Universidade Federal de Alfenas, \hspace{200pt} 37715-400 Po\c{c}os de Caldas-MG, Brazil}

\author[18,19,20]{Andr\'e V. Giannini}
\affil[19]{Faculdade de Ci\^encias Exatas e Tecnologia, Universidade Federal da Grande Dourados,\hspace{100pt} Caixa Postal 364, CEP 79804-970 Dourados, MS, Brazil}
\affil[20]{Departamento de F\'isica, Universidade do Estado de Santa Catarina, 89219-710 Joinville, SC, Brazil}

\author[21]{Fr\'ed\'erique Grassi~\textcolor{red}{\small $\dagger$}}
\affil[21]{Instituto de F\'{\i}sica, Universidade de  S\~{a}o Paulo,  Rua  do  Mat\~{a}o, 1371, \hspace{200pt} Butant\~{a},  05508-090,  S\~{a}o  Paulo,  Brazil}

\author[22]{Eduardo Grossi~\textcolor{red}{\small $\dagger$}}
\affil[22]{Dipartimento di Fisica, Universit\`a di Firenze and INFN Sezione di Firenze,\hspace{140pt} via G. Sansone 1, 50019 Sesto Fiorentino, Italy}

\author[16]{Jan Hammelmann~\textcolor{red}{\small $\dagger$}}

\author[1,23]{\\Andreas Kirchner~\textcolor{red}{\small $\dagger$}}
\affil[23]{Department of Physics, Duke University, Durham, NC 27708, USA}

\author[24]{Dean Lee~\textcolor{red}{\small $\dagger$}}
\affil[24]{Facility~for~Rare~Isotope~Beams~and~Department~of~Physics~and~Astronomy,\hspace{100pt} Michigan~State~University, East~Lansing~MI~48824,~USA}

\author[21]{Matthew Luzum~\textcolor{red}{\small $\dagger$}}

\author[25]{\\Hadi Mehrabpour}
\affil[25]{Center for High Energy Physics and School of Physics, Peking University, Beijing 100871, China}

\author[6]{Emil G. Nielsen}

\author[2]{Govert Nijs}

\author[26]{Tamara Nik\v si\'c~\textcolor{red}{\small $\dagger$}}
\affil[26]{Physics Department, Faculty of Science, University of Zagreb, 10000 Zagreb, Croatia}

\author[27]{Jacquelyn Noronha-Hostler~\textcolor{red}{\small $\dagger$}}
\affil[27]{Illinois Center for Advanced Studies of the Universe, Department of Physics,\hspace{100pt} University of Illinois at Urbana-Champaign, Urbana, IL 61801, USA}

\author[28]{Jean-Yves Ollitrault~\textcolor{red}{\small $\dagger$}}
\affil[28]{Universit\'e Paris Saclay, CNRS, CEA, Institut de physique th\'eorique, 91191 Gif-sur-Yvette, France}

\author[29,30]{\\Takaharu Otsuka~\textcolor{red}{\small $\dagger$}}
\affil[29]{Department of Physics, University of Tokyo, Hongo, Bunkyo-ku, Tokyo 113-0033, Japan}
\affil[30]{RIKEN Nishina Center, 2-1 Hirosawa, Wako, Saitama 351-0198, Japan}

\author[21]{Kevin P.~Pala}

\author[7]{Udeshika C. Perera}

\author[31]{\\Luis M. Robledo~\textcolor{red}{\small $\dagger$}}
\affil[31]{Departamento de F\'isica Te\'orica and CIAFF, Universidad Aut\'onoma de Madrid,  E-28049 Madrid, Spain}

\author[31,32,33]{Tom\'as R. Rodr\'iguez~\textcolor{red}{\small $\dagger$}}
\affil[32]{Departamento de Estructura de la Materia, F\'isica T\'ermica y Electr\'onica and IPARCOS, \hspace{100pt} Universidad Complutense de Madrid, E-28040 Madrid, Spain}
\affil[33]{Departamento de F\'isica At\'omica, Molecular y Nuclear, Universidad de Sevilla, E-41012 Sevilla, Spain}

\author[34]{Wouter Ryssens~\textcolor{red}{\small $\dagger$}}
\affil[34]{Institut d’Astronomie et d’Astrophysique, Universit\'e Libre de Bruxelles,\hspace{150pt} Campus de la Plaine CP 226, 1050 Brussels, Belgium}

\author[16]{\\Nils Saß~\textcolor{red}{\small $\dagger$}}

\author[2,35,36]{Wilke van der Schee~\textcolor{red}{\small $\dagger$}}
\affil[35]{Institute for Theoretical Physics, Utrecht University, 3584 CC Utrecht, The Netherlands}
\affil[36]{Nikhef, Science Park 105, 1098 XG Amsterdam, The Netherlands}

\author[4]{Bj\"orn Schenke~\textcolor{red}{\small $\dagger$}}

\author[23,37]{\\Willian M.~Serenone~\textcolor{red}{\small $\dagger$}}
\affil[37]{Instituto de Fisica Gleb Wataghin, Universidade Estadual de Campinas, Campinas, Brasil}

\author[38,39]{Pragya Singh~\textcolor{red}{\small $\dagger$}}
\affil[38]{Department of Physics, P.O. Box 35, 40014 University of Jyv\"askyl\"a, Finland}
\affil[39]{Helsinki Institute of Physics, P.O. Box 64, FI-00014 University of Helsinki, Finland}

\author[40]{Chun Shen~\textcolor{red}{\small $\dagger$}}
\affil[40]{Department of Physics and Astronomy, Wayne State University, Detroit, Michigan 48201, USA}

\author[41]{\\Noritaka Shimizu}
\affil[41]{Center for Computational Sciences, University of Tsukuba, 1-1-1 Tennodai, \hspace{160pt}Tsukuba, Ibaraki, 305-8577, Japan}

\author[42]{Huichao Song~\textcolor{red}{\small $\dagger$}}
\affil[42]{School of Physics, Peking University, Beijing 100871, China}

\author[43]{Seyed Farid Taghavi~\textcolor{red}{\small $\dagger$}}
\affil[43]{TUM School of Natural Sciences, Technische Universit\"at M\"unchen, Garching, Germany}

\author[44]{\\Derek Teaney~\textcolor{red}{\small $\dagger$}}
\affil[44]{Center for Nuclear Theory, Department of Physics and Astronomy, Stony Brook University, \hspace{100pt} Stony Brook, New York, 11794-3800, USA}

\author[29]{Yusuke Tsunoda~\textcolor{red}{\small $\dagger$}}

\author[12]{Kathrin Wimmer~\textcolor{red}{\small $\dagger$}}

\author[30]{Kota Yanase}

\author[45,46]{\\Chunjian Zhang}
\affil[45]{Key Laboratory of Nuclear Physics and Ion-beam Application (MOE), and \hspace{100pt} Institute of Modern Physics, Fudan University, Shanghai 200433, China}
\affil[46]{Shanghai Research Center for Theoretical Nuclear Physics, NSFC, and \hspace{150pt} Fudan University, Shanghai 200438, China}

\author[42]{Shujun Zhao}

\author[47,48]{Wenbin Zhao}
\affil[47]{Nuclear Science Division, Lawrence Berkeley National Laboratory, Berkeley, California 94720, USA}
\affil[48]{Physics Department, University of California, Berkeley,  California 94720, USA}

\date{}

\begin{document}

\begin{textblock*}{50mm}(140mm,0mm) 
\raggedleft
CERN-TH-2025-124
\end{textblock*}

\thispagestyle{empty}

\maketitle

\thispagestyle{empty}

\newpage

\thispagestyle{empty}

\newpage

\begin{center}
    \textbf{Motivation for a Task Force}
\end{center}
High-energy collisions involving the $A=96$ isobars $^{96}$Zr and $^{96}$Ru have been performed in 2018 at Brookhaven National Laboratory's  Relativistic Heavy Ion Collider (RHIC) as a means to search for the chiral magnetic effect in QCD. This would manifest itself as specific deviations from unity in the ratio of observables taken between $^{96}$Zr+$^{96}$Zr and $^{96}$Ru+$^{96}$Ru collisions. Measurements of such ratios (released at the end of 2021) indeed reveal deviations from unity, but these are primarily caused by the two collided isobars having different radial profiles and intrinsic deformations. To make progress in understanding RHIC data, nuclear physicists across the energy spectrum gathered in Heidelberg in 2022 as part of an EMMI Rapid Reaction Task Force (RRTF) to address the following question. Does the combined effort of low-energy nuclear structure physics and high-energy heavy-ion physics enable us to understand the observations made in isobar collisions at RHIC?

\bigskip
\begin{center}
    \textbf{Conclusions of the Task Force}
\end{center}

\begin{itemize}
    \item $^{96}$Ru and $^{96}$Zr have entirely different types of structure in a region of the nuclear chart where nuclei display complex behavior along isotopic chains, for which we have a limited understanding. Data on the shape of $^{96}$Ru is especially scarce.

    \item  Preliminary results from energy density functional calculations of nuclear structure hint at a large octupole deformation in the ground state of $^{96}$Zr. A quantitative understanding of this phenomenon will require the development of  \textit{ab initio} approaches that will rely less on the notion of an intrinsic nuclear frame.

    \item Initial-state and hydrodynamic computations indicate that the observations of the STAR collaboration in isobar collisions are naturally explained if the two nuclei present different intrinsic shapes, with the corresponding differences qualitatively agreeing with the available information from low-energy nuclear experiments.

    \item Any quantitative conclusion about the impact of the strong magnetic field on the measured collective flow in isobar collisions is premature in absence of a proper uncertainty quantification on the knowledge of the nuclear geometries. The results of the STAR collaboration pave the way to unprecedented precision tests of the paradigms underlying our understanding of nuclear collisions around and above the GeV scale.
    
\end{itemize}

\bigskip
\begin{center}
    \textbf{Recommendations of the Task Force}
\end{center}

\begin{itemize}
    \item Future proposals for precision high-energy physics studies with nuclei will require a detailed assessment of the uncertainties related to the knowledge of the relevant species. That should be carried out in collaboration with the low-energy nuclear community.
    \item If an isobar run is repeated to explore the effects of the strong magnetic field, we recommend to use the pair $^{136}_{54}$Xe--$^{136}_{58}$Ce, due to the near-spherical shape of $^{136}$Xe.
    \item The RRTF recommends the investigation of novel science cases exploiting isobar collisions as input for precision physics studies with atomic nuclei.
\end{itemize}

\newpage
\thispagestyle{empty}

\tableofcontents

\thispagestyle{empty}

\newpage

\section{Isobar collisions at RHIC}

\label{sec:1}

\subsection{Search for CME and isobar operation mode}

\begin{figure}[t]
    \centering
    \includegraphics[width=.8\linewidth]{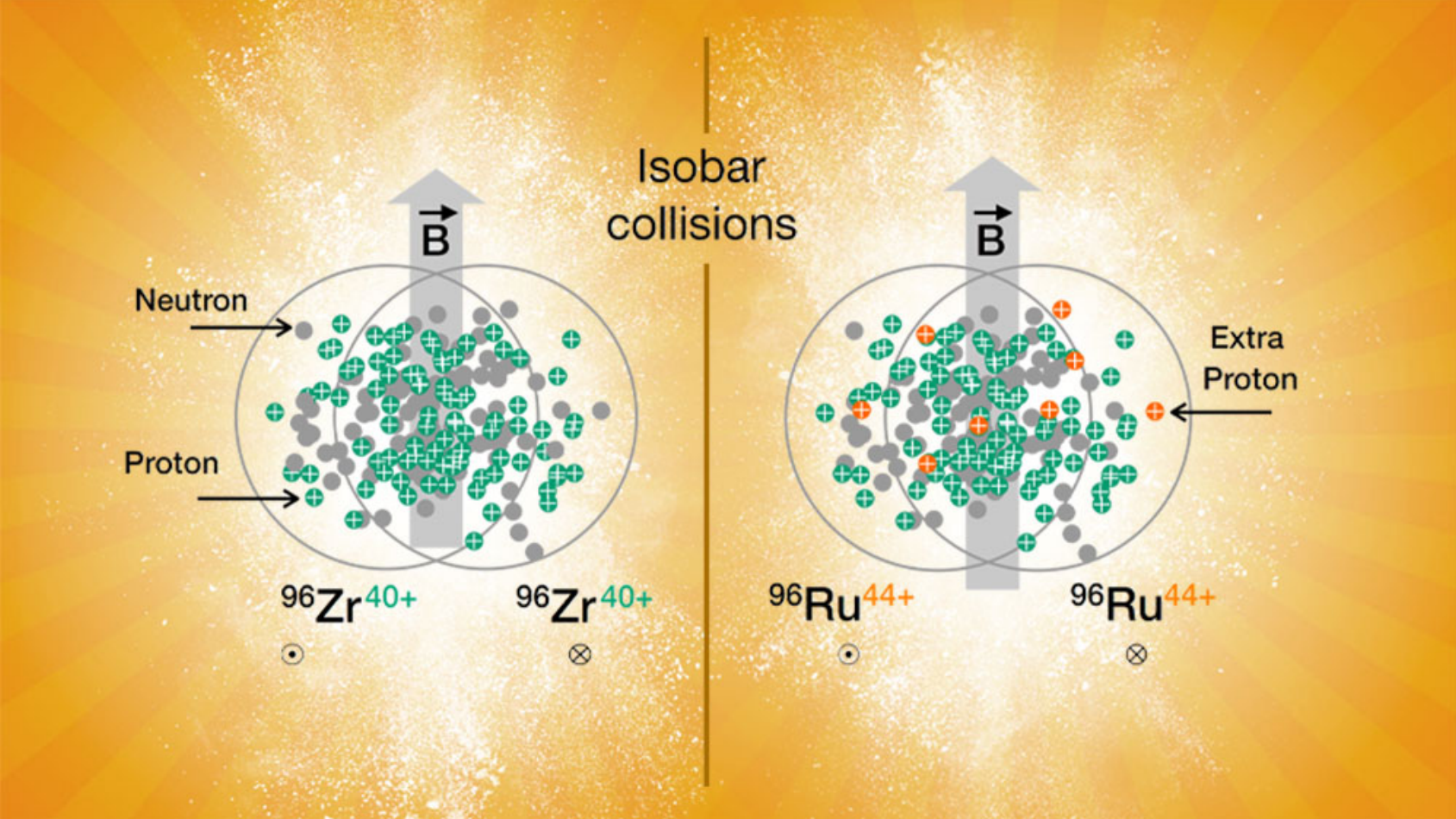}
    \caption{Sketch of isobar collisions. The beam direction is orthogonal to the plane of the figure. A \ruru{} collision (right) involves more protons than a \zrzr{} collisions (left). As a consequence, the magnitude of the magnetic field ($\vec B$) acting on the overlap area is stronger in \ruru{} collisions. As shown, for an off-central collision the direction of the $\vec B$ field will be preferably orthogonal to that of the collision impact parameter. The charge separation due to the CME hence mainly develops in the direction orthogonal to the reaction plane, where the elliptic flow is the strongest. Courtesy of the STAR collaboration.}
    \label{fig:isobar_sketch}
\end{figure}

In high-energy nuclear collisions, there exists the possibility of detecting a transient violation of parity ($\mathcal{P}$) and charge-parity ($\mathcal{CP}$) symmetries by the strong interaction occurring within localized areas in the interaction region \cite{Kharzeev:1998kz,Kharzeev:1999cz}. This would result in an unequal distribution of right- and left-handed (anti)quarks which, in the presence of a strong electro-magnetic field, such as that expected to be sourced by protons in off-central nuclear collisions (see a sketch in Fig.~\ref{fig:isobar_sketch}), would manifest as a charge separation along the magnetic field direction \cite{Kharzeev:2004ey}. This effect is termed the Chiral Magnetic Effect (CME) \cite{Kharzeev:2007jp,Fukushima:2008xe}. A conclusive detection of CME in heavy-ion collisions would substantiate the existence of these $\mathcal{CP}$-violating zones, indicating partial chiral symmetry restoration in the Quark-Gluon Plasma (QGP) medium and the presence of an exceptionally strong magnetic field \cite{Kharzeev:2015znc,Kharzeev:2020jxw}. Over the last fifteen years, rigorously testing the CME signals has been a primary focus of the Relativistic Heavy-Ion Collider (RHIC) program at Brookhaven National Laboratory (BNL).

The CME-sensitive charge separation would emerge primarily in the direction perpendicular to the reaction plane (RP) of the collisions \cite{Huang:2015oca,Li:2020dwr} (as defined by the plane spanned by the impact parameter and the beam direction, see Fig.~\ref{fig:isobar_sketch}). The idea is to decompose the final-state azimuthal particle distribution in Fourier modes,
\begin{equation}
    \frac{dN_\alpha}{d\phi^*} \approx \frac{N_\alpha}{2\pi} \biggl[ 1 + 2v_{1,\alpha} \cos(\phi-\phi_{RP}) + 2a_{1,\alpha} \sin(\phi-\phi_{RP}) + 2v_{2,\alpha} \cos[2(\phi-\phi_{RP})] + \cdots \biggr],
    \label{eq:flow_equation}
\end{equation}
where $\phi$ is the azimuthal angle of the outgoing hadron, and $\phi_{\rm RP}$ is the RP direction. The subscript $\alpha=\pm1$ labels the sign of the charge of the detected particle. The coefficients $v_n$ are the so-called \textit{anisotropic flow} coefficients \cite{Heinz:2013th,Ollitrault:2023wjk}, while the coefficient $a_1$ quantifies the charge separation along the direction of the RP, expected to be correlated with that of the magnetic field. For an experimental event-by-event analysis of the dipole asymmetry, $a_1$, in the charge distribution, one typically measures its mean squared value via the two-particle correlator \cite{Voloshin:2004vk}
\begin{equation}
   \gamma_{\alpha\beta} = \langle \cos (\phi_\alpha-\phi_\beta - 2 \Psi_{\rm RP}),
\end{equation}
where $\alpha$ and $\beta$ denote the charges of the correlated particles. The average is here performed over all pairs of particles, first in one event, and then over all events. The difference between opposite-sign and same-sign $\gamma$ correlators is typically measured in order to eliminate background effects, mainly due to global momentum conservation \cite{Pratt:2010zn}. 

However, measurements involving $\gamma_{\alpha\beta}$ are strongly affected by the underlying flow phenomena, which dominate the low-momentum particle production in heavy-ion collisions. To reduce such backgrounds, the STAR collaboration proposes using \textit{isobar} collisions. The idea is very simple. One looks at the relative variation of the $\gamma_{\alpha\beta}$ correlator between two collision systems involving nuclei that have the same mass number but that differ by 4 units in electric charge, that is, comparing the outcome of
${}^{A}_{Z}\mathrm{X}$+${}^{A}_{Z}\mathrm{X}$ collisions to that of $_{Z+4}^{A}\mathrm{Y}$+${}^{A}_{Z+4}\mathrm{Y}$ collisions. Because of the nearly identical system size in the QGP formed in X+X and Y+Y collisions, these two systems should present a nearly identical collective flow background. However, the CME signal would not be the same in both systems, as ${}^{A}_{Z+4}\mathrm{Y}$+${}^{A}_{Z+4}\mathrm{Y}$ collisions would present a higher magnetic field acting over the interaction region.

The species chosen by the STAR collaboration for this analysis are ${}^{96}_{44}\mathrm{Ru}$ and ${}^{96}_{40}\mathrm{Zr}$.  Isobar collisions were performed at RHIC in 2018, and the data was publicly released in September 2021 following a blind analysis \cite{STAR:2021mii}. The run is very successful. Ratios of observables between Ru+Ru and Zr+Zr collisions are obtained with a relative precision of 0.4\% or lower, deemed small enough to possibly reveal the small charge separation induced by the CME.

The elephant in the room is the low-energy structure of the isobar species. Atomic nuclei are many-body quantum systems whose properties can vary significantly between neighboring isotopes. As we discuss in the next Section, high-energy collision probe the distributions of nucleons, and the correlations therein, in the interaction region. As a consequence, backgrounds to the CME signal are largely impact by nuclear structure effects, invalidating thus the idea that the collective flow is identical in these two systems. In particular, throughout this document we will see that for the chosen isobars these effects are so large that they simply mask any relative variation of observables genuinely driven by magnetic field effects. This makes a sound conclusion about the detection of the CME largely dependent on our knowledge of the ground-state properties of the collided isobars.

We now discuss to what extent these issues related to nuclear structure were predicted or understood before the release of the experimental measurements.

\subsection{The nuclear structure question before the isobar run}

 When precision physics searches involving atomic nuclei are envisaged, one has to deal with a potentially large and irreducible uncertainty coming from the imperfect knowledge of the nuclear ground states (e.g., their shapes or radii). One bright example is the long-standing problem of determining the nuclear matrix element of the neutrinoless double beta decay ($0\nu\beta\beta$) transition \cite{Agostini:2022zub}. These quantities require a detailed understanding of the many-body structure of the candidate isotopes \cite{Yao:2021wst}. For this reason, even in the event of a detection of the $0\nu\beta\beta$ process, a sound connection with physics Beyond the Standard Model would be largely plagued by uncertainties related to nuclear structure. In hindsight, it is only natural that for the characterization of a small CME signal at RHIC one has to deal with an intrinsic uncertainty coming from the imperfect knowledge of the collided species.

\subsubsection{Literature prior to the year 2021}

It seems that the first mention of isobar collisions using $^{96}$Ru and $^{96}$Zr isotopes as a tool to discover the CME appears in a paper by Voloshin at the end of 2010 \cite{Voloshin:2010ut}. It is then pointed out in 2013 in a conference note by Filip \cite{Filip:2013uka} that these two isotopes pose some issues related to their structure. As we shall see in Sec.~\ref{sec:Wimmer}, these two isotopes belong to a transitional region of the nuclear chart where nuclei exhibit highly complex behavior, such that no firm knowledge of their structure can be inferred from low-energy experimental data alone. Different model calculations provide, then, a highly unclear picture of these ground states. In some cases ruthenium is more deformed than zirconium \cite{Pritychenko:2013gwa}, in others ruthenium is spherical while zirconium is deformed \cite{Moller:2015fba}, while essentially nothing is known about the difference in skin thickness between these isotopes.  The actual impact of these uncertainties on the outcome of the collisions was not estimated for a long time.

\begin{figure}[t]
    \centering
    \includegraphics[width=\linewidth]{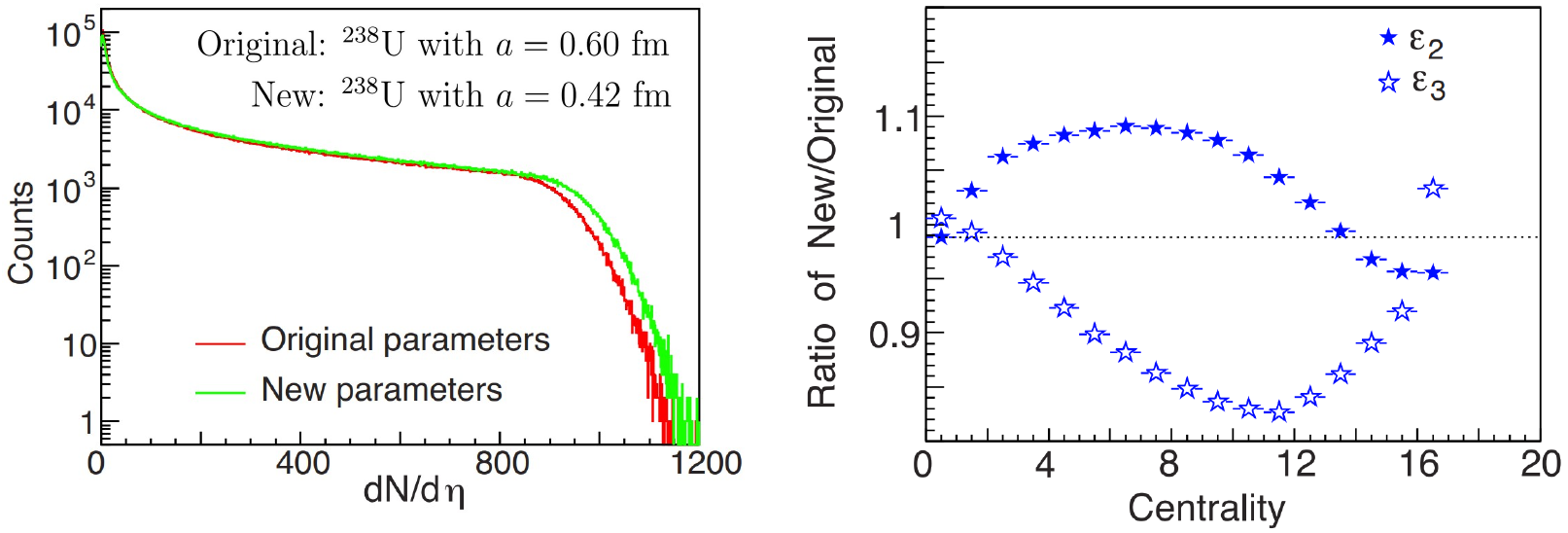}
    \caption{\textit{Left:} Minimum bias charged multiplicity distribution in \uuuu{} collisions at top RHIC energy, for $^{238}$U nuclei with $a=0.60$ fm (Original, red line) and $a=0.42$ fm (New, green line). Right, New/Original ratio for the eccentricities of \uuuu{} collisions as a function of the collision centrality (where each integer number should be multiplied by 5\%). Figures adapted from Ref.~\cite{Shou:2014eya}.}
    \label{fig:1409}
\end{figure}

The first paper that should be mentioned discussing nuclear structure effects relevant for isobar collisions is a letter by Shou \textit{et al.} \cite{Shou:2014eya} which is actually unrelated to the search for the CME or the isobar campaign. In that paper, the authors assess the impact of structural features of $^{238}$U isotopes in the context of \uuuu{} collisions at RHIC \cite{STAR:2015mki}. The colliding nuclei are modeled as bunches of nucleons sampled according to a quadrupole-deformed one-body density of the form: 
\begin{equation}
    \rho(r,\theta) \propto \frac{1}{1+e^{\frac{r-R(\theta)}{a}}}, \hspace{30pt}R(\theta)= R(1+\beta_2 Y_2^0(\theta) + \ldots).
\end{equation}
The authors analyze, in particular, the effect of varying the parameter $a$ in the density of $^{238}$U while leaving all the other parameters essentially unchanged. They consider two cases. The first, denoted Original, uses $a=0.60$ fm, whereas the second, denoted New, uses $a=0.42$ fm. The main effects of this change in diffuseness are displayed in Fig.~\ref{fig:1409}. The left panel shows the minimum bias distribution of charged particles, $dN/d\eta$, produced in \uuuu{} collisions. Because of the sharper nuclear surface, the New parametrization (green curve) leads to an overall lower probability of a hadronic collisions at large values of the impact parameter (low multiplicities) compared to the Old parametrization (red curve). Moreover, in the limit of central collisions the lower value of $a=0.42$ fm leads to a higher density of participant nucleons within the interaction region, causing higher multiplicity values for the New parametrization. The right-hand panel of Fig.~\ref{fig:1409} shows instead the impact of the spatial eccentricities of the overlap area. The effect of changing the skin thickness of the nucleus is manifest mainly in off-central collisions, and, remarkably, its impact changes qualitatively between the ellipticity, $\varepsilon_2$, and the triangularity, $\varepsilon_3$. The corrections are very sizable, with eccentricities overall varying by more than 10\% from a skin thickness variation of 0.18 fm.

Recall now the isotopes under consideration for the isobar campaign:
\begin{equation}
    ^{96}{\rm Ru}~(Z=44, N=52), \hspace{40pt} ^{96}{\rm Zr}~(Z=40, N=56).
\end{equation}
Based on the simple fact that $^{96}$Zr contains four extra neutrons, the skin thickness of this nucleus should be significantly larger than that of $^{96}$Ru. This will lead to variations in the observables along the lines of those shown in Fig.~\ref{fig:1409}.

Only with the advent of the isobar experimental campaign have comprehensive dynamical calculations of the outcome of Ru+Ru and Zr+Zr collisions appeared in the literature. These calculations are mainly performed with the A Multi Phase Transport (AMPT) model \cite{Lin:2004en}. In the very first study by Deng \textit{et al.} \cite{Deng:2016knn}, the parameter $a$ is kept the same between the two isobars, but the value of $\beta_2$ is varied. Two scenarios are considered, one where the deformation parameter is higher for $^{96}$Ru, and vice versa, reflecting the uncertain knowledge of this parameter in the literature on low-energy nuclear structure. The authors compute, in particular, an \textit{isobar ratio} of elliptic flow coefficients, defined by
\begin{equation}
    R(v_2) = \frac{v_{2,{\rm Ru+Ru}}}{v_{2,{\rm Zr+Zr}}},
\end{equation}
which is found to be either above or below unity depending on whether ruthenium is more deformed than zirconium or not. The authors are led to conclude: \textit{"$v_2$ measurements in central collisions
will discern which information source (case 1 or 2) is more reliable regarding the deformity of the Ru and Zr nuclei"}.  That said, deformation mainly impacts central collisions that are not relevant for searching the CME signal. 

A major advance is then achieved in the subsequent calculations by Xu \textit{et al.} \cite{Xu:2017zcn,Li:2018oec}. By using nuclear densities and parameters consistent with the predictions of energy density functional calculations, these simulations focus on the impact of the larger skin diffuseness of $^{96}$Zr, naturally predicted by any realistic nuclear model. In particular, it is shown that the smaller value of $a$ for the $^{96}$Ru nucleus leads to an enhancement of the eccentricity in \ruru{} collisions with the collision centrality, precisely consistent with the observation made in the right-hand panel of Fig.~\ref{fig:1409}. This finding essentially invalidates the idea that the background to the CME signal is both under control and the same for both collision systems. Not only does this background exist because the two isobars have different shapes and skins, but it also presents the same multiplicity dependence as the signals induced by the CME (as the strength of the magnetic field grows with the impact parameter \cite{Giacalone:2021bzr}). Moreover, it is realized by Hammelmann \textit{et al.} \cite{Hammelmann:2019vwd} that properly taking into isospin-dependent effects, that is, the fact that the periphery of the nuclei is mainly populated by neutrons rather than protons (neutron skin) can reduce by as much a factor of two the expected magnetic field strength difference between \ruru{} and \zrzr{} collisions.

\begin{figure}[t]
    \centering
     \includegraphics[width=\linewidth]{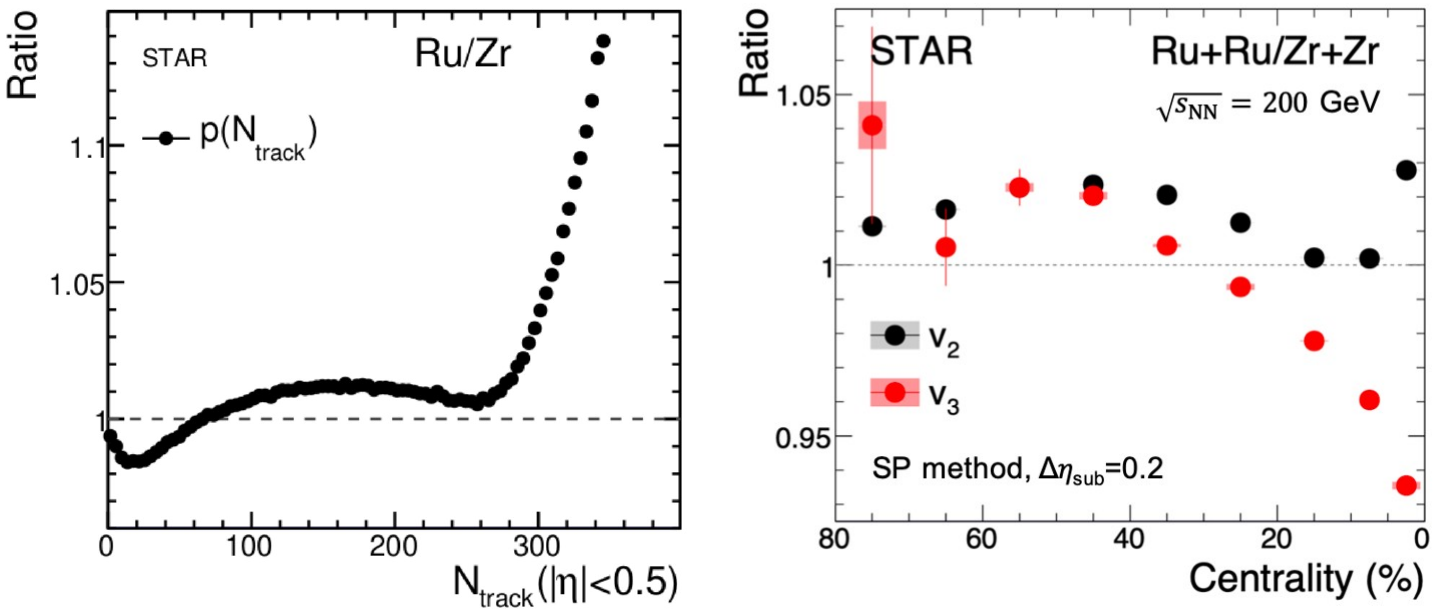}
    \caption{\textit{Left:} Ratio of multiplicity distributions, $P(N_{\rm track})$ taken between \ruru{} and \zrzr{} collisions as measured by the STAR collaboration. \textit{Right:} Isobar ratio of anisotropic flow coefficients as a function of the collision centrality. Black points are for $v_2$, while red points for $v_3$. Note that the limit of central collisions is in both panels on the right-hand side of the plots. The data is from Ref.~\cite{STAR:2021mii}.}
    \label{fig:iso_ratio_star}
\end{figure}

In 2018, the isobar collision campaign finally takes place. The data is publicly released following a three-year-long blind analysis \cite{STAR:2021mii}. The main results we are interested in reported by the STAR collaboration can be found in Fig.~\ref{fig:iso_ratio_star}.

The left-hand side of the figure shows the ratio of multiplicity distributions taken between \ruru{} and \zrzr{} collisions. As expected, the plot follows closely the multiplicity curves shown in Fig.~\ref{fig:1409}, where \ruru{} collisions correspond to the New parametrization (smaller $a$), while \zrzr{} collisions correspond to the Original parametrization (larger $a$). We see, in particular, that collisions of $^{96}$Ru nuclei have a lower probability for the lowest values of $N_{\rm track}$, while their probability is higher at intermediate and large multiplicity values. In agreement with the theoretical predictions, this phenomenon is immediately explained by the fact that $^{96}$Ru nuclei have a smaller $a$ parameter. The sharp rise in the ratio of $P(N_{\rm track})$ is also predicted by the theoretical calculations \cite{Li:2019kkh}. 

We move on, then, to the right-hand side of Fig.~\ref{fig:iso_ratio_star}, showing the isobar ratio of the anisotropic flow coefficients $v_2$ and $v_3$. We see distinct patterns in this plot. {\bf (i)} \ruru{} collisions have a larger elliptic flow in the central limit. This is consistent with the previous statement that this nucleus has a larger quadrupole deformation than $^{96}$Zr, which is in fact also suggested by low-energy nuclear structure experiments (Sec.~\ref{sec:Wimmer}). {\bf (ii)} Moving leftward from small to large centralities, the ratio of elliptic flows first decreases toward unity, and then increases again showing the characteristic bump structure predicted by theoretical calculations that implement a difference in skin thickness parameters (fully consistent, e.g., with the right-hand panel in Fig.~\ref{fig:1409}). {\bf (iii)} The isobar ratio of $v_3$ coefficients is completely nontrivial, with a very strong departure from unity even in high-multiplicity events, where the skin thickness has little impact. The $v_3$ of \zrzr{} collisions is indeed higher by about 10\% in the central limit. This was not predicted by any kind of calculations, and points to an effect that is akin to the presence of a large octupole deformation in the ground state of $^{96}$Zr. It is worth noting that, in the choice of species to use for the isobar collision campaign, the possibility of such an effect to manifest was not even remotely considered, even though it was known for a long time that the nucleus $^{96}$Zr has one of the lowest-lying $3^-$ states among all isotopes in the nuclear chart, indicating strong octupole collectivity \cite{ISKRA2019396}. As we shall see in the upcoming sections, very little is understood at present about this phenomenon.

In summary, the isobar collision campaign thus reveals an unprecedentedly powerful tool for obtaining precise signatures of the fine properties of the structure of the colliding nuclei. Arguably, even in the absence of a discovery of the CME, this finding represents one of the main legacies of the RHIC machine.

\subsubsection{Relation to a previous BNL Task Force}

It is insightful to compare this state of affairs with the discussion of an earlier BNL-based Task Force from 2016 appointed to deliver a recommendation concerning isobar collisions as means of searching for the CME. Their analysis and conclusions are in Ref.~\cite{Koch:2016pzl}. 

The BNL Task Force acknowledges at length that there is an issue associated with the highly uncertain knowledge of the quadrupole deformation of the $A=96$ isobars. Their assessment is that:
\begin{displayquote}
    \textit{\ldots the effect of the different possible $\beta_2$ sets is very small in mid-central collisions where the magnetic field is largest. Since the geometry of the nuclei only weakly affects the initial eccentricity and the magnetic field, it does not seem to represent a challenge to the success of the isobar program.}
\end{displayquote}
There is no mention of the seemingly large octupole collectivity in the ground state of $^{96}$Zr, nor of the more problematic issue related to the (entirely unknown) difference in skin thickness between these two isotopes. Similarly, with regard to the choice of isotopes to employ (the candidates are $^{96}$Ru-Zr, $^{124}$Xe-Ba, $^{130}$Ba-Te, $^{136}$Xe-Ce) they state:
\begin{displayquote}
\textit{ \ldots any of the four isobar pairs is similarly suited for the study and the deciding factor should be the practicality of using them in RHIC.}
\end{displayquote}
 In fact, along with this paper we argue that the choice of the isobar pair in the region $A=96$ turns out to be the most suboptimal for the pursuit of precision studies where the quantification of theoretical uncertainties related to the nuclear structure is essential.
 
Therefore, while the statements of the BNL Task Force accurately reflect the understanding and perceived impact of nuclear structure effects prior to 2017, a substantially refined assessment of the importance of these features is in order. Thanks to the dramatic insights brought by the successful isobar collision campaign of 2018, we are today able to provide a statement on this matter based on the judgment of both the high- and the low-energy nuclear physics communities. This is the goal of this document.  

\subsection{The RRTF and organization of the document}

These discussions should have clarified why a new Task Force was called in 2022, nine months after the publication of the experimental results. The crucial novelty of our effort is that about half of the Task Force is composed of low-energy nuclear physicists. \textbf{The EMMI Rapid Reaction Task Force (RRTF) \textit{``Nuclear Physics Confronts Relativistic Collisions of Isobars''} \cite{EMMI_RRTF_First,EMMI_RRTF_Second} represents the first official gathering where experts on nuclear structure and heavy-ion collisions collaborate with a common goal.} This report provides a summary of the results of the extensive discussions and computations of the RRTF. They give an overview of our current understanding of isobar collisions, of the involved isotopes, and of the connection between nuclear structure and heavy-ion collisions in general.

To start with, in Sec.~\ref{sec:2} we shall recall the foundational principles related to the implementation of the low-energy structure of nuclei in the simulation of high-energy nuclear processes. The remainder of the document is devoted to the following questions.
\begin{itemize}
    \item How much do we know about the low-energy structure of $^{96}$Ru and $^{96}$Zr?
    \item Within the hydrodynamic framework of heavy-ion collisions, is it possible to generate signals akin to those observed by the STAR collaboration by means of model features unrelated to nuclear structure? 
\end{itemize}
The first question is addressed in Sec.~\ref{sec:3}, while the second one is addressed in Sec.~\ref{sec:4} and Sec.~\ref{sec:5}. In Sec.~\ref{sec:6}, we summarize our results and comment on the new research avenues opened by isobar collisions as a tool for precision nuclear physics studies.

\newpage

\section{Nuclear structure input to high-energy collisions}

\label{sec:2}

\subsection{Foundational principles}

The current state-of-the-art understanding of the initial condition of heavy-ion collisions has emerged around the year 2005, with the discovery by the PHOBOS Collaboration of sizable elliptic-flow fluctuations in central \auau{} and \cucu{} collisions \cite{PHOBOS:2006dbo}. This led to the development of the Glauber Monte Carlo approach \cite{Miller:2007ri,Loizides:2014vua,dEnterria:2020dwq}, which we shall briefly review hereafter, according to which the relevant degrees of freedom for a nuclear collision at high energy are the positions of the nucleons that populate the colliding ions at the time of interaction. This picture became fully established, and in a sense unquestionable, in 2010 when Alver and Roland \cite{Alver:2010gr} pointed out that fluctuations induced by nucleon positions provide both a natural and a quantitative explanation for the sizable triangular flow observed at RHIC and the LHC.

This Glauber Monte Carlo picture, or ``Alver and Roland'' relies in essence on a high-energy position-space factorization enabled by a separation of scales between the internal dynamics of the colliding wave functions and the time scale of the interaction processes \cite{Gelis:2015gza}. For a nucleus, collective and single-particle excitations are at best of the order of $\mathcal{O}(10)$ MeV in energy, corresponding to timescales of order 10$^{-21}\,s$ \cite{RevModPhys.88.045004}. At ultrarelativistic speeds, the interaction between two nuclei takes place on a time scale of 0.1 fm/$c$ or so (10$^{-24}$s), that is, thousands of times faster.  Therefore, it is understood that the interaction process takes an instantaneous snapshot of the nucleon positions sampled from the ground-state wavefunction, say $|\Psi|^2$ \cite{Giacalone:2023hwk}. More specifically, each colliding nucleus is composed of a number of constituents, i.e., partons closely attached to the nucleon centers, which are flying parallel to each other along the beam direction. Once two nuclei prepared in this way overlap in the interaction point, collisions occur between individual partons (or nucleons). This provides a unique experimental tool for the imaging of the nuclear ground state, in a way that is akin to imaging methods recently developed for finite quantum systems in the context of ultra-cold atomic gases \cite{2021NatPh..17.1316G,Parish}, where quantum correlations among constituents are measured from \textit{images} of their positions \cite{Brandstetter:2023jsy,Xiang:2024isi,Yao:2024rew,deJongh:2024pmo}.

To understand the subsequent connection between QGP and nuclear structure, we start from the Glauber picture and then follow the calculations of Ref.~\cite{Giacalone:2023hwk}. We consider collisions at zero impact parameter. The nucleus participating in a high-energy collision is associated with a two-dimensional map of \textit{density} (e.g., of gluons) in the plane transverse to the beam direction, ${\bf x}=(x,y)$, where the center of coordinates, ${\bf x}=0$ corresponds to the center of mass of the nucleus. This map is commonly referred to as the \textit{thickness} function,
\begin{equation}
    t({\bf x}) = \sum_{i=1}^A g({\bf x}-{\bf r}_{i\perp}),
\end{equation}
where $A$ is the nuclear mass number, $g({\bf x})$ is a form factor associated with the density of matter (gluons) within nucleon $i$ (typically $g({\bf x})$ is a 2D Gaussian with a width, $w$, of order 0.5 fm), while ${\bf r}_{i\perp}$ is the transverse component of the three-dimensional coordinate of this nucleon, ${\bf r}_i=({\bf r}_{i\perp},r_z)$. The random variable in the problem is this set of coordinates ${\bf r}_{i\perp}$.

Now, we collide one nucleus with another one represented by a second thickness function, $t^\prime({\bf x})$, and consider a simple binary-collision-type Ansatz for the mid-rapidity energy density, $\epsilon({\bf x})$, at the initial time
\begin{equation}
\label{eq:tatb}
    \epsilon({\bf x}) = t({\bf x})t^\prime({\bf x}) = \sum_i \sum_j g({\bf x}-{\bf r}_{i\perp}) g({\bf x}-{\bf r}_{j\perp}).
\end{equation}
Note that the coordinates $r_i$ and $r_j$ belong to different nuclei and are therefore independent random variables. Observables are then constructed in the final state from averages that involve particles that are emitted after the expansion of this density profile.
 
 In heavy-ion collisions, observables such as elliptic and triangular flow are dominated by the geometric structure of the produced matter on large scales, e.g., its ellipticity or triangularity, which are lowest-order anisotropies in a Fourier expansion of the medium geometry \cite{Teaney:2010vd,Sousa:2024msh}. Event-by-event, the relevant statistical fluctuations of ellipticity and triangularity are largely encoded in the 1-point function (local average) and the 2-point function (local variance) of the density field $\epsilon({\bf x})$ \cite{Blaizot:2014nia,Gronqvist:2016hym}. Assuming Eq.~(\ref{eq:tatb}), and averaging over events, one obtains the following key result:
\begin{align}
\label{eq:1p2p}
\nonumber     \langle \epsilon({\bf x}) \rangle_{\rm ev} &=  \left ( A  \int_{{\bf r}_1} \rho^{(1)}({\bf r}_1) g({\bf x}-{\bf r}_{1\perp}) \right )^2 \\
\nonumber    \langle \epsilon({\bf x}) \epsilon({\bf y}) \rangle_{\rm ev} &= \biggl ( A \int_{{\bf r}_1} \rho^{(1)} ({\bf r}_{1}) g({\bf x}-{\bf r}_{1\perp})g({\bf y}-{\bf r}_{1\perp})\\
    &\hspace{30pt} + (A^2-A)\int_{{\bf r}_1 \, {\bf r}_2} \rho^{(2)} ({\bf r}_{1},{\bf r}_{2}) g({\bf x}-{\bf r}_{1\perp} )g({\bf y}-{\bf r}_{2\perp})  \biggr)^2,
\end{align}
and so on for higher-point functions. In these expressions, the averages over events correspond to statistical averages computed with respect to the (potentially $n$-body) probability density functions of the nucleon positions. In this model, nuclear structure is the only source of fluctuations, such that the latter are given by the $n$-body densities associated with the nuclear ground state
\begin{equation}
\rho^{(n)}({\bf r}_{1},\ldots,{\bf r}_{n})  \equiv \int_{{\bf r}_{n+1},\ldots,{\bf r}_A} | \Psi({\bf r}_{1},\ldots,{\bf r}_{A}) |^2 \, ,\hspace{40pt} \Psi({\bf r}_1, \ldots, {\bf r}_A) \equiv \langle {\bf r}_1, \ldots, {\bf r}_A | \Psi \rangle  .
\end{equation}
Note that the averages in Eq.~(\ref{eq:1p2p}) always imply a marginalization over the $z$ component of the nucleon coordinates, taken as the beam direction along which the nuclei are boosted.

We conclude that fluctuations and correlations in the final states of heavy-ion collisions are determined by fluctuations and correlations that are fundamentally governed by one- and many-body distributions of nucleons in the colliding ground states. Specifically in the case of interest for the present document, quantities such as mean squared eccentricities, which are reconstructed in the final states from two-particle correlation measurements, give access to the two-body density, $\rho^{(2)}({\bf r}_1, {\bf r}_2)$, of the nuclear ground state \cite{Duguet:2025hwi}. This is, ultimately, the information needed to understand the non-trivial observations made in isobar collisions.

\subsection{Density-based picture of the collisions}

\begin{figure}[t]
    \centering
    \includegraphics[width=0.45\linewidth]{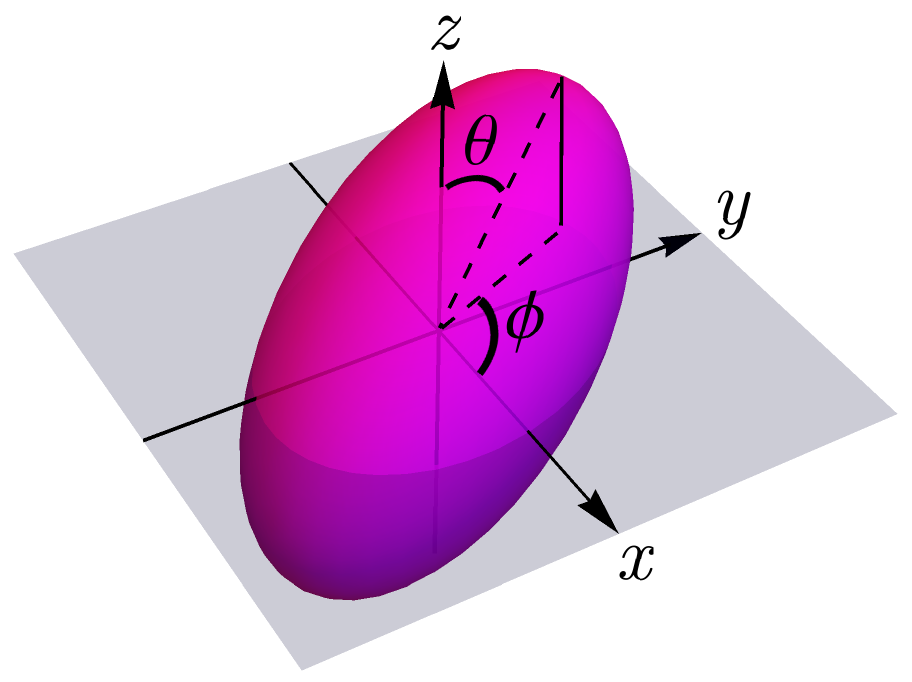}
    \caption{Illustration of an intrinsic nuclear density presenting an ellipsoidal deformation that is rotated with respect to the lab frame $(x,y,z)$ by a polar tilt, $\theta$, and an azimuthal spin, $\phi$. Figure from Ref.~\cite{Giacalone:2020awm}.}
    \label{fig:rhoO}
\end{figure}

One in general expects that a function such as $\rho^{(2)}({\bf r}_1, {\bf r}_2)$ presents a high degree of complexity, because atomic nuclei are strongly-correlated quantum systems characterized by up to $A$-body correlations of nucleons that show up in a wide spectrum of length scales, depending on the mass number. A powerful method to account for correlations of nucleons in the ground states is to introduce the notion of an intrinsic nuclear shape \cite{Verney:2025efj}. The nucleus is represented as a deformed, or clustered density of matter in a fictitious intrinsic nuclear frame (see Fig.~\ref{fig:rhoO} for an illustration) which is rotated in space with respect to the lab frame. The average over such orientations (symmetry restoration), a necessary step to evaluate lab-frame observables, then generates many-body spatial correlations of nucleons to all orders \cite{RevModPhys.75.121}. 

 For a detailed account of how such a picture can be more formally constructed, we refer the reader to the recent Ref.~\cite{Blaizot:2025scr}. Here, we illustrate this phenomenon through a simple application. We assume that the density of nucleons (both protons and neutrons) in the intrinsic frame of the nucleus is given by a Woods-Saxon distribution [${\bf r}=(x, y, z)$]:
\begin{equation}
\label{eq:Ws}
    \rho ({\bf r}) = \frac{1}{1+e^{\frac{r-R(\Theta)}{a}}},\hspace{30pt}r=\sqrt{x^2+y^2+z^2},
\end{equation}
where the nuclear surface is expanded in spherical harmonics and presents an axial \textit{quadrupole deformation} quantified by a $\beta_2$ parameter, that is,
\begin{equation}
    R(\theta) = R_0 [1 + \beta_2 Y_2^0 (\Theta)], \hspace{30pt} Y_2^0(\Theta) = \sqrt{\frac{5}{16\pi}} (3 \cos^2 \Theta - 1 ) = \sqrt{\frac{5}{16\pi}} \frac{3 z^2- r^2}{r^2}.
\end{equation}
Note that in the Woods-Saxon formula we align the intrinsic $(x,y,z)$ frame with the lab frame. The coordinate $z$, that is, the symmetry axis of the nucleus, is typically taken as the beam direction in a high-energy collider experiment. Now, we consider that the intrinsic density is rotated by some random set of Euler angles, $\vec \Omega$, in its intrinsic frame with respect to the lab frame,
\begin{equation}
    \rho_\Omega({\bf r}) ~\equiv ~ \rho({\bf r}) ~{\rm rotated~by~angles~\vec \Omega~in~space}.  
\end{equation}
An intrinsic deformed density with a quadrupole deformation that is rotated with respect to the lab-frame is depicted in Fig.~\ref{fig:rhoO}.
\begin{figure}[t]
    \centering
    \includegraphics[width=\linewidth]{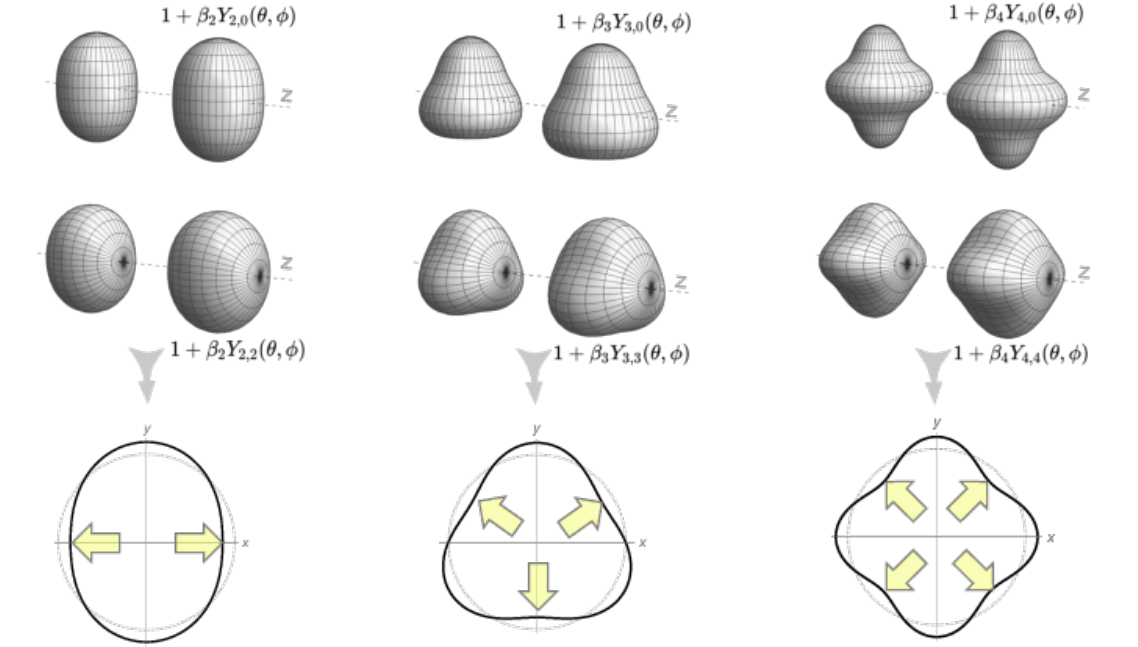}
    \caption{The figure illustrates nuclear collisions involving different types of deformation: quadrupole (left), octupole (center), and hexadecapole (right). Each case considers either the $Y_{n}^0$ mode (top row) or the $Y_{n}^n$ mode (middle row), with a fixed deformation parameter $\beta_n = 0.25$. Lorentz contraction along the $z$-axis is not shown. The bottom row represents the initial condition in the transverse plane of the quark-gluon plasma (QGP) formed after the collision. The arrows indicate the directions of maximal pressure gradients, along which the medium undergoes the most rapid expansion. This leads to the final-state harmonic flow coefficients $v_n$, which exhibit $n$-fold symmetry. Figure adapted from Ref.~\cite{Jia:2021tzt}.}
    \label{fig:2}
\end{figure}

The effect of many-body correlations in the evaluation of final-state observables emerges, then, from the average over all possible orientations of the intrinsic nuclear density. First, the spherical lab-frame one-body density (as in the case of the ground states of $J=0$ nuclei such as $^{96}$Ru or $^{96}$Zr) is obtained by performing the angular average:
\begin{equation}
\label{eq:rho1d}
    \rho^{(1)}({\bf r}) = \frac{1}{4\pi} \int_{\Omega} \rho_\Omega({\bf r})
\end{equation}
More important, though, nontrivial two-body correlations arise from this averaging process. In this classical model, the two-body density is given by
\begin{equation}
\label{eq:rho2d}
    \rho^{(2)}({\bf r}_1, {\bf r}_2) = \frac{1}{4\pi} \int_\Omega \rho_\Omega({\bf r}_1) \rho_\Omega ({\bf r}_2) .
\end{equation}
In the spherical limit with $\beta_2=0$, one recovers:
\begin{equation}
    \rho^{(2)}({\bf r}_1, {\bf r}_2)_{\beta_2=0} = \rho^{(1)}({\bf r}_1) \rho^{(1)}({\bf r}_2).
\end{equation}
However, as soon as $\beta_2 \neq 0$, genuine two-body correlations appear. Similarly, one can obtain the higher-order many-body densities. The notion of the shape generalizes beyond the quadrupole deformation and the $\beta_2$ parameter.  Within a given isotopic chain, the nuclear shape can change dramatically as a function of the nuclear mass number.

This discussion exhibits a connection between the many-body properties of the nuclear ground state and the notion of an intrinsic nuclear shape. Within the shape-based picture, the subsequent link to the geometry of the initial condition of the QGP is immediate\footnote{It was indeed pointed out already decades ago with the start of the BNL RHIC program \cite{Shuryak:1999by}.}. Once again, the idea is that the collision process is so fast that the orientations of the two incoming nuclei are frozen during their overlap. Depending thus on the orientation and deformation of the colliding ions, one can witness large-scale distortions of the region where the QGP is formed. This affects the subsequent hydrodynamic expansion (pressure gradient forces) and the observables in the final state \footnote{Reference~\cite{Kolb:2000sd} provides the first ever hydrodynamic simulation of this phenomenon.}. We refer to Fig.~\ref{fig:2} for an illustration of this effect.

In the context of isobar collisions at RHIC, the ``nuclear structure problem'' is that the intrinsic nuclear shapes that are probed in high-energy experiments cannot be directly measured by low-energy scattering experiments, but rather have to be evaluated within theoretical models (usually self-consistent mean-field calculations \cite{Ryssens:2023fkv,Yan:2024ggt}) that are tuned to low-energy nuclear structure data. This turns out to be especially problematic for the nuclei $^{96}$Ru and $^{96}$Zr, for which we have very little understanding, as discussed in the next section.

\newpage


\section{The intrinsic shapes of \texorpdfstring{$^{96}$Ru and $^{96}$Zr}{96Ru and 96Zr}}
\label{sec:3}

\subsection{Knowledge from low-energy experiments}
\label{sec:Wimmer}
\newcommand{\nuc}[2]{$^{#1}$#2}
\newcommand{\betwo}{$B(E2)$\xspace}
\newcommand{\bev}{$B(E2)$ value\xspace}
\newcommand{\bevs}{$B(E2)$ values\xspace}
\newcommand{\bevup}{$B(E2;\;0^+ \rightarrow 2^+)$ value\xspace}
\newcommand{\bevups}{$B(E2;\;0^+ \rightarrow 2^+)$ values\xspace}
\newcommand{\bevoup}{$B(E2;\;0_1^+ \rightarrow 2_1^+)$ value\xspace}
\newcommand{\bevoups}{$B(E2;\;0_1^+ \rightarrow 2_1^+)$ values\xspace}
\newcommand{\bevodo}{$B(E2;\;2_1^+ \rightarrow 0_1^+)$ value\xspace}
\newcommand{\bevodos}{$B(E2;\;2_1^+ \rightarrow 0_1^+)$ values\xspace}

\subsubsection{Spectroscopic probes of nuclear deformation}

Experiments at low excitation energy probe single-particle and collective degrees of freedom of atomic nuclei. The deformation of nuclei is quantified by the electric quadrupole moment of the ground state. One distinguishes the intrinsic and the spectroscopic quadrupole moment, the latter is measured in the laboratory system.
The nuclear spectroscopic electric quadrupole moment can be extracted from measurements of the hyper-fine splitting of the atomic levels. This requires however a nuclear total angular momentum $I>1/2$ and is thus not applicable to the even-even isotopes of interest \nuc{96}{Ru} and \nuc{96}{Zr}.
Assuming axial deformation, the intrinsic transitional quadrupole moment $Q_0$ can be extracted from the $B(E2)$ value
\begin{equation}
  e^2Q_0^2 = \frac{16\pi}{5} \frac{B(E2;\; I_\text{i},K \rightarrow I_\text{f},K)}{ \langle I_\text{i} K 2 0 | I_\text{f} K \rangle ^2} = \frac{16\pi}{5} \frac{\langle I_\text{f} || M(E2)|| I_\text{i}\rangle}{ \langle I_\text{i} K 2 0 | I_\text{f} K \rangle ^2}
  \label{eq:q0_be2}
  \end{equation}
where $K$ is the projection of the angular momentum $I$ onto the symmetry axis of the nucleus~\cite{bohr75}. $Q_0$ is positive for prolate and negative for oblate nuclei. It can be expressed in terms of the quadrupole deformation parameter $\beta_2$
\begin{equation}
  Q_0 = Z R_0^2 \frac{3}{\sqrt{5\pi}} \left( \beta_2 + \frac{2}{7} \sqrt{\frac{5}{\pi}} \beta_2^2\right) + \mathcal{O}(\beta_4^2,\beta_2\beta_4,\beta_2^3)
  \label{eq:Qbeta} 
\end{equation}
ignoring higher order deformations. The shape of a nucleus as viewed in the laboratory system is different and is expressed in the spectroscopic quadrupole moment $Q_S$:
\begin{equation}
  Q_S (I) = \sqrt{\frac{16\pi}{5}} \frac{\langle II20 | II\rangle }{\sqrt{2I+1}} \langle I ||E2|| I\rangle = \frac{3K^2 - I(I+1)}{(I+1)(2I+3)}Q_0 (I).
  \label{eq:spQmom} 
\end{equation}
$Q_S$ is zero for $I=0$ and therefore no measurements via atomic spectroscopy are possible for the even even nuclei \nuc{96}{Ru} and \nuc{96}{Zr}. Spectroscopic quadrupole moments and \bevs , while related to collectivity and shapes, do not give a model independent measure of the shape of the nucleus. For $K=0$, the signs of intrinsic and spectroscopic quadrupole moments are opposite. Experimentally the \bevs are accessed through measurements of the level lifetime and branching ratios, or through the excitation of the nucleus in an electromagnetic field, called Coulomb excitation. Coulomb excitation experiments below the Coulomb barrier are sensitive to both the transitional and diagonal $E2$ matrix elements and thus give good indications of the nuclear shape. Note that these quantities relate to the excited state, typically the first $2^+$ state, and that there is no a priori rule that the deformation of the ground state is identical.

Low-energy Coulomb excitation experiments also allow to determine the ground state shape in a model independent way from Kumar-Cline sum rules~\cite{kumar72,cline86}. This technique requires to sum over all $E2$ matrix elements connecting states, which is experimentally very challenging and the sums have to be truncated. The method makes use of rotationally invariant zero-coupled products, which give access to the shape through the intrinsic frame electric quadrupole moments, parameterized as $Q$ and $\delta$, as well as their variances which characterize the softness. $Q$ is then related to the magnitude of the deformation, while $\delta$ describes the triaxial degree of freedom. Moreover, the method also gives access to the variances and thus the degree of softness of the deformation.
\subsubsection{Nuclear structure in the \texorpdfstring{$A\sim 100$}{A~100} region}
Nuclei with mass number $A\sim 100$ are extremely interesting from a nuclear structure point of view. In this region a sharp transition between the sub-shell closures at neutron number $N=56$ and $58$ with spherical ground states and strongly deformed ground states at $N=60$ and further has been observed in Zr and Sr nuclei. 

\nuc{96}{Zr} at $N = 56$ and $Z = 40$ exhibits properties of a doubly-magic nucleus with a relatively high excited $2^+_1$ state.
\begin{figure}[t]
    \centering
    \includegraphics[width=\textwidth]{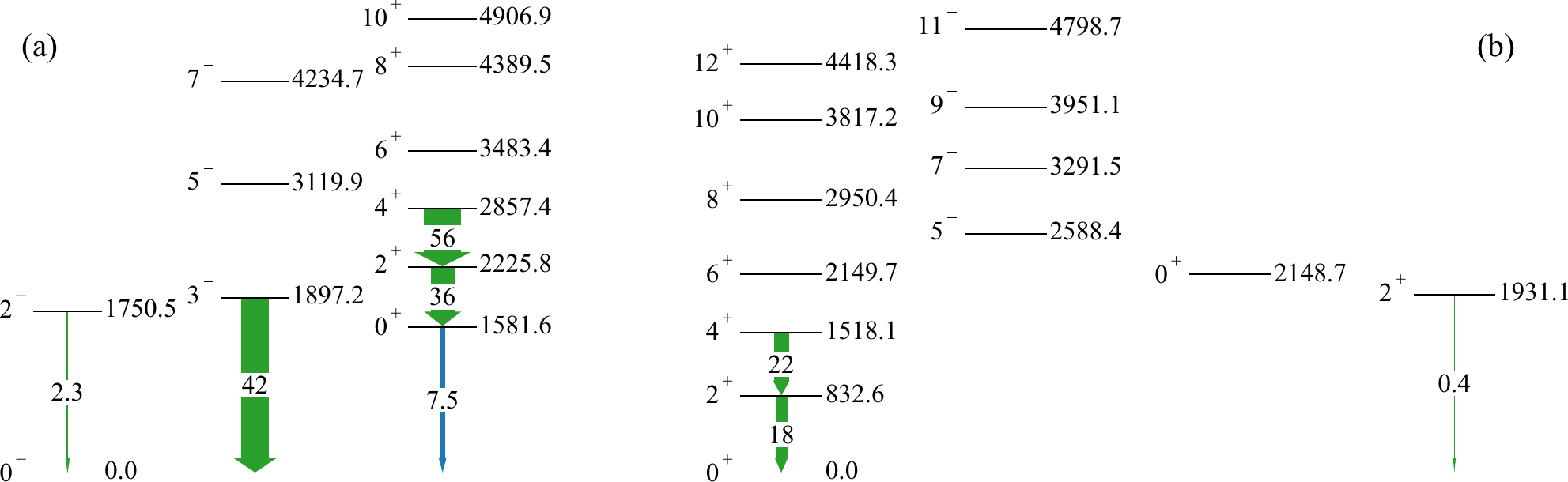}
    \caption{Level schemes of (a) \nuc{96}{Zr} and (b) \nuc{96}{Ru} at low excitation energy. States are labeled with their excitation energy in keV, known transitions are given in Weisskopf units for $E2$ and $E3$ transitions and electric monopole transition strength $\rho^2(E0)$ are given in $10^{-3}$. }
     \label{fig:level}
\end{figure}
For the ground state band, the evaluation~\cite{ensdf} only one measurement of the \bev~\cite{kumbartzki03} which determined a small value of $B(E2;\;2_1^+ \rightarrow 0_1^+)=2.3(3)$~W.u. from the measured lifetime $\tau(2^+) = 0.82(10)$~ps. The value, albeit with larger uncertainty has been more recently confirmed~\cite{peters13}. In an axial rotor model this translates to $\beta_2 = 0.061(3)$ consistent with the expectation of a spherical nucleus. The compilation cites also another measurement with a factor two higher value~\cite{gangrskii65} but the result is not adopted by the evaluators. The quadrupole moment of the $2^+_1$ state is not known, therefore detailed high statistics Coulomb excitation experiments are called for. Regarding the octupole degree of freedom, the $E3$ strength has been determined from the lifetime of the $3^-$ state~\cite{horen93} and its branching ratio to the ground and $2^+_1$ state~\cite{ISKRA2019396}. These together give a $B(E3;\;3^-_1 \rightarrow 0^+_1)=42(3)$~W.u. and assuming axial symmetry translate to $\beta_3=0.253(5)$. In the level scheme shown in Fig.~\ref{fig:level}, also the shape coexisting excited band is shown. Here the intra-band $B(E2)$ values are much more collective, the \bev of 36(11)~W.u. for the $2^+_2 \rightarrow 0^+_2$ transition~\cite{kremer16} corresponds to a deformation of $\beta_2=0.24(4)$. The deformed nature of this state is also highlighted by the rotational band built on top of it. The structure of \nuc{96}{Zr} at low energy is thus governed by the shape coexistence of two configurations with different shape. The rather small electric monopole strength, $\rho^2(E0;\;0^+_2\rightarrow 0^+_1) = 7.5\cdot10^{-3}$~\cite{burch72}, as well as the interband \bevs ($B(E2;\;2_2^+ \rightarrow 0_1^+)=0.26(8)$~W.u.~\cite{kremer16}) point to very weak mixing of these structures.

Turning now to \nuc{96}{Ru}, the structure changes completely. As shown in Fig.~\ref{fig:level}, the level scheme features a, slightly irregular, ground state rotational band, which extends up to angular momentum $J=18$. Lifetime measurement as well as Coulomb excitation measurements for transitions and states in the ground state band yield moderate values. The adopted $B(E2;\;2_1^+ \rightarrow 0_1^+)$ amounts to 18.2~W.u. a value almost a factor of ten larger than for the isobar \nuc{96}{Zr}. In an axial rotor model, this translates to a deformation parameter $\beta = 0.154$. Further evidence for deformation comes from the fragmentation of the single-particle strength observed in neutron adding and removing transfer reactions on \nuc{96}{Ru}, see e.g. Refs.~\cite{ball71,hollas77}. The electric quadrupole moment was determined from a Coulomb excitation experiment~\cite{landsberger80}, the value $Q=-0.13(9)$~$e^2$b$^2$ supports the moderate prolate deformation. Limited experimental information and sensitivity of the experiment prevented a more conclusive result on the nature of the deformation, prolate or oblate as well as the analysis of sum rules. High statistics experiments with this goal are planned in the future.

In summary, the two nuclei \nuc{96}{Ru} and \nuc{96}{Zr} exhibit very different properties at low excitation energy. The experimental data on \nuc{96}{Zr} is consistent with a spherical closed shell nucleus in the ground state, with a deformed coexisting excited configuration which mixes only weakly with the ground state. \nuc{96}{Ru} is less studied, in particular observables related to the $\gamma$ and octupole degrees of freedom are lacking in precision. For both isobars, future experiments will result in more precise extraction of the deformation parameters to be able to compare with the results of the analysis of heavy ion collisions at relativistic energies.

\subsection{Mean field results with covariant energy density functionals}

\label{CDFT-MF}

    The detailed investigation of octupole deformation in the mass regions of interest has 
been performed within the covariant density functional theory (DFT)  at the mean field
level with four covariant energy density functionals (EDF) in Refs.\ \cite{AAR.16,CAANO.20}. 
None of  these studies show the presence of static octupole deformation in $^{96}$Zr. It is 
important to mention that no static octupole deformation has also been found in this nucleus 
in the mean field calculations with six Skyrme energy density functionals (EDF) (Ref.\ 
\cite{EN.17,CAANO.20}) and Gogny D1S EDF (Ref.\ \cite{RB.11}) as well  in the mean-field 
calculations based on the microscopic+macroscopic method (Ref.\ \cite{Moller:2015fba}).

   These calculations reveal that at mean field level 
there are differences in model predictions even for the nuclei located in well established 
regions of octupole deformation such as lanthanides with $Z\approx 60, N\approx 88$ and 
actinides with $Z\approx 90, N\approx 134$ (see Fig. 4 in Ref.\ \cite{CAANO.20}). For 
example, a shift in the position of the regions of the nuclei with octupole deformation
(by two to four neutron numbers) 
is seen when comparing the results of Skyrme EDFs and covariant EDFs. Note that the location of 
the regions of strong octupole collectivity and their size in $(Z,N)$ are defined by the presence 
of close-lying proton and  neutron shells with $\Delta l = \Delta j = 3$ (see Ref.\ \cite{BN.96}).  
Thus, these theoretical uncertainties in the predictions of octupole deformed nuclei are traced 
back to underlying  single-particle structure.  Indeed there are some differences in the predictions of 
the single-particle structure not only between different classes of the models but also between 
different functionals for a given type of DFT. This is systematically studied for quadrupole deformed 
nuclei in actinides in Skyrme, Gogny and covariant DFTs in Refs.\ \cite{DABRS.15,AS.11}. The 
observed differences in the predictions of octupole  deformed nuclei indicate that similar 
uncertainties in the underlying single-particle structure exist also in this type of nuclei.

   Based on the analysis of the results presented in Refs.\ \cite{AAR.16,Moller:2015fba,CAANO.20} one 
can conclude that theoretical uncertainties in the prediction of octupole collectivity are especially large 
for the nuclei which are characterized by the potential energy surfaces which are
soft in octupole deformation. For such nuclei, moderate changes in underlying
single-particle structure can either eliminate or create octupole deformation.
Existing results suggest  that the $^{96}$Zr nucleus belongs to this group of nuclei. 
It is important to remember that the accounting of correlations beyond mean 
field will not completely eliminate these uncertainties.

   In such a situation, it is  important to search for experimental fingerprints which would 
undeniably indicate the presence of octupole deformation. In even-even nuclei, there are 
two such fingerprints, namely, significantly enhanced (as compared with neighboring nuclei) 
strength of the $B(E3, 3^- \rightarrow 0^+)$ transition rate and the formation of 
alternating parity rotational bands the rotational states of which are connected by strong 
E1 transitions (see Ref.\ \cite{BN.96}). No such band built on the ground state has been 
observed in $^{96}$Zr (see Refs.\ \cite{96Zr-endsf,kremer16,ISKRA2019396}).  In transitional
nuclei the strength of the E3 transition is affected by the possibility that the ground  $0^+$
and excited $3^-$ states can have different  ($\beta_2, \beta_3$) deformations 
\cite{Rohozinski_1988,BN.96}. In addition, there is no a simple experimental procedure 
to verify whether these states have equal deformations or not. Note that octupole deformation 
$\beta_3$ of the  ground state  can be extracted from $B(E3, 3^- \rightarrow 0^+)$ only 
assuming equal deformations of the $0^+$ and $3^-$ states (see Refs.\ 
\cite{Rohozinski_1988,BN.96}).   The beyond mean field projection-after-variation calculations 
of $^{96}$Zr based on multidimensionally-constrained relativistic Hartree-Bogoliubov approach 
suggests that the deformations of these two states are very similar (see Ref.\ \cite{RWLY.23}).
However, these calculations overestimate the excitation energy of the $3^-$ state by a factor 
of $\approx 2.5$.  Moreover, this is the first application of this approach and thus neither
reliability of this model in the reproduction of the experimental data nor related theoretical
uncertainties are well understood and quantified.

    Because of the time scale of the collision process only the deformation of the ground state
is relevant in high-energy processes. At present, only the charge radii of the ground states can provide such 
information in undisturbed way since this observable is not dependent on the properties of 
excited nuclear states (like in the case of the $B(E3, 3^- \rightarrow 0^+)$ transition rate). In 
particular, the differential mean-square charge radii defined as 
\begin{eqnarray}
\delta \left<r^2\right>_p^{N,N'} =r^2_{ch}(N)-r^2_{ch}(N'), 
\end{eqnarray}
where $N'$ in the neutron number of reference nucleus and $r_{ch}(N)$ is the
charge radius of the nucleus with neutron number $N$, provide a very sensitive
measure of the evolution of nuclear shapes in isotopic chains (see Ref.\ \cite{PAR.21}
and references therein). They are measured by means of precise laser 
spectroscopy \cite{CMP.16} which does not excite the nucleus from the ground 
state.
\begin{figure}[t]
\centering
\includegraphics[width=.6\linewidth]{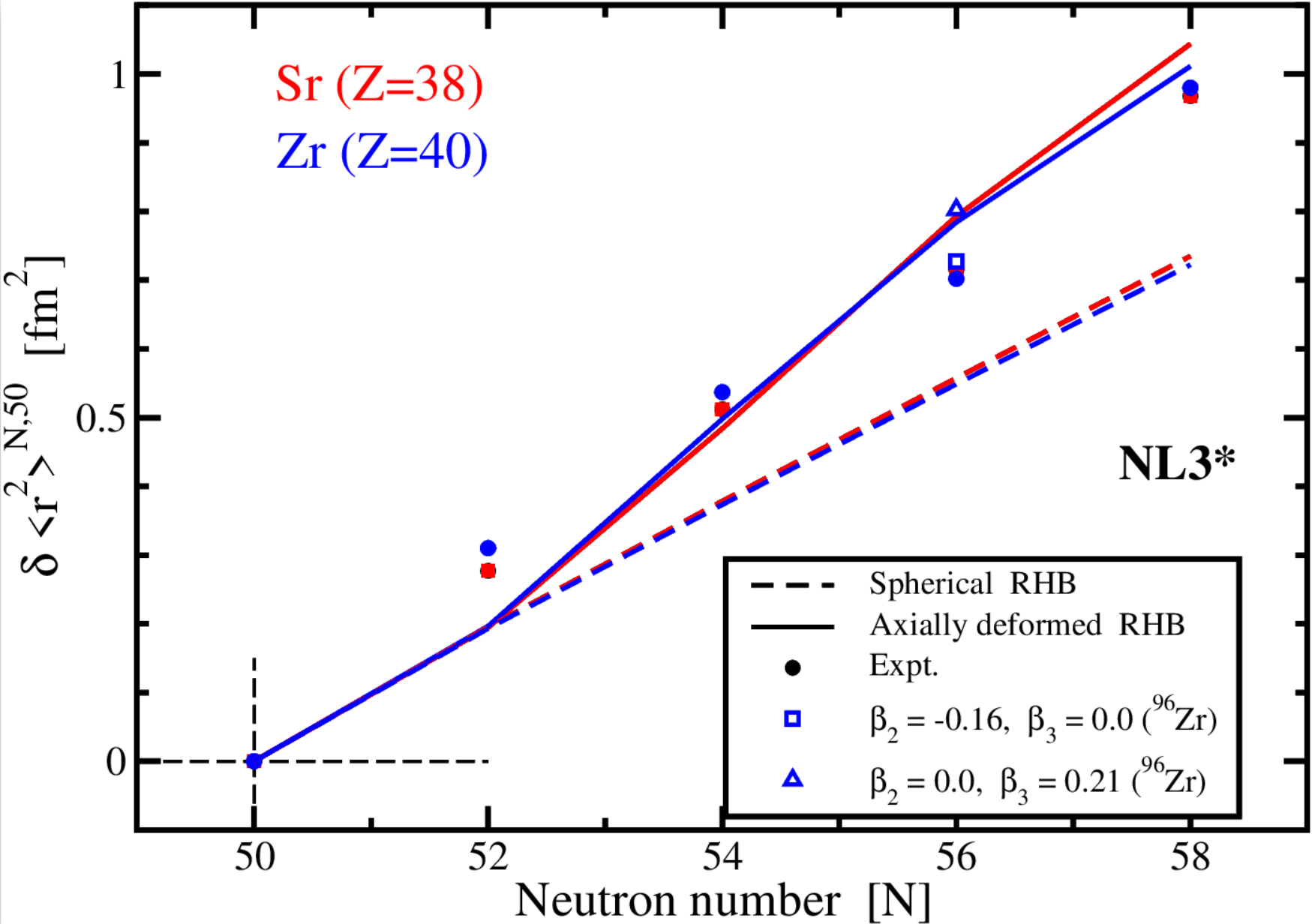}
\caption{The comparison of experimental and calculated differential charge radii of the $N=50-58$
even-even Sr and Zr isotopes. The experimental data are taken from Ref.\ \cite{Zr-radii-PRL-02}
and shown by solid symbols.  The results of spherical and axially deformed RHB calculations are 
shown by dashed and solid lines, respectively.   Red and blue colors are used for Sr and Zr results, 
respectively.
\label{fig-dif-radii}
}
\end{figure}

    The experimental values of  $\delta \left<r^2\right>_p^{N,N'}$ for the $N=50-58$
isotopes in the Sr and Zr isotopic chains are shown in Fig.\ \ref{fig-dif-radii}. They are very similar 
for a given value of neutron number $N$ which suggest very similar deformed shapes
for the  respective Sr and Zr isotopes.  This similarity is rather well reproduced in the
relativistic Hartree-Bogoliubov (RHB)  calculations with the NL3* covariant EDF, but the 
calculations fail to reproduce the transition to weakly deformed
shapes at $N=52$ and somewhat overestimate the magnitude of oblate deformation
in the $N=56$ and $N=58$ Sr and Zr isotopes.  Note that these calculations are
restricted to axial reflection symmetric shapes (no octupole and higher order
odd-multipolarity deformations) and that detailed analysis of differential charge radii 
of the Sr isotopes with five different covariant EDFs is carried 
out in Ref.\ \cite{PAR.21}.  In agreement with experiment, the differential charge radii 
of the $N=56$ Sr and Zr isotopes are very similar in the RHB calculations with these 
functionals.

    It is interesting to compare these results with those obtained in Ref.\  \cite{RWLY.23} 
with PC-PK1 covariant EDF.  These calculations predict $(\beta_2 =-0.16, \beta_3=0)$ and 
$(\beta_2 =0.0, \beta_3=0.21)$ deformations for the ground state of $^{96}$Zr at and beyond 
mean field level, respectively. The differential charge radii for this nucleus obtained with these 
deformations are shown by open orange square and open orange triangle in Fig.\ 
\ref{fig-dif-radii}. One can 
see that the former result comes very close to experimental data. Note that the RHB calculations
with NL3* bring somewhat larger  $\delta \left<r^2\right>_p^{N,N'}$ value because of
larger oblate deformation $\beta_2 \approx -0.18$.  However, the solution with 
$(\beta_2 =0.0, \beta_3=0.21)$ brings  $\delta \left<r^2\right>_p^{N,N'}$ value which
is by $\approx 0.1$ fm$^2$ larger than the one corresponding to oblate solution in 
Ref.\ \cite{RWLY.23}.  This is significantly larger than the difference between experimental
differential mean-square charge radii of the $N=56$ Sr and Zr isotopes. Considering
that the deformations of the $N=56$ Sr and Zr isotones are similar at the mean field
level, the reproduction of near equal values of differential charge radii seen in experiment
would most likely require the presence of octupole deformation also in $^{94}$Sr. However,
to our knowledge, there are no experimental indications of strong octupole collectivity in the 
ground state of  $^{94}$Sr. Thus, it remains to be seen whether the calculations beyond 
mean field can reproduce the similarity of differential charge radii of these two isotones 
assuming  quadrupole and octupole deformed shapes in $^{94}$Sr and $^{96}$Zr, respectively.

\subsection{Mean field results based on Skyrme-type EDFs}

Among energy density functionals (EDF), the Skyrme EDF is arguably the most 
widely used~\cite{RevModPhys.75.121}, in part because of its low computational
complexity compared to other types of EDFs. Here we present the results of 
mean-field Hartree-Fock-Bogoliubov (HFB) calculations for both $^{96}$Zr 
and $^{96}$Ru using 17 different Skyrme parameterizations of the traditional 
form: SLy4/6~\cite{chabanat1997,chabanat1998,chabanat1998a}, 
SV-mas07/mas08/min/mas10~\cite{klupfel2009}, the SLy5s1-8~\cite{jodon2016} 
and BSkG1/2/3~\cite{scamps2021, ryssens2022a,grams2023}. Although the form of
the underlying EDF is not exactly identical in all cases, the differences are 
not relevant for the purpose of this study. As we will see, 
there is an important aspect of BSkG1/2/3 that separates them from 
the rest: where all other models we consider were intended for use in pure mean-field 
calculations, the BSkG models include a phenomenological correction for spurious 
collective motion that aims to mimic the effect of beyond-mean-field corrections
on the nuclear binding energy without incurring their enormous computational cost. 

We use the \texttt{MOCCa} code~\cite{RyssensThesis} to solve the self-consistent 
Skyrme-HFB equations, a tool that is particularly adept for the task at hand due to its 
flexibility in terms of the geometry of the nuclear configurations and its 
coordinate space representation of the single-particle wavefunctions. The 
latter is also the reason why this survey does not include other widely-used 
models such as the UNEDF-series~\cite{kortelainen2010}; We limit ourselves to 
EDF parameterizations that either include pairing terms optimized using a similar
numerical representation (the SV- and BSkG-series) or for which default values
are available (SLy4/6 and the SLy5s1-8)~\cite{ryssens2019}. This selection of models
thus certainly does not exhaust the rich literature on the subject, but it does 
include both extremely widely used parameterizations and very recent efforts 
of several groups, offering us an idea on the range of predictions for structure of the isobars.

\subsubsection{Multipole moments}
Before discussing the EDF results, let us recall the precise definition of 
the multipole moments used to characterize the deformation of a nucleus in both EDF-based approaches and low-energy nuclear
experiments. For a pair of integers $(\ell, m)$ that satisfy $\ell \geq m \geq 0$, 
the dimensionless multipole moment $\beta_{\ell m}$
is defined as
\begin{align}
\beta_{\ell m} = \frac{4 \pi }{ 3 R^{\ell}_0 A^{\ell/3+1} }\int d^3 r \rho_0(\mathbf{r}) \text{Re} \left[Y_{\ell m}(\theta, \phi)\right] \, , 
\label{eq:betalm}
\end{align}
where $R_0 = 1.2$ fm, $\rho_0(\mathbf{r})$ is the \emph{matter} density and $Y_{\ell m}$ is 
a spherical harmonic\footnote{Technically speaking, the imaginary parts of the
spherical harmonics could also play a role. Such deformation modes, to 
the best of our knowledge, have never been shown to have a meaningful impact on
nuclear ground states. They could be included by extending the
definition Eq.~\eqref{eq:betalm} to include negative values of $m$.}.
Despite the degeneracy in notation, the deformation parameters typically used 
to generate Woods-Saxon (WS) densities for the purposes of simulations of heavy
ion collisions are \emph{not equivalent} to these multipole moments: inserting
a WS form into Eq.~\eqref{eq:betalm}, one finds that the resulting 
$\beta_{\ell m}$ are different from the deformation parameters 
$\beta^{\rm WS}_{\ell m}$ used to generate the nuclear density. The difference
between both notions of deformation is small when (i) the WS parameters
are small and (ii) not too many deformation modes coexist. When one or both of 
these assumptions do not hold, one should exercise care when comparing the 
nuclear deformation extracted from simulations of heavy-ions collisions that rely
on WS parameterizations and low-energy theory or experiment 
that employ Eq.~\eqref{eq:betalm}. For $^{238}$U for example, the large 
quadrupole deformation and coexisting hexadecapole deformation induce a 
substantial difference $\beta_{2 0} - \beta^{\rm WS}_{2 0} \approx 0.04$
which impacts the description of central U+U collisions~\cite{Ryssens:2023fkv}.
\begin{figure}[t]
\centering
\includegraphics[width=.55\textwidth]{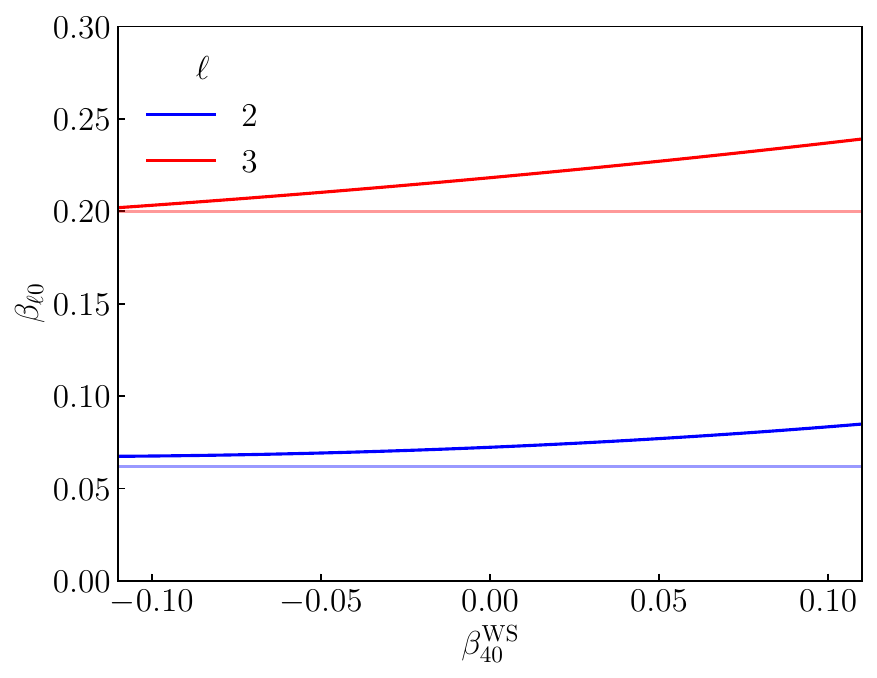}
\caption{ Multipole deformations $\beta_{20}$ and $\beta_{30}$ for a WS 
          density for $^{96}$Zr with $\beta_{20}^{\rm WS} = 0.062, \beta_{30}^{\rm WS} = 0.2$
          and varying hexadecapole deformation. The faint horizontal lines
          indicate $\beta_{\ell 0} = \beta_{\ell 0}^{\rm  WS}$
           to guide the eye.
        }
\label{fig:WSdemonstration}        
\end{figure}

Although the deformations involved are smaller than those of $^{238}$U, the 
study of the $A=96$ isobars is not immune to this subtle issue. We provide 
Fig.~\ref{fig:WSdemonstration} as an illustration: it shows the values of $\beta_{20}$ and 
$\beta_{30}$ obtained by numerically integrating Eq.~\ref{eq:betalm} for a 
WS density of $^{96}$Zr for various values of the hexadecapole deformation 
parameter $\beta_4^{\rm WS}$ but taking $\beta_{20}^{WS} = 0.062$ and 
$\beta_{30}^{\rm WS} = 0.2$ as in Ref.~\cite{Zhang:2021kxj}. It is immediately clear
that the quadrupole and octupole moments are only equal to the corresponding WS
parameters $\beta_{\ell 0}^{\rm WS}$ for large negative values of the 
hexadecapole deformation. For vanishing hexadecapole moment, i.e. when taking 
exactly the WS density of Ref.~\cite{Zhang:2021kxj}, we find $\beta_{30} - \beta^{\rm WS}_{30} = 0.018$. 
If one includes other deformation modes such as $\beta_{22}$ or $\beta_{32}$
that are discussed elsewhere in this report, the situation becomes even more
complicated and the differences between both notions of deformation can 
potentially become larger. Although corrections to the nuclear deformation on 
the order of 10\% make little difference for qualitative conclusions, 
accounting for them becomes important when aiming at a quantitative 
comparison between high- and low-energy nuclear physics.

\subsubsection{Pure mean-field models}

The mean-field minima obtained with SLy4, the SV- and SLy5sX-series are axially
symmetric: all multipole moments $\beta_{\ell m}$ vanish when $m\not = 0$. We list
the corresponding quadrupole and octupole deformations in 
Tab.~\ref{tab:deformations}\footnote{We remind the reader that, although they
are not listed explicitly here, higher order multipole moments with $\ell > 3$
are naturally included in self-consistent approaches. The non-spherical 
configurations in Tab.~\ref{tab:deformations} generally have small but non-vanishing
hexadecapole deformations for instance.
}. The predictions can be grouped by model family:
(i) SLy4/6 predict spherical configurations for both isobars; 
(ii) the SV-series predict finite octupole deformation for $^{96}$Zr with 
     (nearly) vanishing quadrupole deformation, while $^{96}$Ru is reflection
     symmetric with moderate prolate quadrupole deformation; 
(iii) the SLy5sX predict a spherical $^{96}$Ru and an oblate $^{96}$Zr without
octupole deformation. 

\begin{figure}[t]
\centering
\includegraphics[width=.95\textwidth]{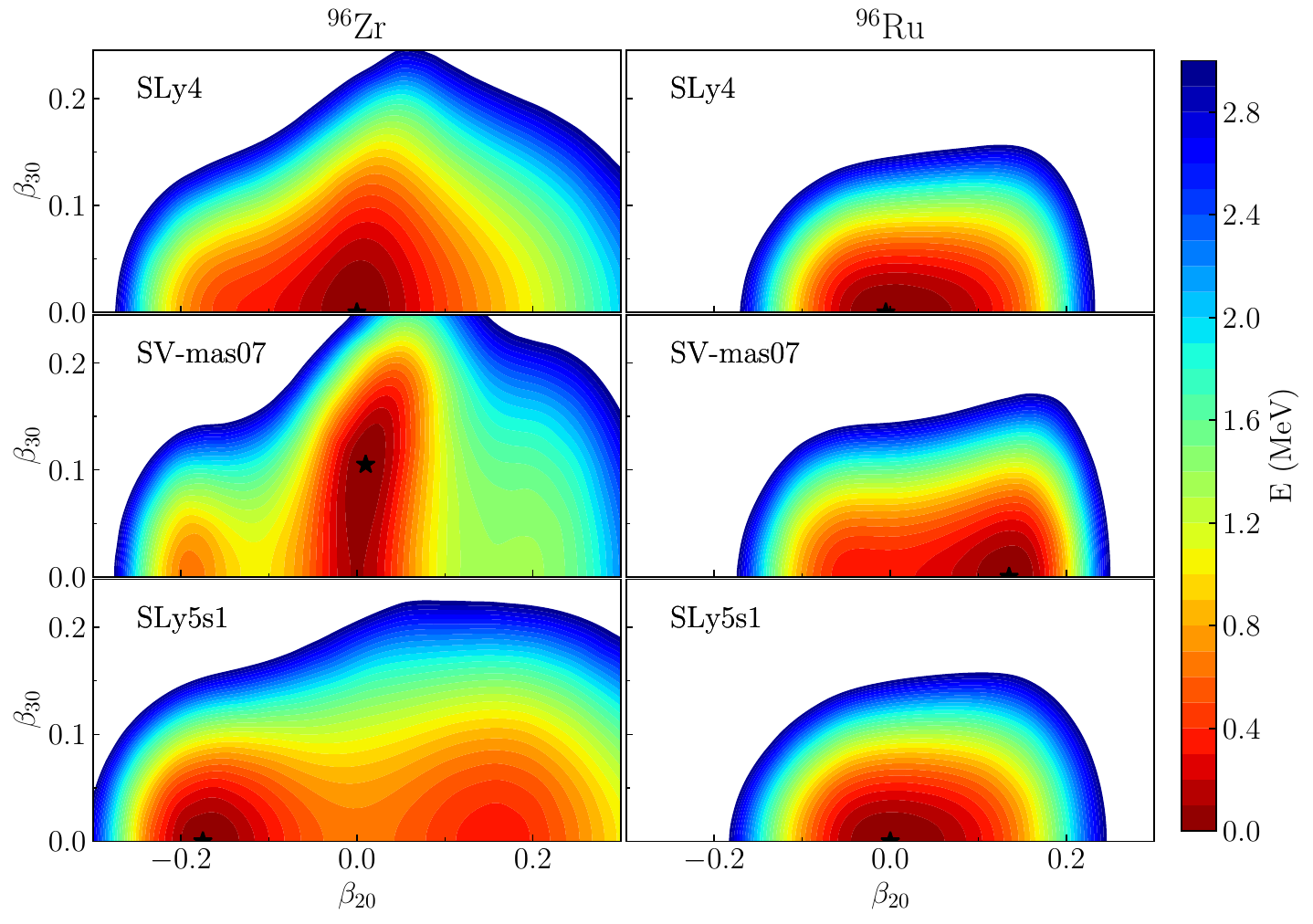}
\caption{Hartree-Fock-Bogoliubov total energy surfaces as a function of 
         quadrupole ($\beta_{20}$) and octupole ($\beta_{30}$) deformation
         for $^{96}$Zr (left column) and $^{96}$Ru (right column), as obtained with 
         the SLy4~\cite{chabanat1998} (top row), SV-mas07~\cite{klupfel2009} 
         (middle row) and SLy5s1~\cite{jodon2016} (bottom row) 
         parameterizations of Skyrme's EDF. 
         All panels are normalized to their respective minima (indicated by 
         black stars) and are symmetric under reflection about the horizontal
         axis.
        }
        \label{fig:2DPES}
\end{figure}

For illustrative purposes, we show the total energy surfaces for both isobars 
obtained with one model out of each family in Fig.~\ref{fig:2DPES}. What is most
striking about Fig.~\ref{fig:2DPES} is the extreme softness of the SV-mas07 
energy of $^{96}$Zr in the octupole direction near the spherical saddle point: the latter lies less than 
100 keV above the octupole deformed minimum near $\beta_{30} \sim 0.1$. 
This octupole softness near the spherical point for is also present, though 
less pronounced, for SLy4. The situation is somewhat different for SLy5s1: 
although the surface is somewhat soft in octupole deformation near both the 
prolate and oblate minima, it gets stiffer near zero quadrupole deformation.
The general octupole softness of the left column of fig.~\ref{fig:2DPES} is 
one indication that a pure mean-field description is of limited use to match
the details of nuclear structure for these isobars.

Of all models, SV-mas07 provides the best match with the analysis of isobar
collisions: octupole deformation in $^{96}$Zr with small quadrupole deformation 
and a prolate, reflection symmetric $^{96}$Ru. The reason for the comparative 
success of the SV-mas family is likely to be found in the details of its shell 
structure; the energy spacing of single-particle orbitals of different parity is
a primary (but not sole) ingredient leading to finite octupole deformation~\cite{chen2021}. 
A detailed investigation of shell structure is outside the scope of this 
contribution, but we note the trend of increasing $\beta_{30}$ among the 
the SV-mas07/08/10 parameterizations. These parameterizations were constructed
with a constraint on the effective mass, $m^*/m = 0.7,0.8$ and $1.0$, such 
that (within a given series of models) octupole deformation decreases with 
increasing effective mass. A similar trend for a different set of parameterizations
was already noted in Ref.~\cite{DaCosta23}.

Ignoring for now the subtleties discussed in the previous subsection, the predicted octupole deformation 
($\beta_{30} = 0.1$) remains modest compared to the value proposed by the 
authors of Ref.~\cite{Zhang:2021kxj}($\beta^{\rm WS}_{30} = 0.2$). It seems unlikely 
that any reasonable pure mean-field based model would be able to achieve a mean-field minimum with $\beta_{30} \sim 0.2$; global surveys with 
Skyrme~\cite{grams2023}, Gogny~\cite{RB.11} or relativistic~\cite{AAR.16} 
EDFs find at most $\beta^{\rm max}_{30} = 0.17-0.19$ on the mean-field level and 
reach such values only for (super)heavy nuclei. The very size of the octupole
deformation suggested by isobar collisions is thus an indication that a mean-field
description of these nuclei is likely insufficient.

\begin{table}
\centering
\begin{tabular}{|l|d{2.2}d{2.2}d{2.2}d{2.2}|l|d{2.2}d{2.2}d{2.2}d{2.2}|}
\hline
                 & \multicolumn{2}{c}{$^{96}$Zr} 
                 & \multicolumn{2}{c|}{$^{96}$Ru}
                 & 
                 &  \multicolumn{2}{c}{$^{96}$Zr} 
                 & \multicolumn{2}{c|}{$^{96}$Ru}\\
Model            & \multicolumn{1}{c}{$\beta_{20}$} 
                 & \multicolumn{1}{c}{$\beta_{30}$} 
                 & \multicolumn{1}{c}{$\beta_{20}$} 
                 & \multicolumn{1}{c|}{$\beta_{30}$} 
                 & Model 
                 & \multicolumn{1}{c}{$\beta_{20}$}
                 & \multicolumn{1}{c}{$\beta_{30}$} 
                 & \multicolumn{1}{c}{$\beta_{20}$} 
                 & \multicolumn{1}{c|}{$\beta_{30}$}  \\
\hline
SLy4     & 0.0    & 0.0  & 0.0  & 0.0 & SLy5s2 & -0.17 & 0.0 & 0.0 & 0.0\\
SLy6     & 0.0    & 0.0  & 0.0  & 0.0 & SLy5s3 & -0.17 & 0.0 & 0.0 & 0.0 \\
SV-mas07 & 0.01   & 0.10 & 0.14 & 0.0 & SLy5s4 & -0.18 & 0.0 & 0.0 & 0.0\\
SV-mas08 & 0.01   & 0.09 & 0.08 & 0.0 & SLy5s5 & -0.17 & 0.0 & 0.0 & 0.0\\
SV-mas10 & 0.0    & 0.05 & 0.0  & 0.0 & SLy5s6 & -0.17 & 0.0 & 0.0 & 0.0\\
SV-min   & 0.0    & 0.0  & 0.10 & 0.0 & SLy5s7 & -0.17 & 0.0 & 0.0 & 0.0 \\
SLy5s1   & -0.17  & 0.0  & 0.0  & 0.0 & SLy5s8 & -0.18 & 0.0 & 0.0 & 0.0\\
\hline
\end{tabular}
\caption{Quadrupole ($\beta_{2 0}$) and octupole ($\beta_{3 0}$) multipole 
         moments of $^{96}$Zr and $^{96}$Ru obtained with multiple pure mean-field
         models (see text).
        }
\label{tab:deformations}
\end{table}

\subsubsection{The BSkG series}

The focus of the BSkG series of models is providing data for astrophysical
applications; since binding energies set the energy scale of all
nuclear reactions, these models are carefully adjusted to essentially all known
nuclear masses. In order to reproduce as best as possible thousands of masses, 
it is necessary to correct for several effects that are impossible to describe 
on a pure mean-field level, see Ref.~\cite{scamps2021}. 
Most important for the discussion at hand is the rotational correction that 
(at least qualitatively) removes the contribution of the spurious rotational 
motion of a deformed mean-field state from the nuclear binding energy. Models 
with such correction will favor deformed minima over spherical ones and 
configurations with triaxial deformation over ones with axial symmetry, as do 
beyond-mean-field approaches that remove the rotational energy by means 
of explicit symmetry restoration~\cite{bender2008b}.

The supplementary material of the BSkG publications~\cite{scamps2021,ryssens2022a,grams2023} 
includes global predictions of ground state deformation. Due to the presence of 
the rotational correction, the predicted ground state configurations for both isobars are no longer axially 
symmetric. In such cases, one can no longer describe the nuclear shape just 
with $m=0$ values and one employs the traditional ($\beta_2, \gamma$)
characterization of triaxial quadrupole deformation:
\begin{align}
\beta_2 = \sqrt{\beta_{20}^2 + 2 \beta^2_{22}} \, , \qquad \gamma = \atan \left( \sqrt{2} \beta_{22}/ \beta_{20}\right)\, .
\end{align}
The ground state deformation of $^{96}$Zr is essentially consistent across all three models, 
with $\beta_2 \sim 0.21$ and $\gamma \sim 30^{\circ}$, i.e. maximally triaxial. 
The tables also consistently list a smaller deformation for $^{96}$Ru,
$\beta_{2} \sim 0.11$, but the three models span between $3$ and $22$ degrees
for $\gamma$. 

However, the original searches for the mean-field minima in 
Refs.~\cite{scamps2021,ryssens2022a,grams2023} were not entirely general: 
for computational reasons the authors only considered configurations with 
$\gamma \in [0^{\circ},60^{\circ}]$. When time-reversal and reflection symmetry
is assumed, there is a six-fold degeneracy due to the freedom of labeling 
Cartesian axes such that such a restriction does not impact the generality of 
the calculation. This is the reason why a significant part of the literature 
discusses plots that are limited to one sextant of the ($\beta, \gamma$)-plane.
This degeneracy is partially lifted when considering reflection-asymmetric
configurations: a non-zero value of an octupole deformation $\beta_{30}$
defines a preferred direction in space such that the remaining degeneracy is
only two-fold and one has to consider a larger interval of $\gamma$. 
\begin{figure}[t]
\centering
\includegraphics[width=.95\textwidth]{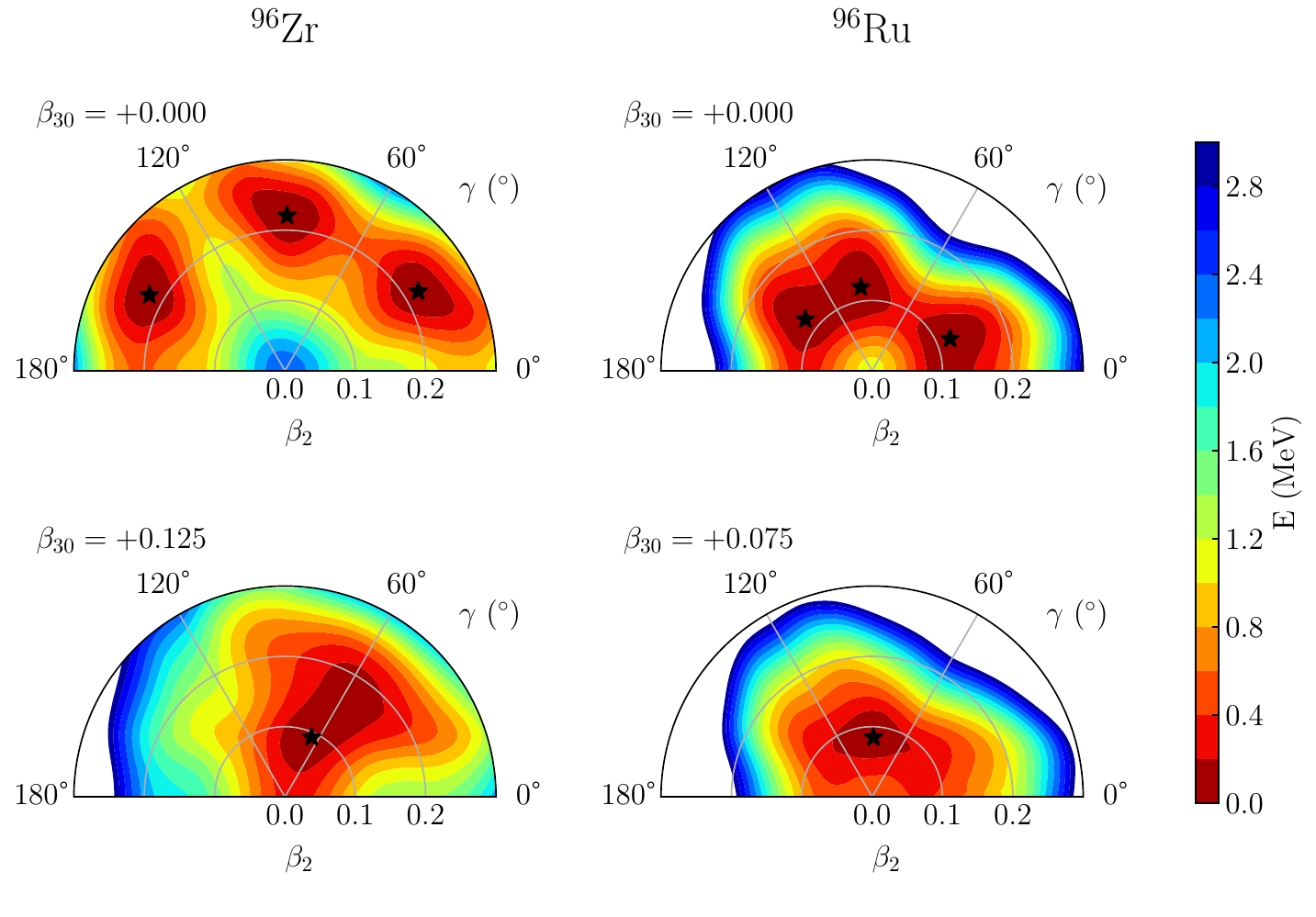}
\caption{ Total energy surfaces as a function of $\beta_2$ and $\gamma$ for
          $^{96}$Zr (left column) and $^{96}$Ru (right column) at fixed 
          values of $\beta_{30}$ as obtained with BSkG2. Top row: calculations
          for reflection symmetric configurations, $\beta_{30} = 0.0$. Bottom
          row: calculations at the value of $\beta_{30}$ corresponding to the
          global minimum. All panels are normalized to the corresponding minima,
          which are indicated by black stars. }
\label{fig:BSkGfull}
\end{figure}

\begin{figure}[t]
\centering
\includegraphics[width=.9\textwidth]{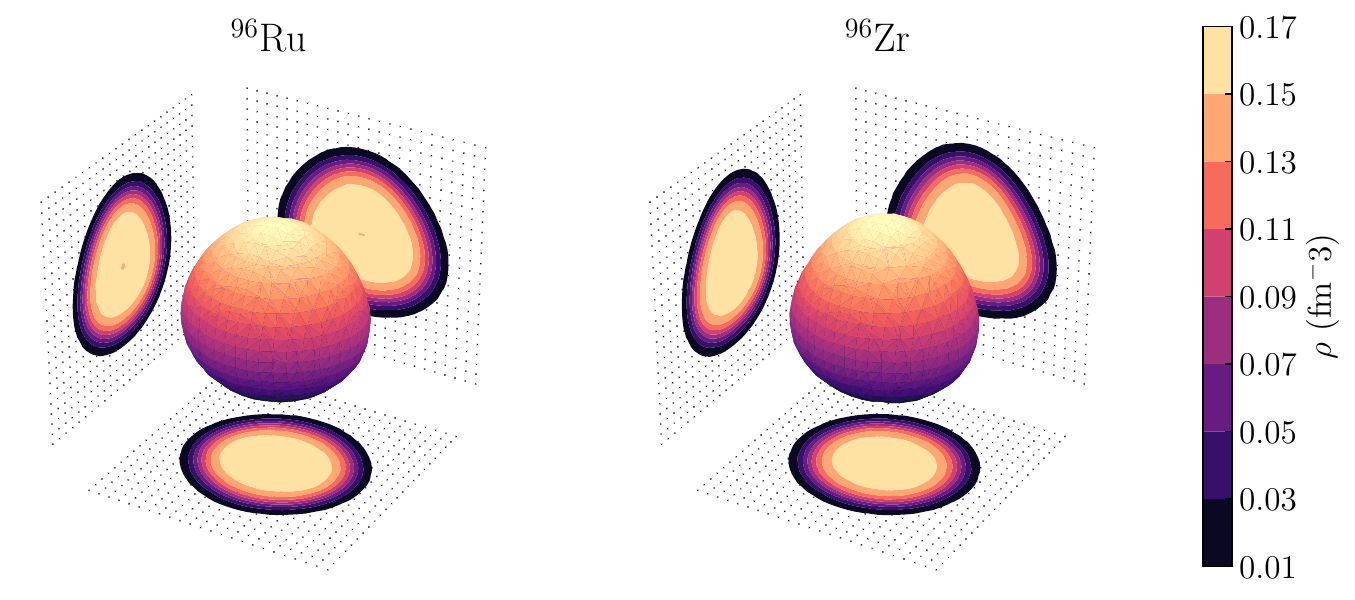}
\caption{Visualization of the nuclear shapes corresponding to the minima depicted 
         in Fig.~\ref{fig:BSkGfull}. The color bar refers exclusively to the two-dimensional 
         contour plots; the colors on the three-dimensional figure are illustrative only.
        }
\label{fig:visualisation}
\end{figure}

For this contribution, we extend the searches for the global energy minima
of the $A=96$ isobars to finite octupole deformation and $\gamma \in [0^{\circ}, 180^{\circ}]$.
The results are shown in Tab.~\ref{tab:BSkG_gs}: the global minima for both 
$^{96}$Zr and $^{96}$Ru are both reflection asymmetric and triaxial with 
large values of $\gamma \sim 80^{\circ} - 90^{\circ}$ except in one case. 
We illustrate the search
for these minima in Fig.~\ref{fig:BSkGfull}. In the top row of this figure, 
we draw the 
total energy surfaces obtained with BSkG2 in the absence of reflection 
asymmetry for both isobars: the triaxial minima with 
$0^{\circ} \leq \gamma \leq 60^{\circ}$ are those tabulated in Ref.~\cite{ryssens2022a}, but
is equivalent with five other minima in the whole $(\beta, \gamma)$ plane.
The degeneracy is partially lifted in the bottom row, which shows the total
energy surface at $\beta_{30}$ fixed to the value of the global minimum. 
Exploratory calculations indicate that the reflection symmetric configurations 
on the top panel are local minima and not saddle points, i.e. they are 
separated from the global minima by small barrier in the space of multipole 
moments. Although this cannot be deduced from the figure, the energy difference between 
global minima and the reflection symmetric local minima is very small; between
150 keV and 300 keV depending on the nucleus and model. Since shapes that 
combine triaxiality and reflection asymmetry have not often been discussed, we 
provide visualizations in Fig.~\ref{fig:visualisation}.

All entries in Tab.~\ref{tab:BSkG_gs} but one concerns rather unconventional 
triaxial shapes with finite octupole deformation The values of $\beta_{30}$ 
and $\beta_{32}$ nevertheless remain modest to the values indicated by 
the analysis of relativistic collisions of Ref.~\cite{Zhang:2021kxj}, although
this comparison ignores again the subtleties related to the precise definition 
of multipole moments. While these results indicate that $^{96}$Zr has a larger 
octupole deformation than $^{96}$Ru, it also indicates that it has a larger
quadrupole deformation. It is worth keeping in mind that these triaxial and reflection asymmetric minima disappear
when the rotational correction is removed. Although the latter is a numerically
efficient way to qualitatively model the effect of beyond-mean-field corrections
for the nuclear binding energy on the scale of the nuclear chart, such 
a phenomenological approach is necessarily more crude than full-fledged multireference
calculations. It is thus an open 
question if this type of unconventional minimum would survive in a more
advanced many-body treatment, particularly since all energy differences are
small and the total energy surfaces soft. Nevertheless, these results indicate that 
exotic configurations comparable to those in Fig.~\ref{fig:visualisation}
could be relevant for these $A=96$ isobars. To verify this claim, more 
advanced methods (such as the PGCM approach discussed elsewhere in this manuscript)
should be extended to cover a collective space of (at least) four different 
multipole moments ($\beta_{20},\beta_{22}, \beta_{30} $ and $\beta_{32}$) 
\emph{simultaneously}.
 
\begin{table}[t]
\centering
\begin{tabular}{|l|d{2.2}d{2.2}d{2.2}d{2.2}|d{2.2}d{2.2}d{2.2}d{2.2}|}
\hline
      & \multicolumn{4}{c|}{$^{96}$Zr} 
                 & \multicolumn{4}{c|}{$^{96}$Ru} \\
Model & \multicolumn{1}{c}{$\beta_{20}$} 
      & \multicolumn{1}{c}{$\beta_{22}$} 
      & \multicolumn{1}{c}{$\beta_{30}$} 
      & \multicolumn{1}{c|}{$\beta_{32}$} 
      & \multicolumn{1}{c}{$\beta_{20}$} 
      & \multicolumn{1}{c}{$\beta_{22}$} 
      & \multicolumn{1}{c}{$\beta_{30}$} 
      & \multicolumn{1}{c|}{$\beta_{32}$}\\
\hline
BSkG1 & 0.05 & 0.08 & 0.125 & 0.09 & 0.11 & 0.00 & 0.000 & 0.00 \\
BSkG2 & 0.04 & 0.06 & 0.125 & 0.03 & 0.00 & 0.06 & 0.075 & 0.04 \\
BSkG3 & 0.05 & 0.07 & 0.125 & 0.03 & 0.00 & 0.06 & 0.075 & 0.04 \\
\hline
\end{tabular}
\caption{ Quadrupole ($\beta_{20}$, $\beta_{22}$) and octupole ($\beta_{30}$, $\beta_{32}$) moments
          of the $A=96$ isobars as predicted by BSkG1, BSkG2 and BSkG3. }
\label{tab:BSkG_gs}          
\end{table}

\subsection{Beyond mean field results with the Gogny density functional}
\label{sec:Gogny}

\subsubsection{Theoretical framework}
Some of the most widely used microscopic descriptions of the atomic nucleus are based on self-consistent mean-field and beyond-mean-field approximations~\cite{RevModPhys.75.121,Niksic11a,Egido16a,Robledo18a}. Particularly, the projected generator coordinate method (PGCM) is a variational approach in which the variational space of wave functions is given by the linear combination of different symmetry-restored mean-field states. In this method, the variational principle is applied in two steps. The mean-field states are obtained self-consistently by minimizing either the Hartree-Fock-Bogolyubov (HFB)~\cite{RS80a} or the particle-number projected HFB (PNVAP)~\cite{Anguiano01a} energy with constraints. These constraints are usually collective deformations (quadrupole, octupole, etc.) and/or other intrinsic degrees of freedom such as pairing fluctuations or intrinsic rotations (cranking). Therefore, the intrinsic deformation of the nucleus is a well-defined theoretical quantity within this framework and is provided by the underlying HFB wave functions. Moreover, since these intrinsic states are not (in general) eigenstates of the particle-number, angular momentum and parity operators, these symmetries must be restored to recover the proper quantum numbers of the nuclear states in the laboratory frame. The second variational step in the PGCM approach concerns the coefficients of the linear combinations that define the PGCM wave functions. Such coefficients and the final energies are obtained by solving the so-called Hill-Wheeler-Griffin (HWG) equations that correspond to the diagonalization of the nuclear Hamiltonian within a reduced and highly correlated Hilbert sub-space. 

The implementation of the PGCM method depends on the kind of HFB states that are subsequently projected onto good quantum numbers and mixed by the HWG equations. In particular, two different types of PGCM calculations have been applied to study the structure of $^{96}$Zr and $^{96}$Ru. In the first one, the intrinsic HFB states are obtained by minimizing the particle number projected energy for different triaxial, parity-symmetric quadrupole deformations, $(\beta_{2},\gamma)$:
\begin{align}
\delta\biggl(\frac{\langle\Phi|\hat{H}P^{N}P^{Z}|\Phi\rangle}{\langle\Phi|P^{N}P^{Z}|\Phi\rangle}-\lambda_{N}\langle\Phi|\hat{N}|\Phi\rangle-\lambda_{Z}\langle\Phi|\hat{Z}|\Phi\rangle&\\
-\lambda_{q_{20}}\langle\Phi|\hat{Q}_{20}|\Phi\rangle-\lambda_{q_{22}}\langle\Phi|\hat{Q}_{22}|\Phi\rangle\biggr)=0
\label{pnvap}
\end{align}
where $\lambda_{X}$ is a Lagrange multiplier that ensures that the expectation value of a generic operator $\hat{X}$ is properly chosen, $\langle\Phi|\hat{X}|\Phi\rangle=x$. Here, the constraints are given by the multipole operators, $\hat{Q}_{\lambda\mu}=r^{\lambda}Y_{\lambda\mu}(\theta,\varphi)$, and the $(\beta_{2},\gamma)$ parameters are defined as:  
\begin{align}
\langle\Phi|\hat{Q}_{20}|\Phi\rangle=&q_{20}=\frac{\beta_{2}\cos\gamma}{C}\\
\langle\Phi|\hat{Q}_{22}|\Phi\rangle=&q_{22}=\frac{\beta_{2}\sin\gamma}{\sqrt{2}C}\\ 
C=&\sqrt{\frac{5}{4\pi}}\frac{4\pi}{3r^{2}_{0}A^{5/3}} 
\end{align} 
where $A$ is the mass number and $r_{0}=1.2$ fm. 
The PGCM wave function is then defined in this implementation as:
\begin{equation}
|\Psi^{JM\sigma}\rangle=\sum_{K=-J}^{+J}\sum_{(\beta_{2},\gamma)}f^{J\sigma}_{K,(\beta_{2},\gamma)}P^{J}_{MK}P^{N}P^{Z}|\Phi_{(\beta_{2},\gamma)}\rangle
\label{pgcm_wf_triax}
\end{equation}
where $J$, $M$, $K$, $N$, $Z$, $\sigma=1,2,...$, $P^{J}_{MK}$, $P^{N}$, and $P^{Z}$~\cite{RS80a} are the angular momentum, its projection onto the $z$-axis of the laboratory and intrinsic frames of reference, the number of neutrons and protons, and the label to identify different states for a given angular momentum, and the angular momentum, neutron-number and proton-number projection operators, respectively. The corresponding HWG equations (one for each value of $J$) read as ($\xi\equiv(K,(\beta_{2},\gamma))$ to shorten the notation):
\begin{equation}
\sum_{\xi'}\left(\mathcal{H}^{J}_{\xi,\xi'}-E^{J\sigma}\mathcal{N}^{J}_{\xi,\xi'}\right)f^{J\sigma}_{\xi'}=0
\label{wf_triax}
\end{equation}  
where the Hamiltonian and norm overlaps are defined as:
\begin{align}
\mathcal{H}^{J}_{\xi,\xi'}=&\langle\Phi_{(\beta_{2},\gamma)}|\hat{H}P^{J}_{KK'}P^{N}P^{Z}|\Phi_{(\beta'_{2},\gamma')}\rangle\\
\mathcal{N}^{J}_{\xi,\xi'}=&\langle\Phi_{(\beta_{2},\gamma)}|P^{J}_{KK'}P^{N}P^{Z}|\Phi_{(\beta'_{2},\gamma')}\rangle
\end{align}
The second type of PGCM calculations is defined by the mixing of parity, particle number and angular momentum projection of HFB states with different axial quadrupole and octupole intrinsic deformations, $(\beta_{2},\beta_{3})$. Hence, the values of $\gamma$ are restricted to $0^{\circ}$ (prolate) and $180^{\circ}$ (oblate), and $K=0$ in this case. Moreover, the intrinsic states are obtained by minimizing the HFB energy instead of the particle number projected one. Therefore, the particle number projectors and the constraint in $\hat{Q}_{22}$ are substituted in Eq.~\ref{pnvap} by identity operators and the operator $\hat{Q}_{30}$, respectively. Taking into account these substitutions, the PGCM wave functions are given in this approach by~\cite{Bernard16a}:
\begin{equation}
|\Psi^{JM\pi\sigma}\rangle=\sum_{(\beta_{2},\beta_{3})}f^{J\sigma}_{(\beta_{2},\beta_{3})}P^{J}_{00}P^{N}P^{Z}P^{\pi}|\Phi_{(\beta_{2},\beta_{3})}\rangle
\label{pgcm_wf_ax}
\end{equation}
where $P^{\pi}$ is the projector onto good parity, $\pi$, and $\beta_{3}=\frac{4\pi}{3r^{3}_{0}A^{2}}q_{30}$ is the axial octupole deformation parameter.

Let us summarize the differences between the two kind of calculations before presenting the results. Thus, the first PGCM implementation includes triaxial shapes, $K$-mixing for states with $J\neq0$ and, because of the use of PNVAP method over plain HFB, incorporates pairing correlations in a more consistent way than the second one. However, it is limited to describe positive parity states and does not explore the octupole degree of freedom unlike the second implementation. Nevertheless, none of the present PGCM calculations scans variationally ground and excited states on an equal footing and a systematic stretching of the spectrum with respect to the unknown exact solution is expected. Such a behavior can be partially corrected by the inclusion of cranking terms and explicit multi-quasiparticle excitations~\cite{Borrajo15a,Chen17a}. However, the computational cost of taking into account all of these degrees of freedom at the same time (triaxiality, octupolarity, cranking, quasiparticle excitations) is still prohibitive. 
\subsubsection{Results}
\begin{figure}[H]
\begin{center}
\includegraphics[width=0.53\linewidth]{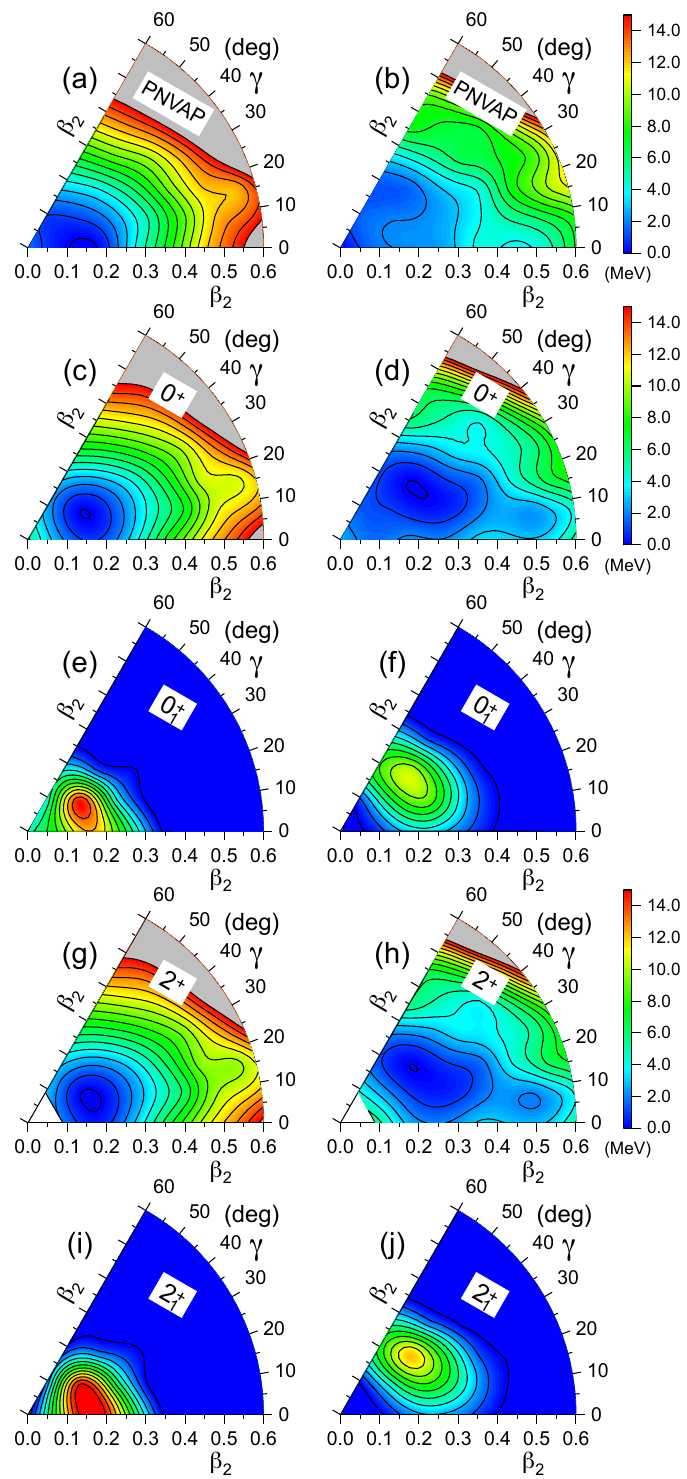}
\caption{(a)-(b) Particle number projected and particle number and angular momentum projected, (g)-(h) $J=2^{+}$, total energy surfaces, as well as (e)-(f) $0^{+}_{1}$ and (i)-(j) $2^{+}_{1}$ collective wave functions in the quadrupole $(\beta_{2},\gamma)$ plane for $^{96}$Ru (left panel) and $^{96}$Zr (right panel). Gogny D1S interaction is used. Energies are normalized to their respective minimum. The sum of the collective wave functions are normalized to 1 and reddish (blueish) areas represent large (small) contributions to the wave function.}
\label{Fig_b2_gamma}
\end{center}
\end{figure}

\begin{figure}[H]
\begin{center}
\includegraphics[width=0.77\linewidth]{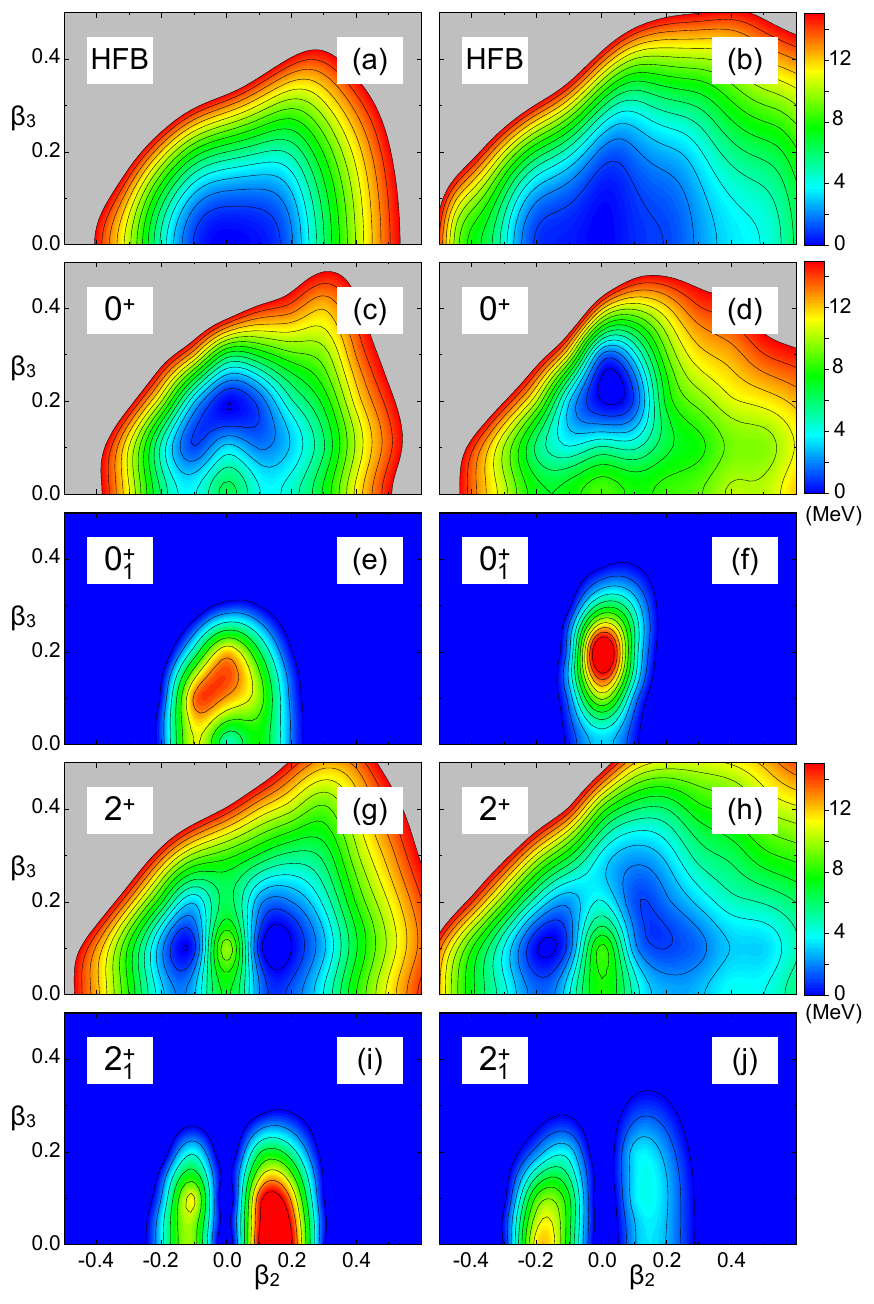}
\caption{(a)-(b) HFB and particle number and angular momentum projected, (g)-(h) $J=2^{+}$, total energy surfaces, as well as (e)-(f) $0^{+}_{1}$ and (i)-(j) $2^{+}_{1}$ collective wave functions in the axial quadrupole-octupole $(\beta_{2},\beta_{3})$ plane for $^{96}$Ru (left panel) and $^{96}$Zr (right panel). Gogny D1S interaction is used. Energies are normalized to their respective minimum. The sum of the collective wave functions are normalized to 1 and reddish (blueish) areas represent large (small) contributions to the wave function.}
\label{Fig_b2_b3}
\end{center}
\end{figure}

\paragraph{Total Energy Surfaces}
A first insight into the collective structure of the nucleus within the PGCM framework comes from the analysis of the ``mean-field'' total energy surface (TES) where either the HFB or the particle number projected energy as a function of the generating coordinates is studied.  These TESs are represented in Fig.~\ref{Fig_b2_gamma}(a)-(b) and Fig.~\ref{Fig_b2_b3}(a)-(b) in the $(\beta_{2},\gamma)$ plane and $(\beta_{2},\beta_{3})$ plane, respectively. For $^{96}$Ru we observe a slightly prolate deformed minimum at $(0.15,0^{\circ})$ with some softness from $(0.20, 0^{\circ})$ to $(0.15,60^{\circ})$ in the PNVAP energy surface. On the other hand, the HFB-TES shows a similar softness around the same values of $\beta_{2}$ as in the triaxial case, the minimum is found at the spherical point and a rather rigid behavior along the octupole degree of freedom is obtained. For $^{96}$Zr the triaxial PNVAP-TES presents two distinct minima close in energy, namely, the spherical and the oblate (at $\beta_{2}\approx0.18$) minima. Additionally, well-deformed prolate (at $\beta_{2}=0.45$) and triaxial (at $(0.4,45^{\circ})$) minima are also visible. These sub-structures are less well-defined in the $(\beta_{2},\beta_{3})$ HFB-TES. Nevertheless, the minimum is still found at the spherical point (due to the $Z=40$ sub-shell closure) and a large softness in $\beta_{3}$ around this minimum is also observed.

The next step in the PGCM method is the simultaneous restoration of the symmetries broken by the intrinsic HFB-like states. We focus our analysis on the positive parity states and the two lowest angular momentum, i.e., $0^{+}$ and $2^{+}$ TESs. In the axial (triaxial) calculation, parity is (not) broken and thus parity projection is (not) needed. In Fig.~\ref{Fig_b2_gamma}(c)-(d), Fig.~\ref{Fig_b2_gamma}(g)-(h), Fig.~\ref{Fig_b2_b3}(c)-(d) and Fig.~\ref{Fig_b2_b3}(g)-(h) we show the projected $0^{+}$ and $2^{+}$ TESs in the $(\beta_{2},\gamma)$ plane and in the $(\beta_{2},\beta_{3})$ plane, respectively. For both nuclei the axial minima are shifted towards more triaxial configurations. In addition, the $0^{+}$ and $2^{+}$ TESs are rather similar within the same nucleus suggesting the appearance of rotational ground-state bands. Hence, the prolate PNVAP minimum is displaced to $(0.15,25^{\circ})$ for $^{96}$Ru. For $^{96}$Zr, the two almost degenerated PNVAP minima have evolved towards a rather soft triaxial single minimum and the prolate deformed minimum is also found now at $(0.5,5^{\circ})$. The energy difference between the PNVAP and  $J=0^{+}$ minima is 3.6 MeV and 2.5 MeV for $^{96}$Ru and $^{96}$Zr, respectively. These correlation energies due to the angular momentum restoration are within the limits obtained in global studies performed with Skyrme~\cite{Bender06a} and Gogny~\cite{Rodriguez15a} interactions. The results are rather different in the $(\beta_{2},\beta_{3})$ plane as it is shown in Fig.~\ref{Fig_b2_b3}. In both nuclei $(\beta_{2}=0,\beta_{3}\neq0)$ minima are obtained for the $J=0^{+}$-TESs with unexpectedly large correlation energies with respect to the HFB minima, namely, 6.6 MeV and 11.4 MeV for $^{96}$Ru and $^{96}$Zr, respectively. This amount of energy gain is also observed with relativistic Lagrangians~\cite{ISKRA2019396}. However, the behavior of the $J=2^{+}$-TESs are consistent with the results in the $(\beta_{2},\gamma)$ plane since the minima found in $(\beta_{2},\beta_{3})$ correspond to the saddle points obtained along $\gamma=0^{\circ}$ and $60^{\circ}$ directions. Hence, two different minima with a certain amount of octupolarity ($\beta_{3}\approx0.1$) are observed at prolate and oblate deformations separated by a barrier at $\beta_{2}=0$. The absolute minimum is the prolate (oblate) one for $^{96}$Ru ($^{96}$Zr). In addition, the prolate minimum presents some $\beta_{3}$-softness and another secondary prolate minimum at $(0.45,0.1)$ is slightly visible in $^{96}$Zr.  

\begin{figure}[H]
\begin{center}
\includegraphics[width=0.5\linewidth]{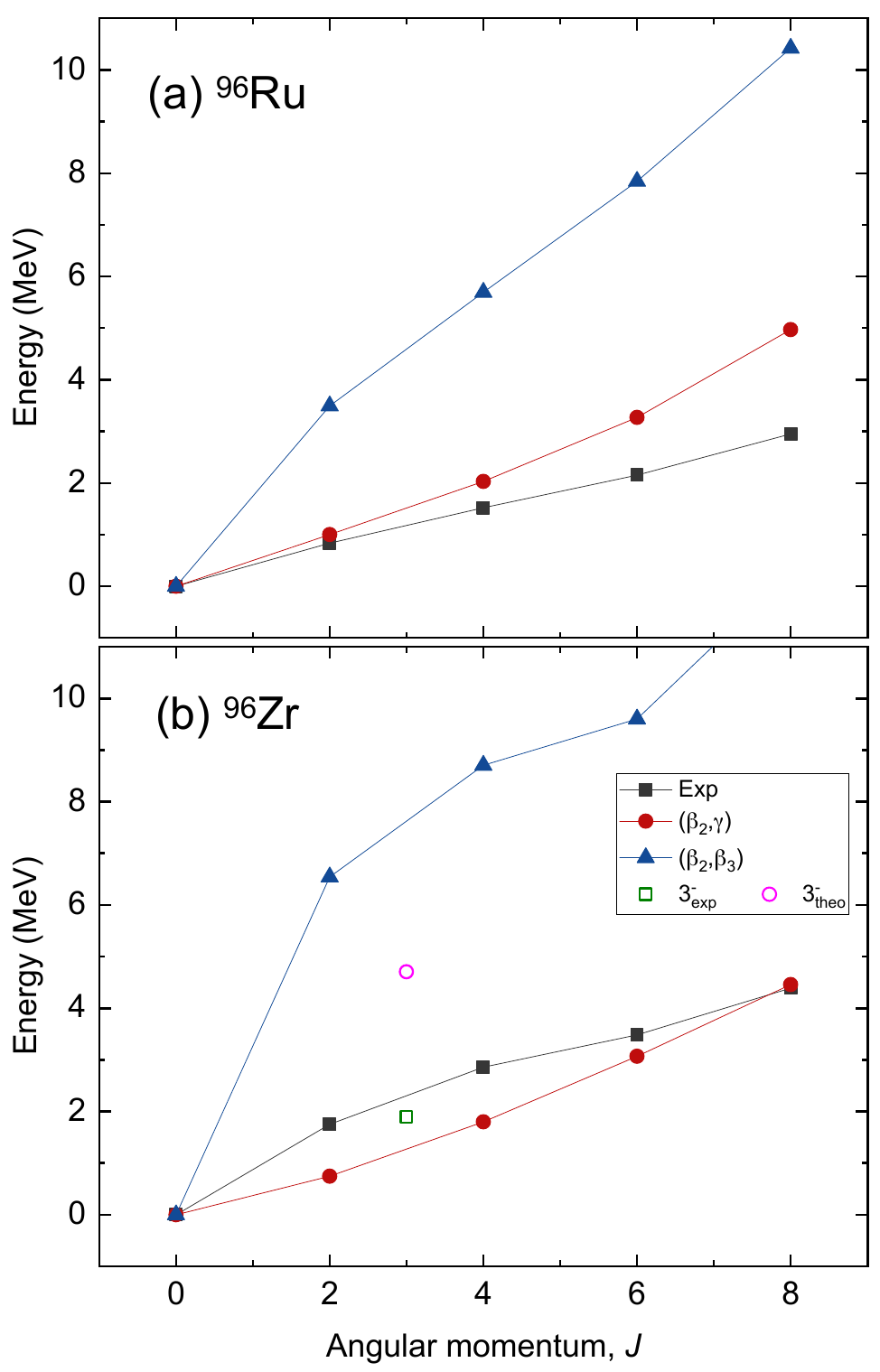}
\caption{Experimental (black squares)~\cite{ensdf} and theoretical excitation energies for the ground state (positive) band computed with the parity-conserving triaxial PGCM method (red bullets) and the parity-breaking axial PGCM method (blue triangles) for (a) $^{96}$Ru and (b) $^{96}$Zr. In the latter, the lowest experimental and theoretical $3^{-}$ excitation energies are also represented.}
\label{Fig_ener}
\end{center}
\end{figure}

\paragraph{PGCM energies and collective wave functions}
The last step in the present EDF method is the mixing of the projected HFB-like states that leads to the HWG equations. The excitation energies of the ground state bands for $^{96}$Ru and $^{96}$Zr calculated with the two PGCM methods described above and the experimental values are shown in Fig.~\ref{Fig_ener}. Apart from the obvious quantitative differences between the theoretical predictions and the experimental data, as well as between the two PGCM implementations themselves, let us discuss first the qualitative behavior of the positive parity states as a function of the angular momentum. The triaxial PGCM calculations predict a parabolic trend for both nuclei that is consistent with a rotational character. In fact, the collective wave functions of those states are rather constant along the same bands as can be seen in Fig.~\ref{Fig_b2_gamma} for the $0^{+}_{1}$ and $2^{+}_{1}$ states. Here, the nucleus $^{96}$Ru is predicted to be deformed ($\beta_{2}\approx0.15$) with some triaxiality in the ground state that evolves towards prolate deformations in the first excited state. The nucleus $^{96}$Zr is slightly more deformed ($\beta_{2}\approx0.20$) and the collective wave functions are more extended in the $(\beta_{2},\gamma)$ plane but with a more oblate character than the other isobar. We also observe that in both nuclei the collective wave functions are found in the minima of the corresponding projected-TESs. The same happens for the $(\beta_{2},\beta_{3})$ PGCM calculations. In this case, the ground state collective wave functions are found at $(\beta_{2}=0,\beta_{3}\neq0)$ configurations (with larger octupolarity in $^{96}$Zr than in $^{96}$Ru) and the $2^{+}_{1}$ are roughly the axial projections of the triaxial counterparts discussed above. This difference between the configuration of the ground and first-excited states and the huge amount of correlation energy attained by the $0^{+}_{1}$ state along the octupole degree of freedom leads to the spectra shown in Fig.~\ref{Fig_ener}. Here, a inverted parabola behavior is obtained in this version of the PGCM approach. The experimental data also show this form (more clear in $^{96}$Zr) but at much smaller excitation energies. In fact, the triaxial calculations reproduce numerically better the experimental results but fail in following the correct trend. Additionally, since these calculations explore variationally better the ground state than the excited states, a systematic stretching is expected. For $^{96}$Zr the predicted $2^{+}_{1}-6^{+}_{1}$ energies are already below the experimental values in the triaxial PGCM and adding cranking will lower the energy further. For $^{96}$Ru these cranking terms could improve the agreement with the experimental data but still the experimental trend seems to be better reproduced if $\beta_{3}$ is included. A final hint of the relevance of the octupole degree of freedom in $^{96}$Zr is the experimental $3^{-}_{1}$ state found at 1.897 MeV almost at the same energy as the $2^{+}_{1}$ state. The axial parity-breaking PGCM also predicts a rather low $3^{-}_{1}$ state (see Fig.~\ref{Fig_ener}) with a collective wave function very similar to the one of the ground state, i.e., peaked at $\beta_{2}=0.0,\beta_{3}=0.2$. According to the calculations this is the first excited state but this is in contradiction with the experiment. This result, together with the disproportionate excitation energies require a better understanding of the mechanism behind the large amount of correlation energy obtained by the $0^{+}$ (and $3^{-}$) states with zero quadrupole and non-zero octupole shapes whenever the symmetry restoration is performed. However, this study is beyond the scope of the present work~\cite{Rodriguez24a}.  

\paragraph{Summary}
We have performed PGCM calculations of $^{96}$Ru and $^{96}$Zr isobars with energy density functionals based on the Gogny D1S interaction. Two different implementations of the method have been used. The results obtained with the mixing of parity-symmetric triaxial quadrupole shapes predict ground state bands with rotational character in both isotopes, being $^{96}$Ru less deformed, more triaxial-prolate like than $^{96}$Zr that is found to be triaxial-oblate deformed. However, neither the quantitative values nor the qualitative trend of the experimental data are well reproduced, although the nucleus $^{96}$Ru is better described than $^{96}$Zr in this PGCM approach. On the other hand, the PGCM method with axial quadrupole-octupole degrees of freedom reproduces the trends of the experimental spectra but the results are too stretched. Both the trend and the stretching are produced by the huge amount of correlation energy attained by the ground states at $(\beta_{2}=0,\beta_{3}\neq0)$ shapes. Whether this energy gain is physically acceptable is still a work in progress. Nevertheless, these PGCM calculations predict that $^{96}$Zr would be more octupole deformed than $^{96}$Ru.

\subsection{Octupole deformation of \texorpdfstring{$^{96}$Zr}{96Zr} in the nuclear shell model}

\subsubsection{Introduction}
As discussed in Sec.~\ref{sec:Wimmer}, the structure of $^{96}$Zr looks rather simple from the experimental spectroscopy viewpoint, with a spherical ground state and high-lying excited states built on this ground state \cite{nudat3}, as shown in Fig.~\ref{fig:level}.  
However, this nucleus is known to exhibit a low-lying $3^-$ state, which is also known to decay to the ground state by a strong $E3$ transition, with $B(E3; 3^-_1 \rightarrow 0^+_1)$ = 53 $\pm$ 6 W.u. \cite{kibedi_2002} and 42 $\pm$ 3 W.u.  \cite{ISKRA2019396}.  This large value is usually interpreted as a result of an octupole vibration around a spherical ground state.

Thus, the structure of $^{96}$Zr attracts certain attention, although it is clear that the quadrupole deformation is not developed in the ground state.  We have conducted systematic studies on Zr isotopes since \cite{togashi_2016} in terms of the Monte Carlo Shell Model (MCSM) \cite{mcsm_1995,mcsm_1998}.   The MCSM has been reviewed in \cite{mcsm_2001,mcsm_2012,mcsm_2017}. While the work for \cite{ISKRA2019396} was made with same Hamiltonian used in  \cite{togashi_2016}, we renewed this study with revised shell-model Hamiltonians so that the theoretical description of Zr isotopes including $^{96}$Zr is substantially improved \cite{yanase_2025}.
This short note is an interim report of a more focused work for $^{96}$Zr \cite{yanase_2025} arising from such systematic studies.

\subsubsection{Results}

\begin{figure}[t]
  \centering
  \includegraphics[width=0.7\linewidth]{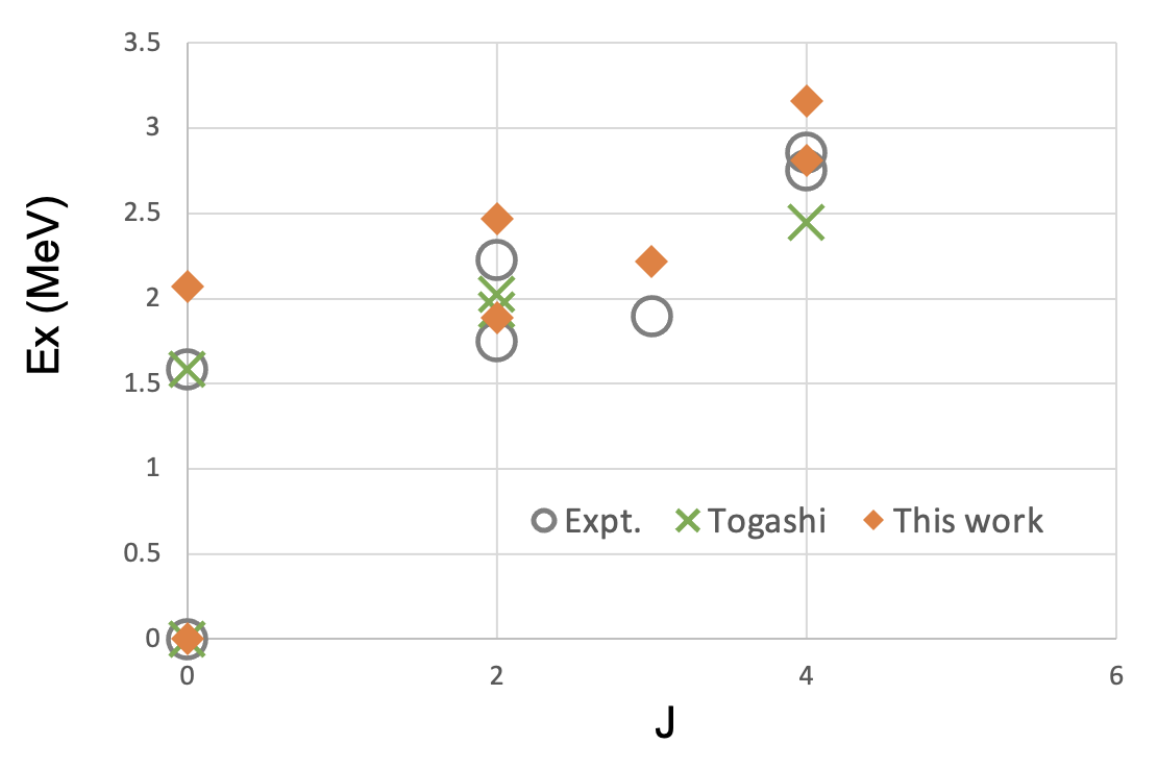}
     \caption{Calculated level energies of some lowest states of $^{96}$Zr in comparison to experimental ones \cite{ensdf}. The horizontal axis is the angular momentum, $J$, of the states.  The states of $J$= 0, 2 and 4 are of positive parity, while that of $J$= 3 is of negative parity.   The calculated results in \cite{togashi_2016} are also shown and labeled as ``Togashi''.} 
  \label{fig:levels}  
\end{figure}

The calculated excitation energies are shown in Fig.~\ref{fig:levels} in comparison to experimental values. This result implies that the level energies of the low-lying states are reproduced reasonably well.  The E2 transition strengths are also reproduced reasonably well in comparison to observed values, although not shown here. It is then quite interesting to see how the $3^-_1$ state is described. The calculated value of $B(E3; 3^-_1 \rightarrow 0^+_1)$ = 52 W.u. with the same E3 effective charges as those used in the MCSM calculation reported in \cite{ISKRA2019396}.  Note that the values of these effective charges are 1.24 $e$ for protons and 0.82 $e$ for neutrons with ``$e$'' being the unit charge, and were originally taken from \cite{bohr75}.

We conclude that a strong E3 transition in $^{96}$Zr arises, thus, in a theoretical calculation that is not specifically adjusted to reproduce this property. From the  $B(E3)$ value, we can deduce the value of so-called octupole deformation parameter, $\beta_3$ by using the standard formulas.  The obtained value is $\beta_3$ = 0.27.  This is consistent with the large values of octupole deformation indicated in theoretical works \cite{Zhang:2021kxj,Nijs:2021kvn} based on STAR data \cite{STAR:2021mii}.  

The present value, 0.27, is obtained based on the assumption that the ground state and the $3^-_1$ state are formed from the same intrinsic state with static octupole deformation.  This assumption has to be further studied, as it is in contrast with the traditional picture of the octupole vibration assuming a spherical ground state.  Such studies are under way, as well as more refined studies of the structure of additional Zr isotopes by the more advanced version of the MCSM, called Quasiparticle Vacua Shell Model (QVSM) \cite{shimizu_2021}. 

\subsection{Explorations with the \textit{ab initio} PGCM approach}
\label{sec:abPGCM}

\subsubsection{Method}

The Projected Generator Coordinate Method (PGCM) is a powerful variational many-body technique that permits to efficiently describe collective phenomena observed in atomic nuclei \cite{Hill53a,Griffin57a}. Within this method\footnote{We consider here the discrete version of the method.}, approximations to the nuclear eigenstates are constructed using the ansatz
\begin{equation}
  \ket{\Psi_\epsilon^{\Lambda M}} = \sum_{q K} f^{\Lambda}_{\epsilon}(qK) P^{\Lambda}_{MK} \ket{\Phi(q)} ,
\end{equation}
where $f^{\Lambda}_{\epsilon}(qK)$ is a weight, $P^{\Lambda}_{MK}$ is a projection operator onto good quantum numbers \cite{Bally21a} (the number of protons $Z$ and neutrons $N$, the angular momentum $J$ and its third component $M$ or $K$, the parity $\pi$) collected under the label $\Lambda \equiv (Z,N,J,\pi)$ and $\ket{\Phi(q)}$ is a reference state labeled by the collective coordinate $q$. The weights are determined by diagonalizing the nuclear Hamiltonian $H$ within the subspace spanned by the projected states $\left\{ P^{\Lambda}_{MK} \ket{\Phi(q)}, q, \Lambda \right\}$, which can be performed for each value of $\Lambda$ separately. Then, the index $\epsilon$ is used to label the different eigenvector of $H$ within that subspace.

The reference states $\left\{ \ket{\Phi(q)}, q \right\}$ are usually taken to be Bogoliubov quasi-particle states and are generated by minimizing $H$ under a set of constraints \cite{RS80a,RevModPhys.75.121} that we label generically: $\bra{\Phi(q)} Q \ket{\Phi(q)} = q$, with $Q$ being a set of one-body operators. The multi-reference aspect of the method makes it well-suited to capture collective correlations and their fluctuations.

In the past, the PGCM method has been employed mostly in combination with phenomenological energy density functionals, see Sec.~\ref{sec:Gogny}. In particular, advanced calculations were performed to provide microscopic inputs to the simulation of heavy-ion collisions in the case of $^{129}$Xe, $^{208}$Pb \cite{Bally:2021qys} and $^{197}$Au \cite{Bally23a}.
More recently, efforts have been made to also use this method within the \emph{ab initio} context using nuclear Hamiltonian derived from Chiral Effective Field Theory \cite{Epelbaum09a,Machleidt11a}.

\subsubsection{Results}
\begin{figure}[t!]
    \centering
    \includegraphics[width=.45\linewidth]{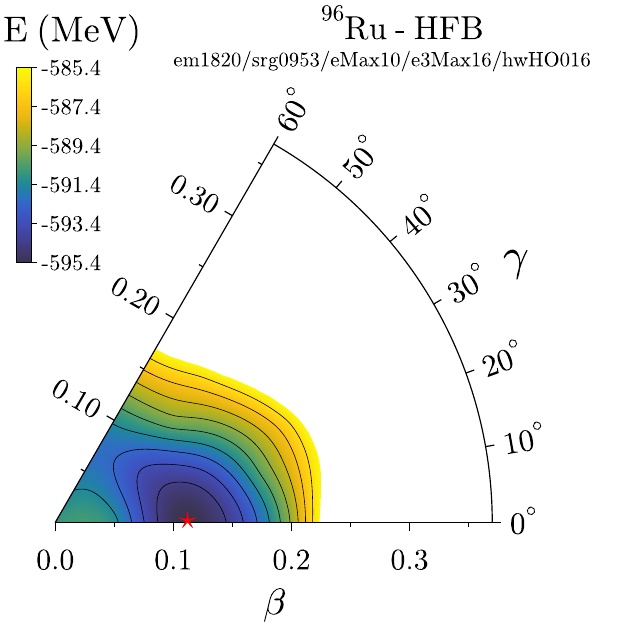}
    \hspace*{0.10cm}
    \includegraphics[width=.45\linewidth]{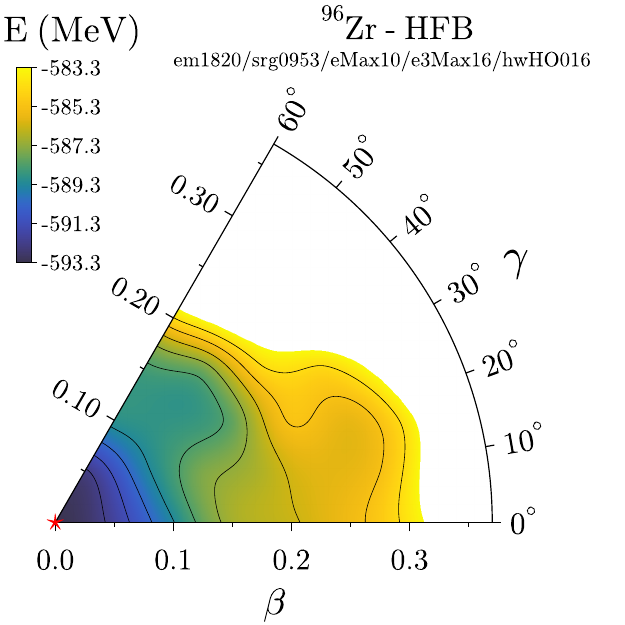}  
    \caption{Hartree-Fock-Bogoliubov (HFB) total energy surfaces for $^{96}$Ru (left) and $^{96}$Zr (right) in the first sextant of the triaxial plane. Black lines are separated by 1 MeV.
    The minimum for $^{96}$Ru ($^{96}$Zr), indicated by a red star, is located at a deformation of $\beta=0.11$ and $\gamma=1^\circ$ ($\beta=0.0$ and $\gamma=0^\circ$).}
     \label{fig:ruzr96_triaxial}
\end{figure}

In this work, we performed exploratory calculations using the chiral Hamiltonian dubbed ``EM1.8/2.0'' that contains two- and three-body interactions \cite{Hebeler11a}. For the single-particle basis, we considered a spherical harmonic oscillator basis with $e_{\text{max}} = 10$ with an additional cut-off $e_{3\text{max}} = 16$ for the three-body matrix elements. While the dimensions of the basis size are not sufficient to fully converge calculations on $^{96}$Ru and $^{96}$Zr, they still allow us to obtain a representative picture of the structure of these nuclei. As a matter of fact, we checked that the topography of the triaxial energy surfaces (see below) was not changing when increasing the size of the basis to $e_{\text{max}} = 12$.

To explore nuclear deformations, we performed mean-field calculations with constraints on the expectation value of the multipole operators $Q_{lm} \equiv r^l Y_{lm}$, where $r$ is the position and $Y_{lm}$ is a spherical harmonics, for $l =2$ (quadrupole) and $l=3$ (octupole).
In Fig.~\ref{fig:ruzr96_triaxial}, we show the energy surfaces at the mean-field level for $^{96}$Ru and $^{96}$Zr as a function of the average quadrupole deformations parametrized using the usual triaxial variables $\beta$ and $\gamma$ \cite{RS80a}.
As can be seen, $^{96}$Ru possesses a deformed axial minimum ($\beta = 0.11$ and $\gamma = 1^\circ$) while the energy surface of $^{96}$Zr exhibits a pure spherical minimum  ($\beta = 0.0$ and $\gamma = 0^\circ$). 

\begin{figure}[t]
    \centering
    \includegraphics[width=.82\linewidth]{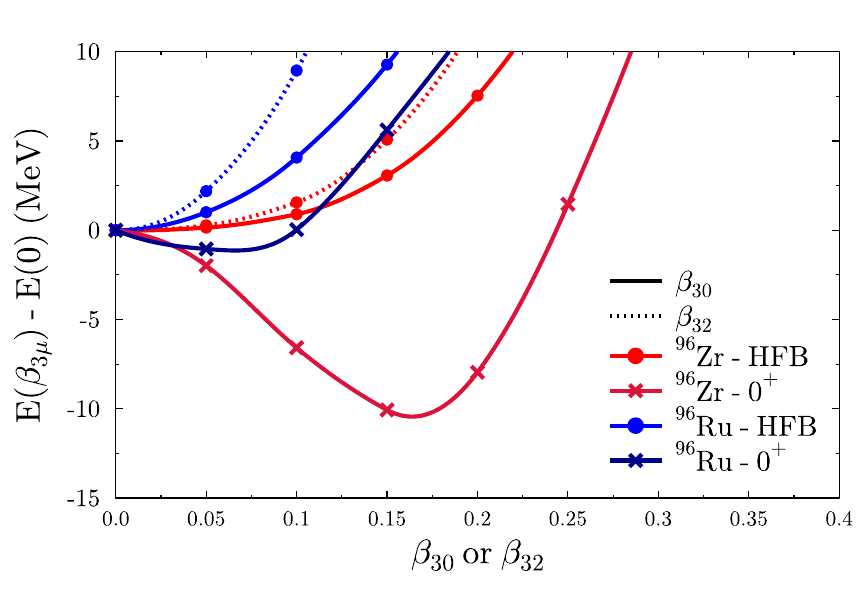} 
    \caption{Differential HFB (circles) and symmetry-projected (crosses) total energy curves with respect to the point with
    vanishing $\beta_{30}$ or $\beta_{32}$ for $^{96}$Ru (blue) and $^{96}$Zr (red). The projected energies are plotted at the deformation of the reference state they are obtained from.}
     \label{fig:ruzr96_proj}
\end{figure}

In a second step, for both nuclei, we fixed the average values of $\beta$ and $\gamma$ to the ones of their respective minimum and varied the average values of $\beta_{30}$ or $\beta_{32}$ to explore the octupole degree of freedom.\footnote{The relation between the parameter $\beta_{lm}$ and the expectation value of $Q_{lm}$ can be found for example in Ref.~\cite{Bally21b}.} The resulting energy curves are displayed in Fig.~\ref{fig:ruzr96_proj}. In both cases, the minimum is found for a vanishing value of $\beta_{30}$ or $\beta_{32}$. Nevertheless, we can observe that the energy curves for $^{96}$Zr are relatively soft against oblate deformations. To go further, for each nucleus and for the different values of $\beta_{30}$, we also represent on Fig.~\ref{fig:ruzr96_proj} the energy curves obtained after symmetry restoration. As can be observed, the minimum after projection is obtained for a reference state with a non-vanishing 
value of $\beta_{30}$ for both nucleus. But while the minimum for $^{96}$Ru is relatively soft and located at a small value of $\beta_{30}$, the one for $^{96}$Zr is well-pronounced at a value of $\beta_{30} \approx 0.175$.

In conclusion, these exploratory calculations seem consistent with the recent analyses of the isobar run performed at RHIC \cite{Zhang:2021kxj}, in particular regarding the importance of octupole deformation in $^{96}$Zr compared to $^{96}$Ru.
Nevertheless, it is important to stress the preliminary character of our calculations that: i) do not explore simultaneously the quadrupole and octupole deformations, ii) do not include the full configuration mixing of reference states with different deformation, iii) are probably not converged as a function of $e_{\text{max}}$ and $e_{\text{3max}}$ and iv) use a chiral Hamiltonian whose performance in this mass region has not been investigated in details.

\subsection{Prospects from Nuclear Lattice Effective Field Theory simulations }

Lattice effective field theory is an \textit{ab initio} approach that combines the framework of effective field theory (EFT) with lattice Monte Carlo methods to allow favorable scaling from few- to many-body systems. Perhaps the most important aspect of lattice EFT is that its strengths and weaknesses are orthogonal
to those of other \textit{ab initio} methods. For example, lattice EFT has little difficulty in probing cluster structures and collective phenomena, which are much more difficult using other methods. 

Nuclear lattice EFT simulations are performed using auxiliary-field Monte Carlo simulations \cite{Lee:2008fa,Lahde:2019npb}.  The auxiliary fields are used to decouple the nucleons from each other.  The interactions are recovered by integrating over all possible configurations of the auxiliary fields.  By sampling
the quantum amplitude for each configuration of auxiliary fields, we
obtain the full set of correlations induced by the interactions.  However,
the amplitude for each auxiliary field configuration involves
quantum states which are superpositions of many different center-of-mass
positions.  Therefore, information about density correlations relative to
the center of mass is lost. The pinhole algorithm is a computational approach that allows for the calculation of arbitrary density correlations
with respect to the center of mass \cite{Elhatisari:2017eno}.

We let $\rho_{i,j}({\bf n})$ be the density operator for nucleons with spin
$i$ and isospin $j$ at lattice site {\bf n},
\begin{equation}
\rho_{i,j}({\bf n}) = a^\dagger_{i,j}({\bf n})a_{i,j}({\bf n}).
\end{equation}
We construct the normal-ordered $A$-body density operator
\begin{equation}
\rho_{i_1,j_1,\cdots i_A,j_A}({\bf n}_1,\cdots {\bf n}_A) = \; :\rho_{i_1,j_1}({\bf
n}_1)\cdots\rho_{i_A,j_A}({\bf
n}_A):.
\end{equation}
In the $A$-nucleon subspace, we note the completeness identity
\begin{equation}
\sum_{i_1,j_1,\cdots i_A,j_A}\sum_{{\bf n}_1,\cdots {\bf n}_A} 
\rho_{i_1,j_1,\cdots i_A,j_A}({\bf n}_1,\cdots {\bf n}_A) \; = A!.
\label{closure}
\end{equation}
The strategy of the pinhole algorithm is to perform Monte Carlo importance sampling of the pinholes as well as auxiliary and/or pion fields in order to provide an unbiased estimate of the ratio
\begin{equation}
 \frac{\langle \Psi_{\rm initial}| e^{-H\tau/2}    
\rho_{i_1,j_1,\cdots i_A,j_A}({\bf n}_1,\cdots {\bf n}_A)e^{-H\tau/2} |\Psi_{\rm initial}\rangle}{\langle \Psi_{\rm initial}| e^{-H\tau}|\Psi_{\rm initial} \rangle}.
 \label{pinholeamp}
\end{equation}

\begin{figure}
\includegraphics[height=6.2cm]{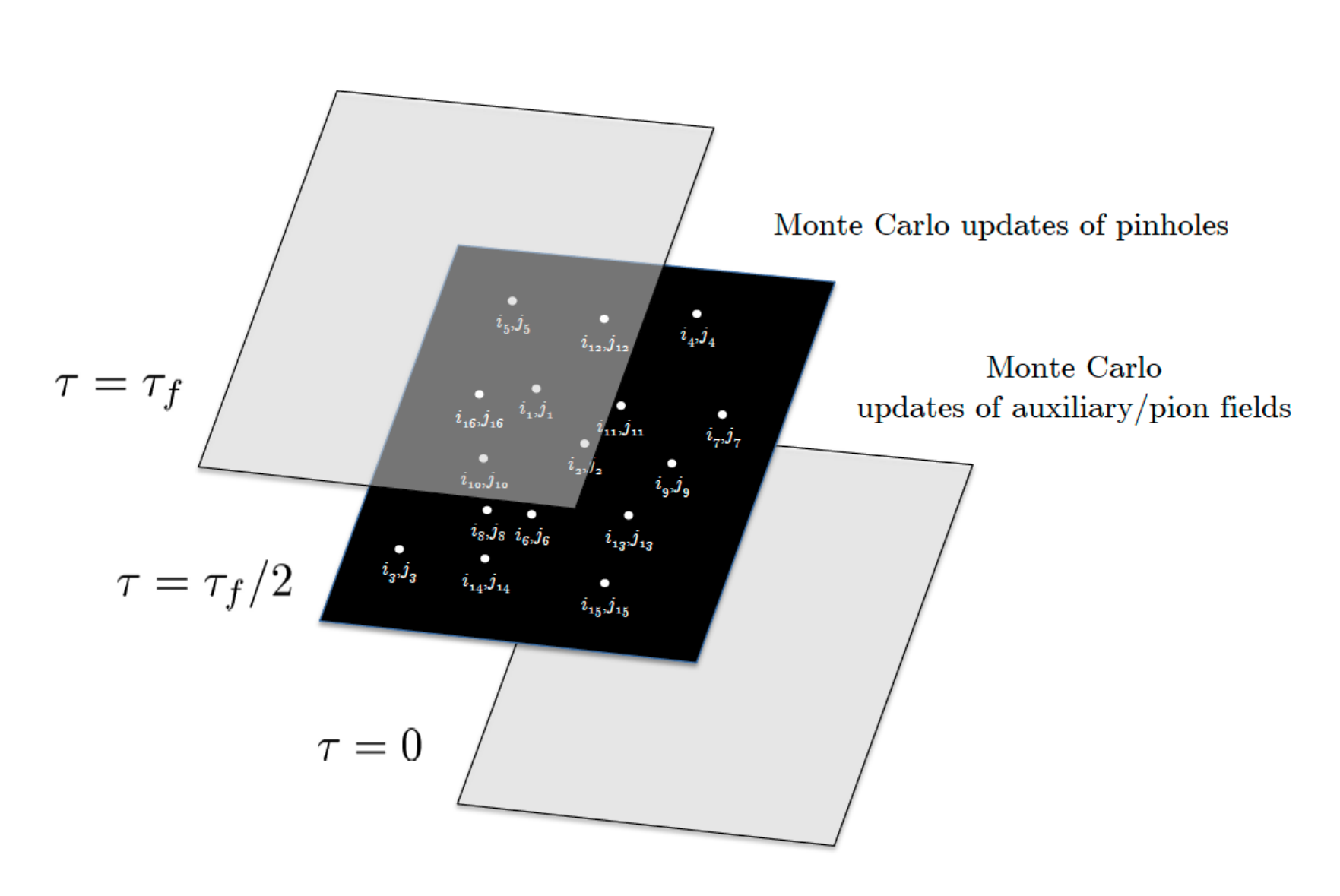} 
\includegraphics[height=6.2cm]{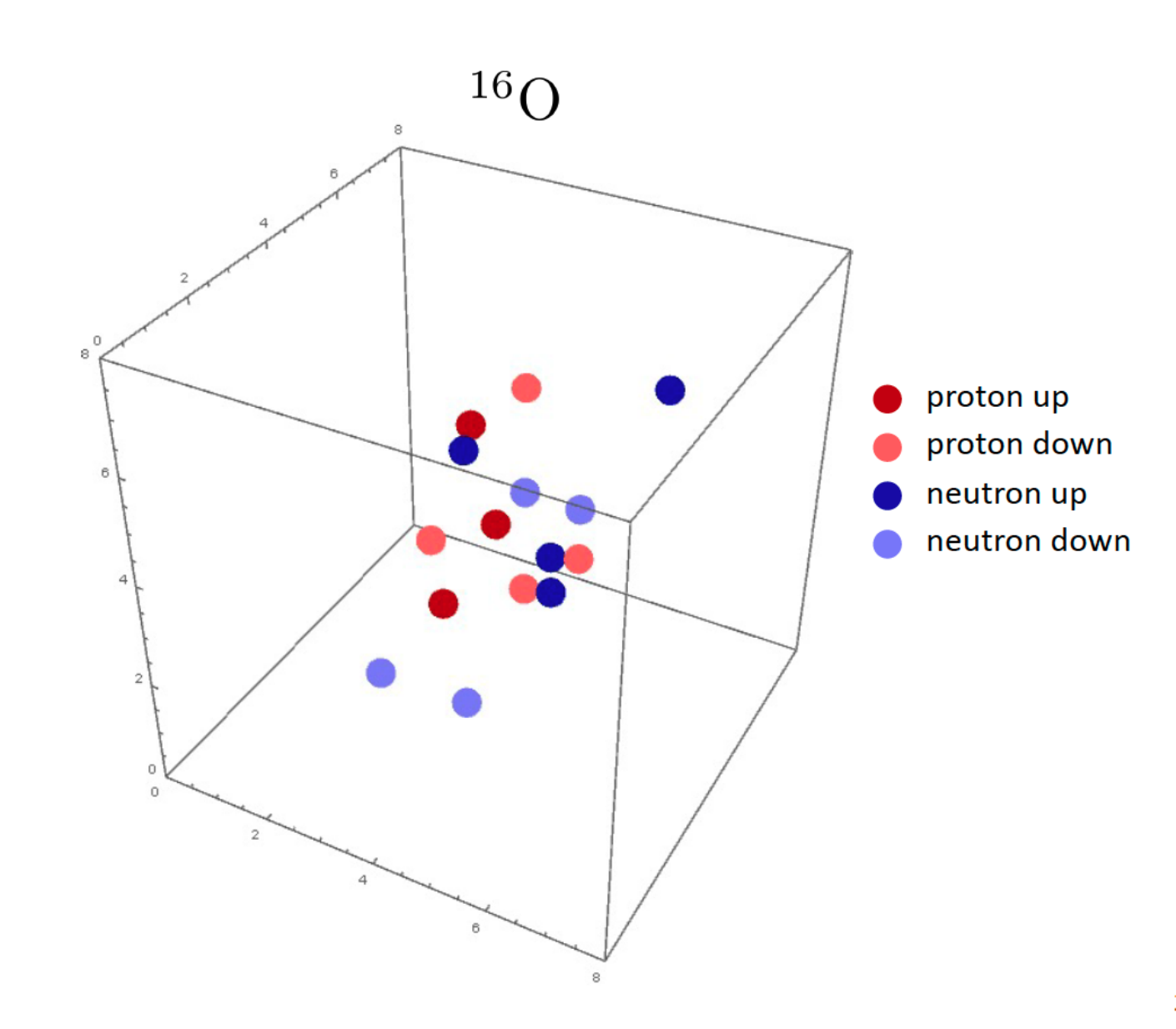}
\caption{The left panel shows the insertion of pinholes with spin and isospin indices in the middle of the Euclidean time evolution. The amplitude vanishes unless all the nucleons pass through the pinholes.  The right panel shows one set of pinhole configurations for the sixteen nucleons comprising $^{16}$O.
 \label{fig:pinholes}}
\end{figure}

The insertion of pinholes in the middle of the Euclidean time evolution is illustrated in the left panel of Fig.~\ref{fig:pinholes}.  The right panel of Fig.~\ref{fig:pinholes} shows one configuration of pinholes for the sixteen nucleons comprising $^{16}$O.  While not quite as pronounced as in $^{12}$C \cite{Shen:2022bak}, the $^{16}$O configurations show evidence of significant alpha clustering.  

The $^{16}$O pinhole configurations are for the first time used in Ref.~\cite{Summerfield:2021oex}, as initial states for simulations of $^{16}$O+$^{16}$O collisions at RHIC and the LHC.  Comparisons are made using initial states derived from the pinhole configurations versus uncorrelated Woods-Saxon nucleon distributions.  These calculations show a clear difference between pinhole and Woods-Saxon results for elliptic flow observables such as $v_4\{2\}/v_2\{2\}$ and $v_4\{4\}^4$ in central collisions at LHC energies of $\sqrt{s_{NN}}=6.5$~TeV.

Configurations for $^{20}$Ne are also available. They are employed in Ref.~\cite{Giacalone:2024luz} to demonstrate \nene{} collisions as a tool for precision studies of collectivity in small system, as well as in Ref.~\cite{Giacalone:2024ixe} to make the first hydrodynamic predictions for $^{208}$Pb+$^{20}$Ne collisions at the SMOG2 system of the LHCb detector. We emphasize that the geometry of collisions involving $^{20}$Ne isotopes obtained from the pinhole configurations is consistent with that obtained starting from an intrinsic nuclear density derived within the \textit{ab initio} PGCM picture, where $^{20}$Ne presents the characteristic shape of a bowling pin.

The recent development of wave function matching \cite{Elhatisari:2022zrb} has enabled lattice calculations at next-to-next-to-next-to-leading order (N3LO) in chiral EFT. These calculations are the first {\it ab initio} results able to provide an accurate description of nucleon-nucleon scattering, binding energies of light and medium-mass nuclei, charge radii of light and medium-mass
nuclei, and saturation properties of symmetric nuclear matter. Work is now in progress to compute pinhole configurations for a range of light and medium-mass nuclei using high-fidelity chiral interactions.  These pinhole configurations will be made publicly available.

\newpage

\section{Geometry of high-energy isobar collisions}
\label{sec:4}

We now move on to the study of nuclear structure effects in isobar collisions. In particular, our goal is to understand whether the observations made by the STAR collaboration at RHIC can be explained on the basis of differences in the structure of the collided isotopes.

\subsection{Initial-state estimators}

  Before running costly dynamical calculations of isobar collisions (which will be done in the next Sec.~\ref{sec:5}), we perform calculations of isobar ratios by means of predictors of final-state quantities that we can conveniently compute in large-scale initial-state model simulations.
  
  We consider that the initial condition of the QGP is given by the entropy density, $s({\bf x}, \tau_{\rm hyd})$, in the transverse plane (${\bf x}$) at some time ($\tau_{\rm hyd}$) where the hydrodynamic evolution starts. Depending on whether the considered model propagates energy from the initial time after the collision ($\tau=0^+$) until the hydrodynamization time, $\tau_{\rm hyd}$, the entropy field can be accompanied by a non-zero velocity field, $u^\mu$. 
  
  With these fields as input and setting $\tau=\tau_{\rm hyd}$, the quantities in which we are interested are the following. The total entropy and total energy,
\begin{equation}
    S \propto \int d^2{\bf x}~ u^0({\bf x},\tau) s({\bf x},\tau), \hspace{30pt}      E = \int d^2{\bf x} ~{\rm EOS}(s({\bf x},\tau)) ,
 \end{equation}
of the initial conditions. Note that the definition of the energy requires the equation of state of the system (EOS), which we shall take either from lattice QCD calculations, or from a simple ideal gas law, $e\propto s^{4/3}$, such that
\begin{equation}
     E \propto  \int d^2 {\bf x} \, e({\bf x},\tau) \propto \int d^2{\bf x} \, s({\bf x},\tau)^{4/3}.
\end{equation}
Note that, as we deal with initial states at very high temperatures, the ideal gas EOS turns out to be an excellent working assumption.

Moving then to features that quantify the anisotropy of the initial geometry,  we introduce the standard complex eccentricity harmonics of the initial entropy profile, defined by:
\begin{equation}
     \mathcal{E}_n = \frac{1}{S \langle r^n \rangle} \int d^2{\bf x}~ |{\bf x}|^n e^{in\phi} u^0 s({\bf x},\tau),
\end{equation}
for $n=2,3$, where $\phi={\rm atan2}(y,x)$, and where we introduce the $n$th power of the initial radius
 \begin{equation}
     \langle r^n \rangle = \frac{1}{S} \int d^2{\bf x} \, |{\bf x}|^n\, u^0 s({\bf x},\tau), \hspace{20pt} {\rm for}~n=2,3,4.
 \end{equation}
The magnitude and phase of the eccentricity vector, $\mathcal{E}_n=\varepsilon_ne^{i\psi_n}$, are given by
\begin{align}
    \varepsilon_n = |\mathcal{E}_n|,  \hspace{20pt}     \psi_n = \frac{1}{n} {\rm atan2} \bigl ( {\rm Im}(\mathcal{E}_n),  {\rm Re}(\mathcal{E}_n)  \bigr ) + \frac{\pi}{n}.
\end{align}

The relation between these initial-state quantities and the final-state observables has been documented at length in the literature (see e.g. Ref.~\cite{Giacalone:2020ymy} for a review). Driven by simple thermodynamic or hydrodynamic considerations, one expects in particular,
\begin{align}
    S &\propto N_{\rm ch} \, , \hspace{20pt} E \propto [p_t] \, ,\hspace{20pt} 
    \mathcal{E}_n \propto V_n \, ,
\end{align}
where $N_{\rm ch}$ is the charge-particle multiplicity within a given considered acceptance, $[p_t]$ is the average transverse momentum of the final-state hadrons, while $V_n$ is the $n$th order complex Fourier coefficient of their azimuthal distribution. These proportionality relations permit us to estimate isobar ratios from initial-state quantities alone, and get an idea of the influence of the various nuclear structure effects.

In regard to the latter, the impact of the properties of the nuclear geometry on the isobar ratios can be understood from a simple Taylor expansion around unity \cite{Giacalone:2021uhj,Jia:2021oyt}. Consider that the intrinsic nuclear density of the collided isotopes is expanded, as done before, following:
\begin{equation}
    \rho(r,\theta,\phi) = \frac{\rho_0}{1+e^{[r-R(\theta,\phi)]/ a }}, \hspace{20pt}  R(\theta,\phi) =  R_0  \left(1+ \beta_2  Y_2^0(\theta,\phi)  +  \beta_3  Y_3^0(\theta,\phi)+\cdots\right).
\end{equation}
For a given observable, $\mathcal{O}$, the corresponding isobar ratio reads \cite{Jia:2021oyt}
\begin{equation}
    \frac{\mathcal{O}_{\rm Ru+Ru}}{\mathcal{O}_{\rm Zr+Zr}} = 1 + c_1 \Delta \beta_2^2 + c_2\Delta\beta_3^2 + c_3\Delta a+c_4 \Delta R_0,
\end{equation}
where $\Delta$ denotes the difference between parameters in $^{96}$Zr and those in $^{96}$Ru, e.g., $\Delta a = a_{\rm Ru}-a_{\rm Zr}$. Note that the deformation parameters enter squared in the expression of the observable under consideration in this document \cite{Giacalone:2021udy,Jia:2021tzt}. The coefficients $c_n$ are determined by the hydrodynamic response of the system and, for central collisions, they vary little with the mass number of the colliding species. Depending, then, on the observable under study there will be more or less sensitivity to a subset of the nuclear density parameters \cite{Nijs:2021kvn,Jia:2022qgl}, allowing us to reconstruct a complete picture of these nuclei from data. This is the unprecedented power of the nuclear imaging method based on isobar collisions.

\subsection{Large-scale \texorpdfstring{\trento{}}{trento} simulations}
\label{sec:trento}

\subsubsection{Objective}
The task of this subsection is to investigate the impact of nuclear structure and of different collision parameterizations on a set of observables. The chosen observables are the average energy $\langle E \rangle$ of the collisions, and the first two eccentricities $\varepsilon_2$ and $\varepsilon_3$. 

\begin{table}
\centering
\begin{tabular}{c | c | c | c | c | c }
Case Label & $R_0$ (fm) & a (fm) & $\beta_2$ & $\beta_3$ & $\gamma$ \\ \hline
Case 1     & 5.09       & 0.46   & 0.16      & 0         & $\pi/6$  \\
Case 2     & 5.09       & 0.46   & 0.16      & 0         & 0 \\
Case 3     & 5.09       & 0.46   & 0.16      & 0.20       & 0 \\
Case 4     & 5.09       & 0.46   & 0.06      & 0.20      & 0 \\
Case 5     & 5.09       & 0.52   & 0.06      & 0.20      & 0 \\
Case 6     & 5.02       & 0.52   & 0.06      & 0.20      & 0 \\
\end{tabular}
\caption{Nuclear structure parameters employed for generating initial conditions of isobar collisions.}
\label{tab:nuclear_param}
\end{table}

Each collision system differs in terms of nuclear radius $R_0$, diffuseness $a$, quadrupole deformation parameter $\beta_2$, octupole deformation parameter $\beta_3$ and triaxiality deformation parameter $\gamma$. We compute each observable in Case 1 to Case 5, as defined in Table \ref{tab:nuclear_param}, and compare it systematically with the default configuration (Case 6) by studying the ratio with respect to the latter case. The initial conditions are produced using the \trento{} model. For each choice of nuclear structure parameters, we take different choices of the \trento{} parameters, namely, the reduced thickness parameter, $p$, the nucleon width, $w$, the entropy fluctuation variance, $k$, the minimum nucleon-nucleon distance, $d$, the number of nucleon constituents, $m$, and the nucleon constituent width, $v$. In the next subsection we provide a detailed explanation of the meaning of each parameter. The different choices of \trento{} parameters which are studied in the following are reported in Table \ref{tab:trento_param}.

\subsubsection{A short description of \texorpdfstring{\trento{}}{trento} parameters}

In the \trento{} model~\cite{Moreland:2014oya,Moreland:2018gsh}, the nucleons are sampled at location $\vec{r}_i$ distributed inside a nucleus based on the  Woods-Saxon distribution (WS),
\begin{equation}
    \rho(r,\theta,\phi) = \frac{\rho_0}{1+e^{[r-R(\theta,\phi)]/ a }},
\end{equation}
where $a$ is the skin thickness, and  $R(\theta,\phi)$ quantifies the overall deformation of the nucleus and is written in terms of (real) spherical harmonics,
\begin{equation}
     R(\theta,\phi) =  R_0  \left(1+ \beta_2  \left[ \cos\gamma Y_2^0(\theta,\phi) +\sin\gamma Y_2^2(\theta,\phi) \right]+  \beta_3  Y_3^0(\theta,\phi)+\cdots\right).
\end{equation}
    The quantity $R_0$ controls the radius and nonvanishing $\beta_2$, $\beta_3$, and $\gamma$ lead to quadrupole, octupole, and triaxial deformation of the nucleus. A two-body short-range correlation is imposed on the distributed nucleons position $\vec{r}_i$ and $\vec{r}_j$ ($i\neq j$) by condition $|\vec{r}_i-\vec{r}_j|>d$.
    
    Nucleons are supposed to have $m$ substructure degrees of freedom whereby their location is sampled based on a Gaussian profile  $\exp\left[-|\vec{r}|^2/2w^2\right]$ with nucleon width $w$, centered at $\vec{r}_i$. The substructure constituent has a Gaussian profile  $\exp\left[-|\vec{r}|^2/2v^2\right]$  with width $v$. In case of no substructure, $m=1$, we have $v=w$. The contribution of each constituent to the thickness function $T_A(x,y)$ is weighted by $\gamma_i$ factor, a random variable sampled by a gamma distribution with a unit mean value and variance $1/k$. The fluctuating $\gamma_i$ factor is included to  explain the large multiplicity fluctuation in proton--proton collisions.

    In an ion-ion collision, some nucleons participate to the interaction with a probability of nucleon-nucleon interaction that depends on inelastic nucleon-nucleon cross section $\sigma_{NN}^{\text{inel}}$ and the parameter $w$. From the coordinates of the participant nucleons, one computes the participant thickness functions $T_{A,B}(x,y)$, as a linear superposition of gamma-fluctuation-weighted nucleonic profiles. Finally, the physics of the entropy production is phenomenologically modeled by the following scale invariant function called reduced thickness function,
\begin{equation}\label{reducedThickness}
    T_R(p;T_A,T_B) \equiv N \left(\frac{T_A^p+T_B^p}{2}\right)^{1/p},
\end{equation}
    where $p$ is the aforementioned reduced thickness parameter, and $N$ is an overall normalization tuned to the desired collision energy. 

\begin{table}
\centering
\begin{tabular}{c | c | c | c | c | c | c}
Batch number & $p$ & $w$ & $k$ & $d$ & $m$ & $v$\\ \hline
Batch 1     & 0.0      & 0.5   & 1      &  0        & 1      & 0.5\\
Batch 2     & 1.0      & 0.5   & 1      &  0        & 1     & 0.5\\
Batch 3     & -1.0     & 0.5   & 1      &  0     & 1     & 0.5\\
Batch 4     & 0.0      & 1.2   & 1      &  0    & 1     & 1.2\\
Batch 5     & 0.0      & 0.5   & 16     &  0     & 1    & 0.5\\
Batch 6     & 0.0      & 0.5   & 1      & 1.0       & 1     & 0.5\\
Batch 7     & 0.0      & 0.5   & 1      &  0     & 4     & 0.1\\
Batch 8     & 0.0      & 0.5   & 1      &  0     & 4     & 0.4\\
\end{tabular}
\caption{Collision parameters employed for generating initial conditions of isobar collisions with \trento{}.}
\label{tab:trento_param}
\end{table}

\subsubsection{Results}
\begin{figure}[H]
    \centering
    \includegraphics[scale=0.7]{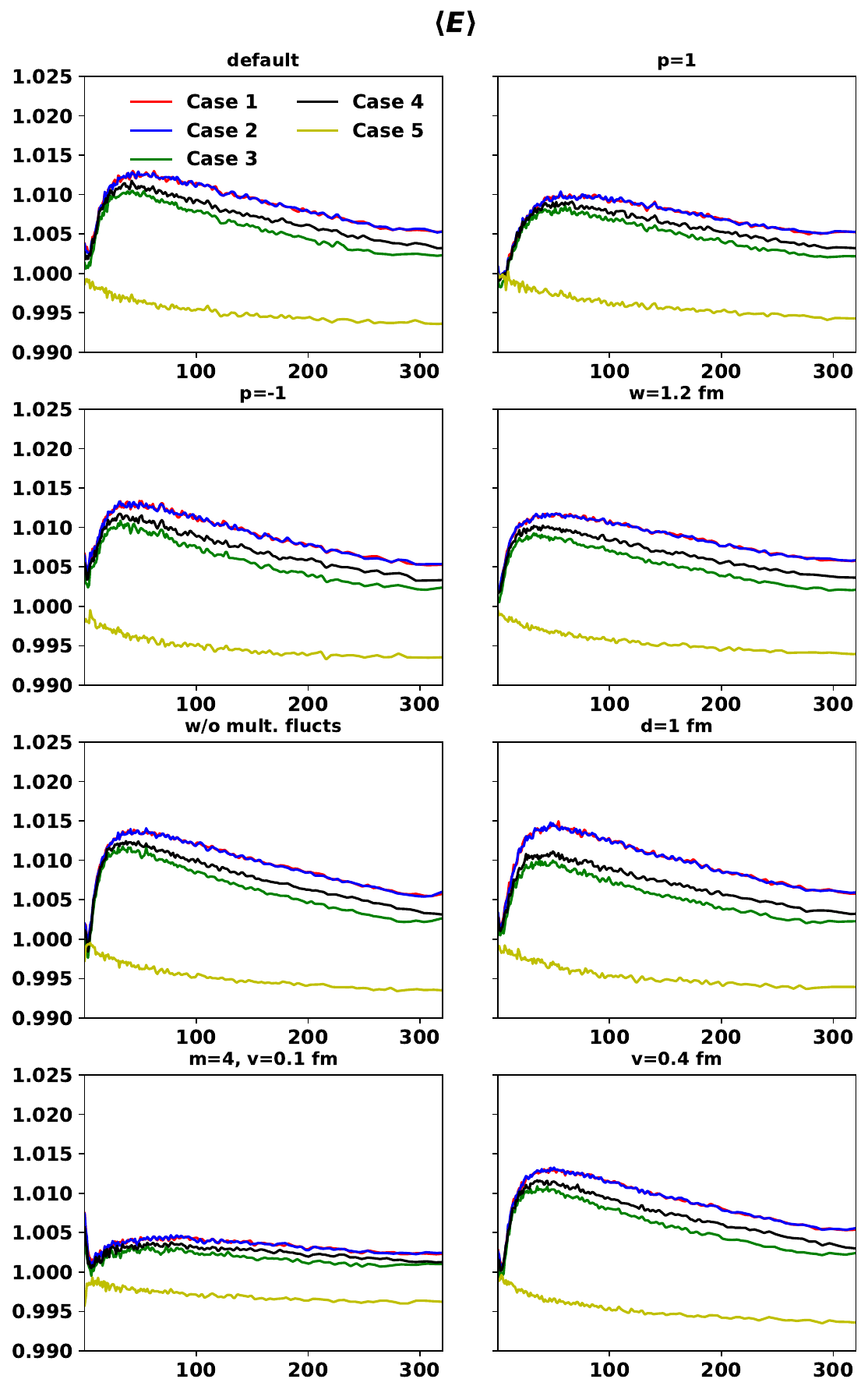}
    \caption{Average energy ratio as a function of entropy for all different configurations of nuclear and \trento{} parameters.}
    \label{fig:E_trento}
\end{figure}

\begin{figure}[H]
    \centering
    \includegraphics[scale=0.7]{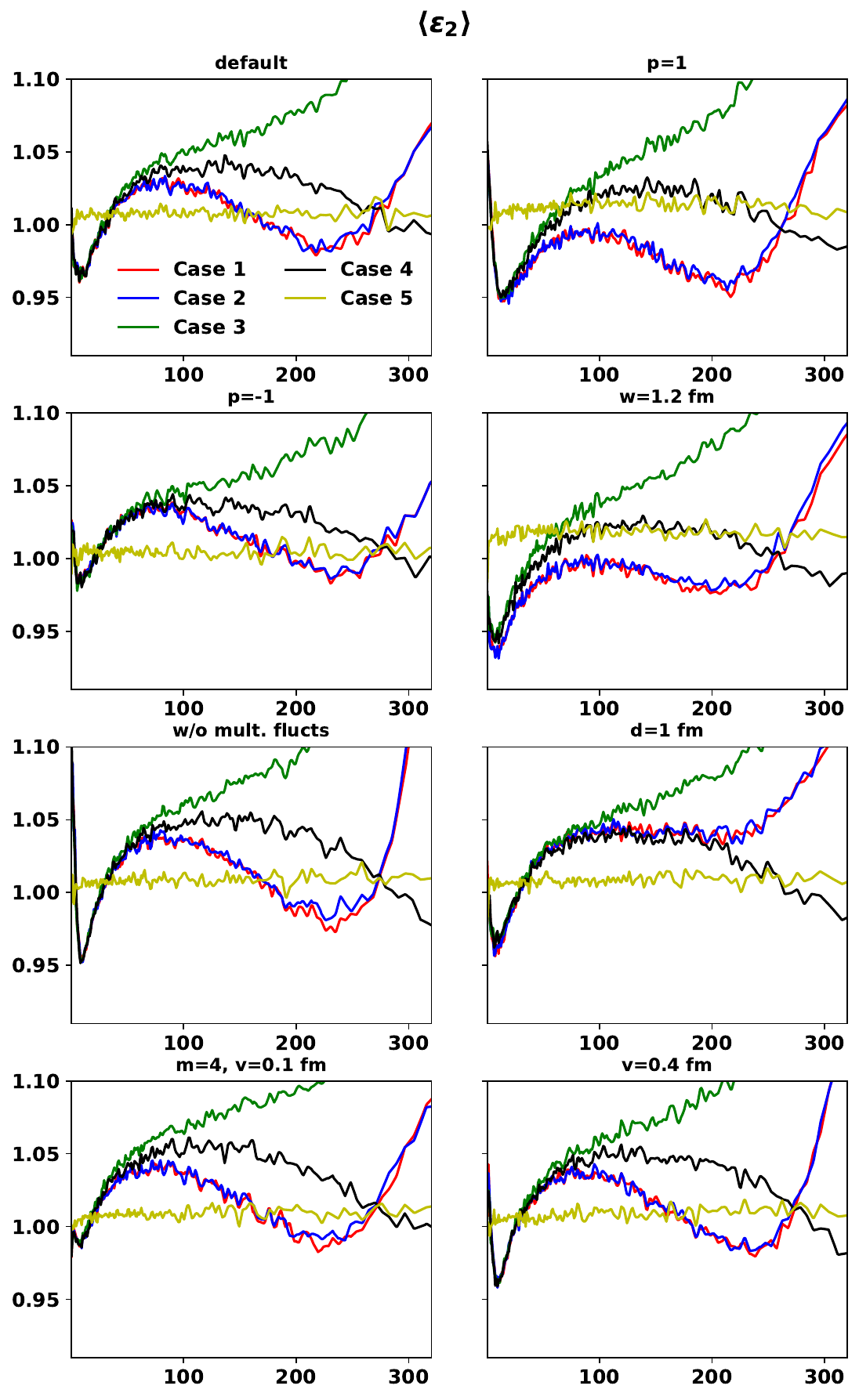}
    \caption{Second harmonic eccentricity $\epsilon_2$ ratio as a function of entropy for all different configurations of nuclear and \trento{} parameters.}
    \label{fig:eps2_trento}
\end{figure}

\begin{figure}[H]
    \centering
    \includegraphics[scale=0.7]{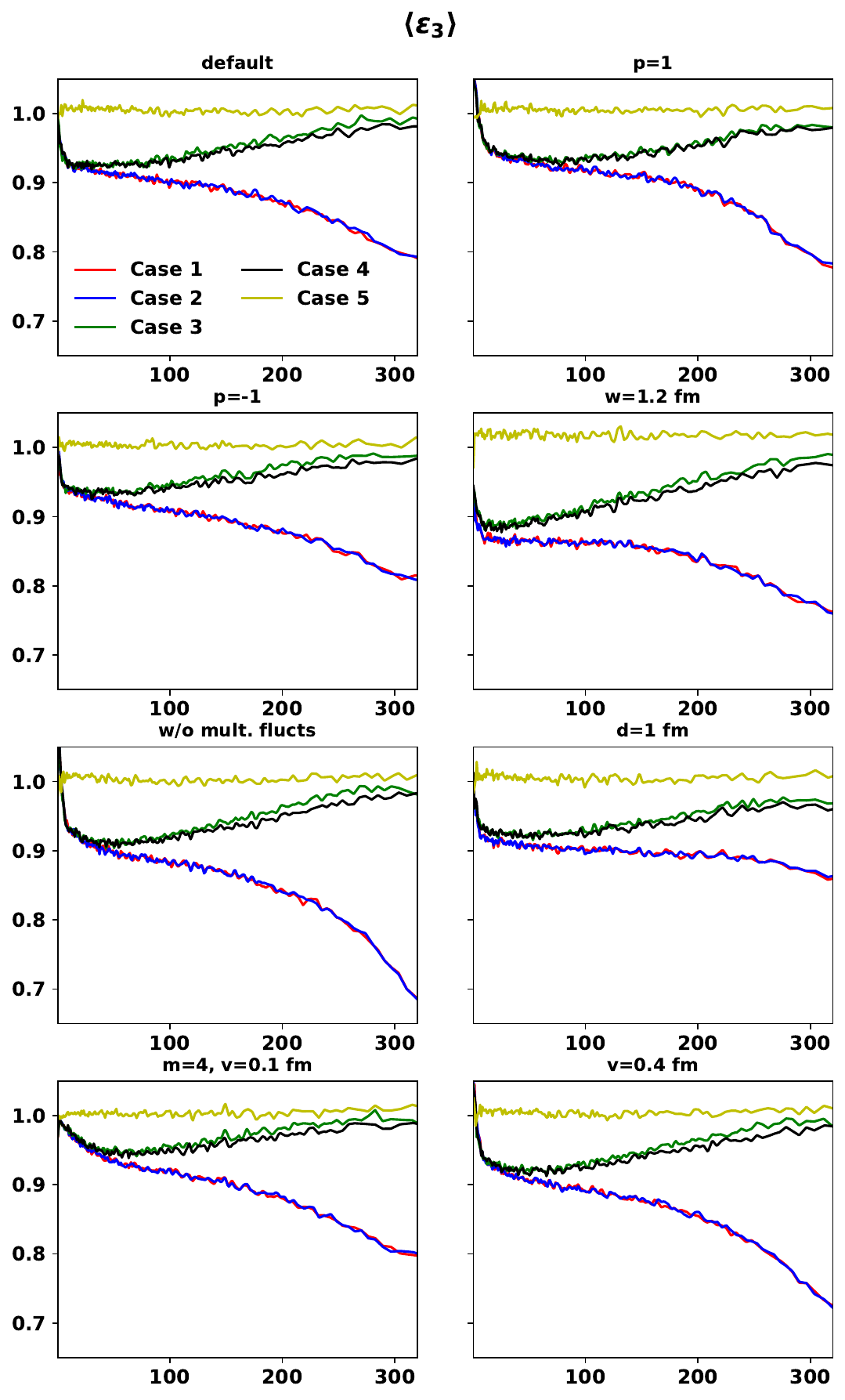}
    \caption{Third harmonic eccentricity $\epsilon_3$ ratio as a function of entropy for all different configurations of nuclear and \trento{} parameters.}
    \label{fig:eps3_trento}
\end{figure}

A total of 20 million minimum-bias events is generated for each combination of nuclear system configurations and \trento{} parametrizations. In Figs. \ref{fig:E_trento}, \ref{fig:eps2_trento}, \ref{fig:eps3_trento} the results for the energy, second and third harmonic eccentricities ratios are presented as a function of the entropy of the system. These plots indicate that different nuclear structure configurations (Case 1 to 5) always induce a certain trend in the observables, irrespective of the specific \trento{} parametrization taken into consideration. Therefore, we will first discuss the impact of each nuclear geometry parameter for each observable under the default \trento{} configuration. Afterwards, we will examine the impact of changing \trento{} parameters. 

\paragraph{Default \trento{} configuration}

\begin{itemize}
    \item \textbf{The role of diffuseness.} To see the effect of changing $a$, one has to look at ratio variations as we move from Case 4 to Case 5. The first observation is that each observable ratio is highly sensitive to the diffuseness $a$. In fact, only in Case 5 -- where we compare systems with equal $a$ -- the ratio is basically 1 throughout the whole entropy range. In the case of the average energy,  $\langle E \rangle$, the residual difference  is a volume effect caused by the smaller radius, $R_0$, that Case 6 presents compared to Case 5, which leads to a reduction of the ratio when going to full overlap, i.e. in the central collisions. Introducing a different diffuseness while keeping the other parameters fixed (Case 4), on the other hand, generates a sensible change in all ratios. Considering the average energy of the system, decreasing the diffuseness causes a increase in the ratio of a few percent. The increase is higher at larger centrality (lower entropy). This is consistent with the fact that the energy of the system is related to the inverse of the average system size. From the effect of the diffuseness alone, we expect, thus, that the isobar ratio of the average transverse momenta of the collisions should present a similar behavior and depart from unity in a similar centrality-dependent manner. This is consistent with both preliminary STAR data and the hydrodynamic calculations discussed in Sec.~\ref{sec:5}. It is noteworthy that the only case in which the impact of varying $a$ is reduced is when the \trento{} parametrization includes multiple nucleon constituents with very peaked distributions. Eccentricities also display a dependence on the value of $a$. In particular, a sharper system (Case 4) enhances the value of the second harmonic $\langle \epsilon_2\rangle$ (Fig.~\ref{fig:eps2_trento}) as the centrality increases, with a peculiar. In peripheral collisions ($S<100$), however, it causes a dip in the eccentricity ratio. Conversely, the third harmonic is suppressed with a smaller $a$, again in a way that has a specific centrality dependence. Note that, since Case 3 and Case 4 share the same octupole deformation as Case 5 and 6, an event with full overlap (central collisions) will result in a very similar third harmonic eccentricity in all four cases. We note that, overall, a change in $a$ parameter (i.e., going from Case 4 to Case 5) gives an effect on $\varepsilon_2$ and $\varepsilon_3$ that is fully consistent with that observed for $^{238}$U in the right-hand side of Fig.~\ref{fig:1409}.

\item \textbf{The dependence on quadrupole and octupole deformation parameters.} While $\beta_2$ and $\beta_3$ do not strongly influence the energy of the system (Fig.~\ref{fig:E_trento}), they highly affect the harmonic eccentricities in Figs.~\ref{fig:eps2_trento} and \ref{fig:eps3_trento}.  A higher quadrupole deformation (Case 3) is correlated with an enhancement in $\langle \epsilon_2 \rangle$ when going to more central collisions. The third harmonic eccentricity, on the other hand, does not display any dependence on $\beta_2$. This is expected since a quadrupole deformation does not affect the triangularity of the system, while it contributes to its ellipticity \cite{Jia:2021tzt}. In Case 2, the ruthenium has no octupole deformation ($\beta_3 = 0$). As expected, this has a strong impact on the third harmonic. The lack of octuple deformation suppresses the triangularity of the system, especially going towards central collisions. This effect is enhanced in the absence of nucleon fluctuations (w/o mult. flucts) in the \trento{} implementation, since these are system-independent and the dominant contribution (larger than that coming from the geometric effect due to $\beta_3$) to $\epsilon_3$ in central collisions in symmetric systems.

\item \textbf{Dependence on half-width radius and triaxiality.} The role played by the half-width nuclear radius is probed by Case 5, where all the parameters are set to default except from $R_0$. We notice that the ratio is flat and close to 1 for all observables in the full entropy range, regardless of the \trento{} parameters, with exception of the energy, as discussed above. We conclude that a relative variation of this parameter of order 1\% does not lead to any visible variation in the considered observables. Triaxiality is taken into account by comparing Case 1 to Case 2. Once again, the two curves overlap independently of the observable type or the \trento{} parameter in the whole entropy range. We conclude that the observables considered here are essentially insensitive to the value of $\gamma$. Note that this is expected from the calculations of Ref.~\cite{Jia:2021qyu}, showing that, to leading order, the mean squared eccentricities do not the depend on $\gamma$. 

\end{itemize}

\paragraph{Varying \trento{} parameters} 

Revisiting the average energy of the collision systems (Fig. \ref{fig:E_trento}), we observe similar ratios in all panels, except for the case of having four small constituents in each nucleon ($m=4$ and $v=0.1$ fm). As mentioned above, this choice of parameters leads to nucleons that consist of four almost point-like subparticles. This can be understood as follows. For highly spiky initial conditions, local density fluctuations [the first term on the right-hand side of Eq.~(\ref{eq:1p2p})] become much more dominant than the effects of deformations and two-body correlations [second line on the right-hand side of Eq.~(\ref{eq:1p2p})]. This naturally drives the ratio toward unity.

The same argument explains why the ratios of harmonic eccentricities in Figs. \ref{fig:eps2_trento} and \ref{fig:eps3_trento} depend somewhat on the amount of fluctuations included in the collision. In the case of no multiplicity fluctuations, Case 1 and 2 seem to deviate the most from unity. Higher local density fluctuations tend to mask the effect of the nuclear geometries, driving the isobar ratios close to unity. This also explains why the isobar ratios deviate more from unity when the $m=4$ partons populating the colliding nucleons have a large width ($v=0.4$ fm), as that dampens local fluctuations.

\subsubsection{Additional scan and comment on the minimum inter-nucleon distance}

\begin{figure}[t]
    \centering
    \includegraphics[width=.9\linewidth]{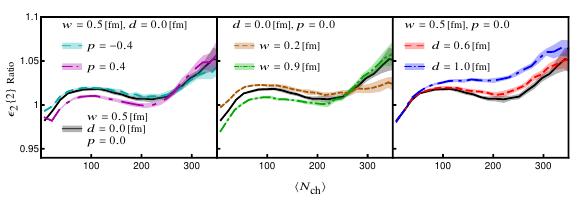}
    \caption{Second harmonic eccentricity ($\varepsilon_2$) ratio as a function of charge multiplicity for seven points in the parameter space $(w,d,p)$. The black curve is the same in all panels. }
    \label{fig:parameterScan_trento}
\end{figure}

We now assume fixed deformations for the colliding nuclei nuclei and perform additional \trento{} parameter variations for structureless nucleons ($m=1$). We focus on Case 2 (for $^{96}$Ru) and Case 6 (for $^{96}$Zr) in Table~\ref{tab:nuclear_param}, and study the ratio of second harmonic eccentricity. In particular, we vary $w$ between 0.2 to 0.9 fm and $d$ between 0.0 to 1.0 fm with 0.1 step size and also $p$ between -0.4 to 0.4 with 0.2 step size. There are 440 points in the parameter space in total, and at each point, we generate fifteen million minimum bias events. The result for seven selected points is depicted in Fig.~\ref{fig:parameterScan_trento}. The black curve is common among all three panels. As seen from the left panel, varying $p$ from -0.4 to 0.4 has a rather small impact on the ratio in the mid-central and central collisions. In central collisions and large $w$, the thickness functions $T_A$ and $T_B$ are approximately similar which reduces the effect of $p$ in Eq.~\eqref{reducedThickness}. The effect of $w$ variation is depicted in the middle panel of Fig.~\ref{fig:parameterScan_trento}. For $w\gtrsim 0.5\,$ fm, the impact of $w$ is small in central collisions where we expect a significant contribution from nuclei deformation. Sharper nucleons increase the effect of fluctuation even in the central collisions. The most interesting case is the variation of $d$ (Fig.~\ref{fig:parameterScan_trento} right panel), indicating a significant dependence in central collisions. Imposing a central impenetrable core for the nucleons modifies the fluctuation behavior. In extreme cases with large $d$, one expects small room for fluctuations. This is also consistent with the fact that $d$ and the fluctuation parameter, $k$, are largely correlated in Bayesian analyses \cite{Nijs:2021clz}. 

Using a chi-square test, we analyzed the sensitivity of the $\epsilon_2$ ratio in the $(w,d,p)$ parameter space, indicating that this ratio has significant sensitivity on the parameter $d$ in central collisions. This remark is particularly interesting because previous global Bayesian analyses constrained parameter $d$ poorly~\cite{Bernhard:2019bmu,JETSCAPE:2020mzn,Nijs:2020ors,Parkkila:2021tqq,Parkkila:2021yha}. Our analysis shows that the second harmonic elliptic flow isobar ratio is an excellent candidate for pinpointing the minimum distance parameter, and the physics of the short-range nucleon-nucleon repulsion in general.

\subsection{Role of the free streaming time}

\subsubsection{Objective}

State-of-the-art simulations of heavy-ion collisions typically consist of four phases. These include the generation of initial conditions, prethermalization, the calculation of fluid expansion using viscous relativistic hydrodynamics, and finally, hadronic transport. While the multiplicity distribution and collective flow are sensitive to the value of viscosity, as well as the inclusion of final state interactions and free streaming, it is expected that ratios of these observables taken between different colliding systems that are close in mass only arise due to differences in the structure of the considered isotopes. Although these assumptions are reasonable, the precision of experimental data requires testing of them. The objective of this subsection is to investigate the effect of the prethermalization time, which was not considered in the previous section where only initial condition parameters within the \trento{} framework are varied. This is done in the scenario where immediately after production, partons free stream until the beginning of the hydrodynamic evolution. We note that such an approach has been used in various recent Bayesian analyses, where one searches for the most likely range of parameter values in the hydrodynamic scenario that reproduce data \cite{Bernhard:2019bmu,JETSCAPE:2020shq,JETSCAPE:2020mzn,Parkkila:2021tqq, Nijs:2020roc,Nijs:2020ors,Liyanage:2023nds}.

\subsubsection{Method}
To achieve our objective, we use the Woods-Saxon distribution as the probability $P(r,\, \theta,\, \varphi)$ of finding a nucleon within a nucleus at position $\vec{r} = (r,\, \theta,\, \varphi)$, given by

\begin{align}
\begin{aligned}
P(r,\, \theta,\, \varphi)  & =  \frac{\rho_0}{1+\exp\left\{[r-\mathcal{R}(\theta,\, \varphi)]/a\right\}} \\
\mathcal{R}(\theta,\, \varphi) & = R_0\left\{1+\beta_2 \left[Y^0_2\left(\theta,\, \varphi\right)\cos\gamma + \sqrt{2} \sin \gamma\, \Re Y^2_2\left(\theta,\, \varphi\right) \right]+\beta_3Y^0_3\left(\theta,\, \varphi\right) \right\}\,,
\end{aligned}
\end{align}
where the nuclear radius $R_0$, diffuseness $a$, quadrupole deformation parameter $\beta_2$, triaxiality deformation parameter $\gamma$ and octupole deformation parameter $\beta_3$ will characterize the nuclei structure. The values of these parameters  are shown in Table \ref{tab:nuclear_param}. For each case, 50\,000 nuclear configurations are created.
These are used as input for the \trento\, model \cite{Moreland:2014oya}, where they are randomly oriented and employed in the generation of the subsequent initial conditions. For each of these cases, 1 million events are generated. Note that, when a free streaming phase is included, the \trento{} model becomes a model for the initial energy density of the system, as opposed to the entropy density which was considered in the previous subsection.

After the collision takes place, one needs to specify how long the produced system spends in the free-streaming phase.
Two cases are investigated: a fixed free-streaming time, $\tau_{FS}=1\,fm$ and a centrality-dependent free-streaming time \cite{JETSCAPE:2020mzn} $\tau_{FS} = \tau_R(\{\epsilon\}/\{\epsilon_R\})^{\alpha}$, where $\{\epsilon\}$ is the average initial energy density of a given collision while $\{\epsilon_R\} = 4.0$ GeV/fm$^2$, $\tau_R = 1.36$ fm/c and $\alpha = 0.031$, as obtained in Ref.~\cite{JETSCAPE:2020mzn}. In practice, this latter free-streaming time varies between $\approx$ 1.35 and $\approx$ 1.49 fm.

Two consequences of free-streaming are expected. First, small scale features are smeared out and the overall geometry is smoothed, resulting in smaller eccentricities $\varepsilon_n$ and larger average radius $R$ at the onset of the hydrodynamic phase. We remind that these quantities are defined as 
\begin{align}
    \label{eq:ecc}
    \varepsilon_n & = \frac{|\{r^ne^{in\phi}\}|}{\{r^n\}}, \\
    R & = \sqrt{\{r^2\}}\,,
\end{align}
where $r=\sqrt{x^2+y^2}$ and $\phi$ are, respectively, the radius and azimuthal angle in the transverse plane (defined after recentering the initial condition such that the center of mass of the energy density is at the origin), while $\{\cdots\}$ denotes an energy density-weighted spatial average (with appropriate Lorentz factors). Second, the Landau matching procedure is used. It connects the free-streaming phase (described in terms of the energy-momentum tensor components) to the hydrodynamic phase (which uses local thermodynamic variables plus fluid flow) \cite{Liu:2015nwa,Broniowski:2008qk}, based upon the assumption of a sudden equilibration of the system.

\subsubsection{Results}

An important factor to consider in this study is the criterion for classifying events. In experiments, multiplicity is often used as a classifier. In our work, we propose using the total energy, computed by integrating the initial \trento\, energy-density profile, as a proxy for total multiplicity\footnote{Preliminary hydrodynamic simulations performed by our group show a linear relation between the total energy for the initial condition and charged multiplicity for the final state, supporting this choice}. Thus, it is reasonable to perform a centrality selection for each case based on this quantity. It is important to note that because of variations in nuclear geometry, events with similar total energies are obtained with slightly different probabilities. Although the difference is small, implementing a centrality selection independently for each case means comparing systems that may have different final multiplicities. To address this concern, we adopt a joint centrality selection approach by considering events generated in all six cases from table \ref{tab:nuclear_param}. The resulting bin edges of energy are then used for each of the 6 separate cases. In the centrality range 0-5\%, we employ a narrow bin width of 1\% to provide higher resolution for central events, where nuclear structure effects are expected to be more prominent. For the remaining range, we utilize a bin width of 5\% to reduce statistical error.

\begin{figure}[t]
       \begin{center}
  \includegraphics[width=\linewidth]{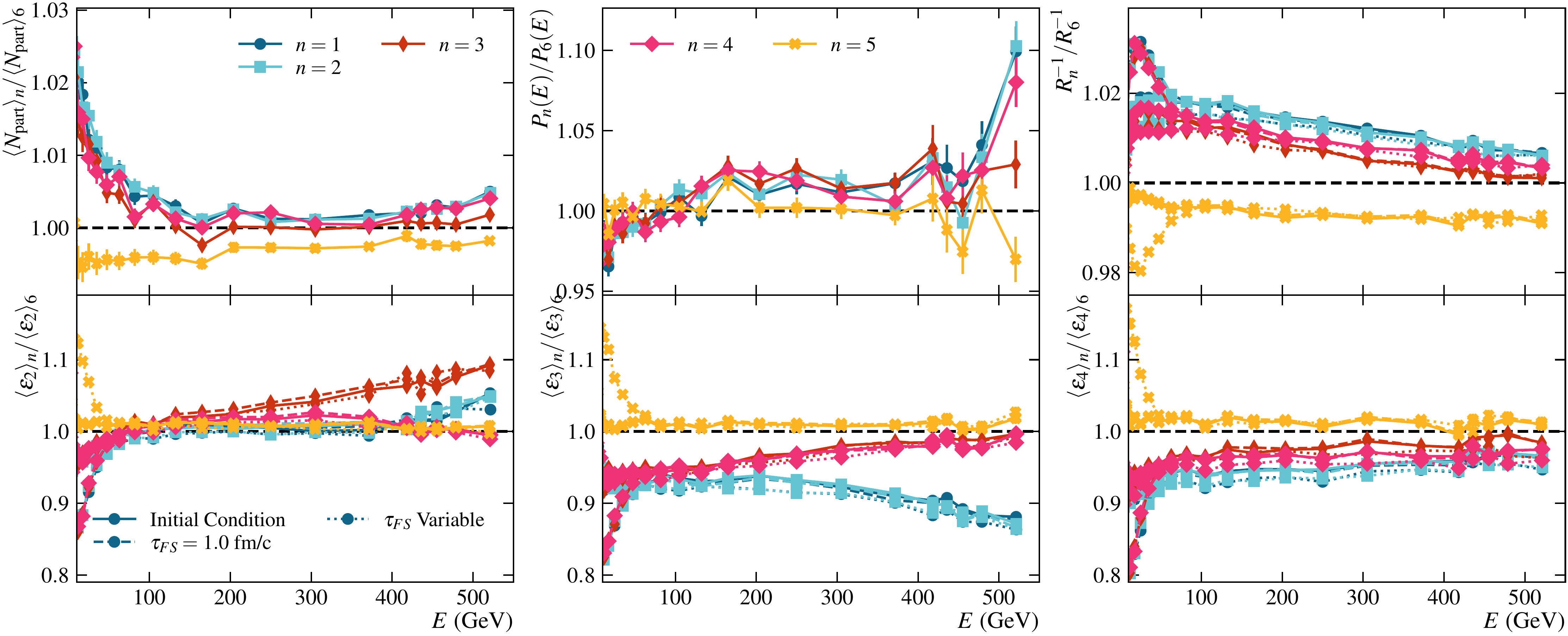}
\label{fig:freestreaming}
   \caption{Isobar ratios of various quantities for different nuclear structure parametrizations. Results for cases 1 to 5 are divided by case 6 as function of total initial energy for
   initial-conditions-only (solid line) in top leftmost panels, for initial conditions with fixed free-streaming time of 1 fm/c (dotted line) and for initial conditions with variable free-streaming time (dashed line) in the remaining panels. See text for discussion.}
  \end{center}
\end{figure}

In figure \ref{fig:freestreaming}, various quantities are shown for the results obtained from cases 1 to 5 and then divided by the results obtained in case 6. This is done for initial conditions alone (first two panels in the top row), and for initial conditions with either fixed free-streaming time of 1 fm/c or variable free-streaming time (remaining  panels). For each case and free-streaming time, one million events are generated. 

In the top left panel, ratios of participant number are close to one, with some slight deviations. More precisely, the number of participants for cases 1 to 4 compared to 6 are similar among themselves, but differs from case 6 for non-central collisions. As expected, the largest effect is observed when we move from case 4 to vase 5, induced by the change in skin thickness. For cases 5 and 6, where only the value of $R_0$ differs, the ratio is constant, but always below 1, due to a volume effect: in a given energy bin, for a larger $R_0$, there are less participants.

In the upper-mid panel, the probability ratios to get a certain value of $E$ are shown. Small departures from unity are observed, except for central collisions. This result is consistent with the previous discussion on the effect of the skin thickness parameter, $a$ (e.g., consistent with the interpretation of the STAR results in Fig.~\ref{fig:iso_ratio_star}). 

In the top right panel, ratios for the inverse mean radius are shown. Case 3, a non-zero $\beta_3$ together with a large $\beta_2$ makes this ratio go toward unity faster when compared to the cases where only the quadrupole deformation is present, that is, cases 1 and 2. Also, the trends are very close to those found for the $N_{\rm part}$ ratio, with the most dramatic effect coming from the increased diffuseness parameter in case 5, causing the isobar ratio to go below unity. The free-streaming time has no significant effect on the displayed ratio, the difference between the curves being at most 0.3\% though the radii can vary much more ~\cite{Serenone:2023zbn}.

The bottom left panel shows ratios for the mean ellipticity, $\varepsilon_2$.  Configurations 1 and 2 have very similar ellipticity as they differ only due to triaxiality, $\gamma$. Configuration 3 shows a higher ellipticity than configuration 2 even though both configurations only differ by the inclusion of an octupole deformation $\beta_3$, indicating the non-trivial interplay between $\beta_2$ and $\beta_3$ discussed in Ref.~\cite{Jia:2021tzt}. The decrease in ellipticity from configuration 3 to 4 is expected due to the decrease of the quadrupole deformation. Configurations 4 and 5 have similar ellipticity, with ratios close to 1, indicating that this quantity is insensitive to nuclear radius and diffuseness. The free-streaming time has no significant effect on ellipticities, being at most a 1\% effect.

The bottom-mid-panel shows ratios for triangularity, $\varepsilon_3$. The picture here is fully consistent with that of the \trento{} calculations reported in the previous section, with a dramatic impact of the octupole deformation of $^{96}$Zr and a depletion of the ratio in peripheral collision driven by the skin thickness. Once more, the effects of the free streaming evolution are only at the percent level and barely visible. 

Finally, the bottom right panel shows the ratios for $\epsilon_4$. Monte Carlo calculations indicate that this quantity is not affected by the quadrupole and octupole deformation parameters in central collisions \cite{Jia:2021tzt}. Indeed, in our results we do not see any visible impact of the deformations of the isobars. The departures from unity are mainly sourced by skin effects, which are essentially identical to those observed in the case of $\varepsilon_3$. 

\subsubsection{Discussion}
In conclusion, the isobar ratios that involve properties of the geometry of the collisions, as shown in Fig.~\ref{fig:freestreaming}, display very little dependence on the dynamics of the first fm/$c$ of the collisions. This corroborates the results of Refs.~\cite{Nijs:2021kvn,Zhang:2022fou}, showing in addition that corrections to the dynamics of the evolution of the system (e.g. viscosities or parton-parton cross sections) have little effects on the isobar ratios. One possible discriminant is the isobar ratio of average transverse momenta, $[p_t]$ \cite{Zhang:2022fou}, which will be discussed in the next section.

Furthermore, and as expected, there is no sensitivity to the triaxiality, $\gamma$. One observable able to probe such parameter is the Pearson correlator, $\rho_n$, between anisotropic flow and the mean transverse momentum~\cite{Bozek:2016yoj}, which is a highly sensitive probe of the deformation of the collided ions \cite{Giacalone:2019pca,Giacalone:2020awm,Jia:2021wbq,Bally:2021qys,ALICE:2021gxt,ATLAS:2022dov,Samanta:2023qem,Fortier:2023xxy,Fortier:2024yxs,Wang:2024vjf}. In the case of isobars, this correlator was discussed by \cite{Jia:2021qyu} (with a different estimator for the mean momentum). In \cite{Serenone:2023zbn}, we computed correlators $\rho_{2,3}$ with the approach of \cite{Giacalone:2020dln}, by considering that the anisotropic flow, $v_n$, is sourced by the eccentricity, $\varepsilon_n$, and that the mean transverse momentum is linearly proportional to the ratio of total energy to total entropy in a very narrow centrality bin. Our results in Ref.~\cite{Serenone:2023zbn} indicate that the isobar ratio of $\rho_n$ correlators enables one, indeed, to access triaxiality and potentially finer properties of the nuclear ground states.

\subsection{Results from SMASH initial conditions}
    \subsubsection{Model}
        We calculate the initial conditions for the hydrodynamical simulation within the hadronic transport approach SMASH \cite{Weil:2016zrk,SMASH_doi}. Such transport approaches are built to model heavy-ion collisions (HIC) at lower collisional energies (e.g. energies on the order of the GeV, as collected by e.g., the HADES spectrometer or at the fixed-target program at the STAR detector) or the late stages of an ultrarelativistic HIC.
        In a recent work \cite{Schafer:2021csj}, an infrastructure for a hybrid approach was built to fully describe the HIC in the energy range from $\sqrt{s} = 4.3$ GeV to $\sqrt{s} = 200$ GeV.
        Starting with the initial conditions from SMASH, the hot dense medium is then evolved using the hydrodynamical code vHLLE and the dilute stage is again modeled in SMASH.
        
        We only consider the first stage of the hybrid approach, with the goal of analyzing the initial condition for hydrodynamical modeling.
        The degrees of freedom in SMASH are all hadrons of the Particle Data Group (PDG) with masses up to $m\approx 2.35$ GeV reported with a certain confidence level \cite{ParticleDataGroup:2018ovx}.
    
        Transport codes rely on information on the constituents of matter, in our case hadrons and their masses and decay channels, as well as the cross sections between those particles. Whenever experimental measurements are available they serve as a basis for those inputs. In the energy regime between $\sqrt{s}\approx 2 - 4$ GeV the interactions are described by resonance formation or $2\leftrightarrow 2$ (in)elastic processes.
        For higher energies, which are more relevant to this study, SMASH employs PYTHIA8 \cite{Sjostrand:2006za, Sjostrand:2007gs} to perform the string fragmentation and hard scatterings during the evolution.
        In addition to the cross sections, a criterion is needed to check whether two particles collide. Here we employ a covariant form of the geometric collision criterion \cite{Hirano:2012yy} which uses the geometric interpretation of the cross section to decide if two particles collide or not.
    
        The shape of the two colliding nuclei is described by the Woods-Saxon distribution
        \begin{equation}\label{Eq:WoodsSaxon}
            \rho(r, \theta, \phi) = \frac{\rho_0}{\mathrm{exp}\left\{\frac{r - r(r_0, \theta, \phi)}{a}\right\} + 1} \, .
        \end{equation}
        Here $\rho_0$ is the ground state density, $a$ the diffuseness and $r(r_0, \theta, \phi)$ a function that describes the surface of the nucleus.
        A common ansatz is to expand the surface in terms of spherical harmonics
        \begin{equation}\label{Eq:RadialExpansion}
            r(r_0, \theta, \phi) = r_0\left\{ 1 + \beta_2\left[\mathrm{cos}\,\gamma Y_2^0(\theta, \phi) + \sqrt{2}\,\mathrm{sin}\gamma \, \mathrm{Re}(Y_2^2(\theta, \phi)) \right] + \beta_3 Y_3^0(\theta, \phi) \right\} \, ,
        \end{equation}
        where $r_0$, $\beta_{2,3}$ as well as $\gamma$ are parameters taken from experiments or calculations to describe the deformation of the nucleus \cite{Hammelmann:2019vwd}.         On an event-by-event basis, the positions of the nucleons are sampled from Eq. (\ref{Eq:WoodsSaxon}) and both nuclei are rotated randomly around all axes. They are then moved according to the impact parameter and in such a way that the outer shell of the spheres including the Lorentz contraction $(R + d) / \gamma$ touch each other at $t=0$.
    
        The nuclei are then propagated and the collisions are performed until the hadrons cross a hypersurface of constant $\tau_0$. The value of $\tau_0$ is determined from the time the two nuclei overlap but with a minimal value of $\tau_0 = 0.5$ fm (for $\sqrt{s_{NN}} = 200$ GeV).
    
        \begin{table}[hbt!]
            \centering
            \begin{tabular}{|l | c c c c c |}\hline
             name & $r_0\,{\rm [fm]}$ & $a\,{\rm [fm]}$ & $\beta_2$ & $\beta_3$ & $\gamma\, \mathrm{[^\circ]}$ \\\hline
             Case 1 $^{96}$Ru (full) & 5.09 & 0.46 & 0.16 & 0 & 30 \\
             Case 2 & 5.09 & 0.46 & 0.16 & 0 & 0 \\
             Case 3 & 5.09 & 0.46 & 0.16 & 0.20 & 0 \\
             Case 4 & 5.09 & 0.46 & 0.06 & 0.20 & 0 \\
             Case 5 & 5.09 & 0.52 & 0.06 & 0.20 & 0 \\\hline
             Case 6 $^{96}$Zr (full) & 5.02 & 0.52 & 0.06 & 0.20 & 0 \\\hline
            \end{tabular}
            \caption{Woods-Saxon parameter variations used in this work to initalize the SMASH simulations.}
            \label{Tab:WSCases}
        \end{table}
        As done in the previous sections, in order to systematically study the influence of shape of the colliding nuclei, the results are calculated by means of a range of Woods-Saxon parametrizations, which we recall for simplicity in Tab.~\ref{Tab:WSCases}.
        
        In addition to these six different cases, we thoroughly investigate the impact of short-range nucleon-nucleon correlations.         It is known from experimental measurements that there exist nucleon-nucleon (NN) short-range correlations (SRC) in momentum space, see e.g. \cite{Subedi2008,CLAS:2018xvc} which translate also into short-range correlations in coordinate space \cite{Alvioli:2013qyz}.
        These correlations are not captured sampling Eq. (\ref{Eq:WoodsSaxon}) as that does not forbid two nucleons from sitting very close in coordinate space. 
        As a result, the radial two-body density $\rho^{(2)}(r_{12})$, which describes the density of states between nucleon 1 and nucleon 2 does not vanish at small distances $r_{12}$ \cite{Alvioli2019}. Although the effect of short-range correlations may be captured to some extent by the inclusion of the $d_{\rm min}$ parameter that we discuss in Sec.~\ref{sec:trento}, to more realistically capture the NN SRC effects we use a method proposed by Alvioli \textit{et al.} \cite{Alvioli:2009ab} to generate nuclear configurations in agreement with a two-body nuclear density $\rho^{(2)}$ calculated independently \cite{Alvioli2008}.

\begin{figure}[t]
            \centering
            \includegraphics[width=0.9\textwidth]{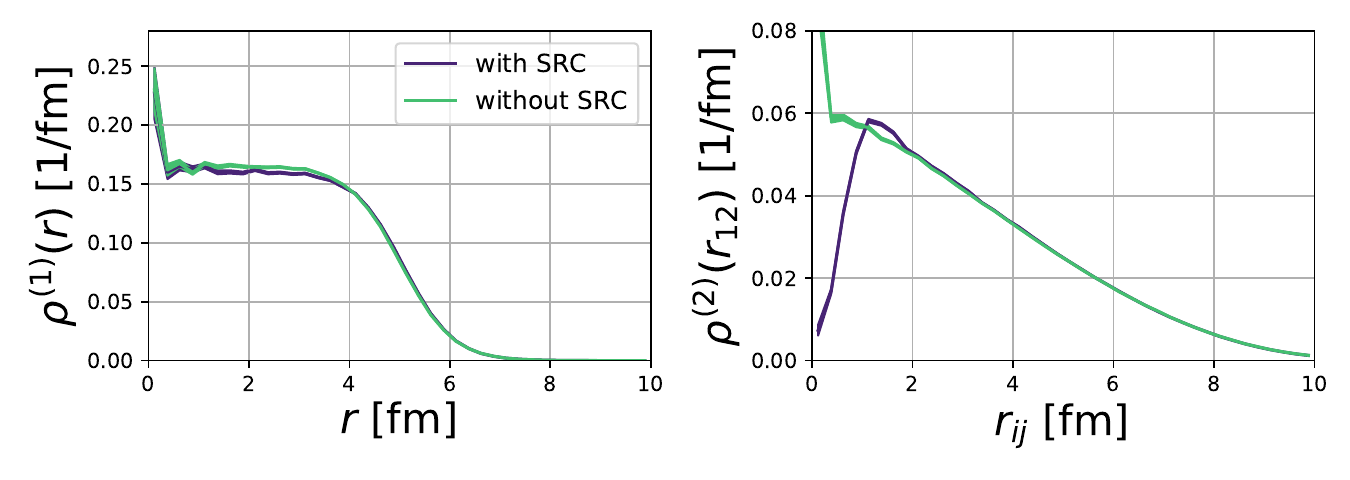}
            \caption{One (left) and two (right) body density of nuclear configurations of Ruthenium (case 1). The purple line correspond to the results obtained with configurations by Alvioli \textit{et al.} \cite{Alvioli:2009ab,Alvioli2019}, while the green line to configurations obtained from sampling Eq. (\ref{Eq:WoodsSaxon}).}
            \label{Fig:RuDensity}
        \end{figure}

        Fig. \ref{Fig:RuDensity} shows the one and two body density of $\approx 10^4$ configurations of Ruthenium nuclei with the Woods-Saxon parameter case 1 (see Tab. \ref{Tab:WSCases}).
        Whereas the one-body density yields the same results of the two descriptions, one can see a clear difference between them on the level of $\rho^{(2)}(r_{12})$. At small distance between the nucleon pairs, the configurations including NN SRC takes into account the short-range repulsion among nucleons. The calculations with short range correlations rely on externally provided configurations for the nucleons, otherwise the propagation and first collisions happen in SMASH as described above until the time $\tau_0$ is reached. 

    \subsubsection{Observables}
        The starting point of the calculations is the energy density of the collision.
        The energy density profile is calculated by summing over the contributions of all participants on the iso-$\tau$ hypersurface with their coordinates $\mathbf{x_i}$ and energy $p_i^0$ respectively. This yields
        \begin{equation}\label{Eq:Edens}
            e(\mathbf{x}, \tau) = \sum_{i=0}^{N_{\mathrm{part}}} p_i^0 K(\mathbf{x} - \mathbf{x}_i, \mathbf{p}_i) \, .
        \end{equation}
        A participant is by our definition a particle that has collided at least once or is not part of the original set of nucleons. 
        Here $K(\mathbf{x} - \mathbf{x}_i, \mathbf{p}_i)$ is the Lorentz invariant smearing kernel \cite{Oliinychenko:2015lva}
        \begin{equation}\label{Eq:SmearingKernel}
            K(\Delta\mathbf{x}_i = \mathbf{x} - \mathbf{x}_i, \mathbf{p}_i) = \frac{\gamma_i}{(2\pi\sigma)^{3/2}} \mathrm{exp}\left\{ -\frac{\Delta\mathbf{x}_i + \gamma_i^2 (\Delta\mathbf{x}_i\cdot\mathbf{\beta}_i)^2}{2\sigma^2} \right\} \, .
        \end{equation}
        Where $\gamma_i$ is the Lorentz factor of each particle and $\mathrm{\beta_i} = \mathbf{v}_i / c$ the velocity. In this work the width of the Gaussian is set to $\sigma = 1$ fm.
        Since we are only interested in midrapidity observable we incorporate a rapidity cut of $|y| < 0.5$.
    
        We quantify the properties of the initial state from the energy density profile in the transverse plane, which we recall for completeness. We compute, in particular, the total energy of the collision
        \begin{equation}\label{Eq:Etot}
            E = \int d^2\mathbf{x}\, e(\mathbf{x}, \tau_0) \, ,
        \end{equation}
        and the entropy density $s(\mathbf{x}, \tau)$, obtained by using the ideal gas equation of state $s(\mathbf{x}, \tau)\sim e(\mathbf{x}, \tau)^{3/4}$, from which a total entropy is derived
        \begin{equation}\label{Eq:Stot}
            S = \int d^2\mathbf{x}\, s(\mathbf{x}, \tau_0) \, .
        \end{equation}
        Additionally we compute the average radius of the initial conditions and its moments via
        \begin{equation}\label{Eq:r_n}
            \langle r^n \rangle = \frac{1}{S}\int d^2 \mathbf{x} |\mathbf{x}|^n s(\mathbf{x}, \tau_0) \, .
        \end{equation}
        In order to quantify the shape of the initial condition, the eccentricity harmonics are used. They describe how far the system geometry is from being circular, and can serve of a proxy of the flow observables in the final states.
        We thus calculate the following vector.
        \begin{equation}\label{Eq:VecEps_n}
            \mathcal{E}_n = \varepsilon_n e^{i\psi_n} = \frac{1}{S\langle r^n\rangle} \int d^2\mathbf{x} |\mathbf{x}|^n e^{in\phi} s(\mathbf{x}, \tau_0) \, ,
        \end{equation}
        with $\phi = \mathrm{atan2}(y,x)$.

    \subsubsection{Results}
        We start with some remarks on the system that is shaped by the simulations. The initial states of the participating nucleons at $\sqrt{s} = 200$ GeV are fed into Pythia which, after performing the hard processes, creates particles from the string fragmentation mechanism. The underlying picture is the so called yo-yo model where created quark-antiquark pairs create newly formed hadrons.
        Since these new hadrons are formed on a constant proper time themselves, at the time of the iso-$\tau$ hypersurface of the simulation they are technically not formed but will of course add to the total energy density of the system. Technically this is done by creating the new hadrons directly after the initial collision but without assigning any cross sections to them such that they cannot collide with other particles \cite{Mohs:2019iee}.
        \begin{figure}[t]
            \centering
            \includegraphics[width=0.5\textwidth]{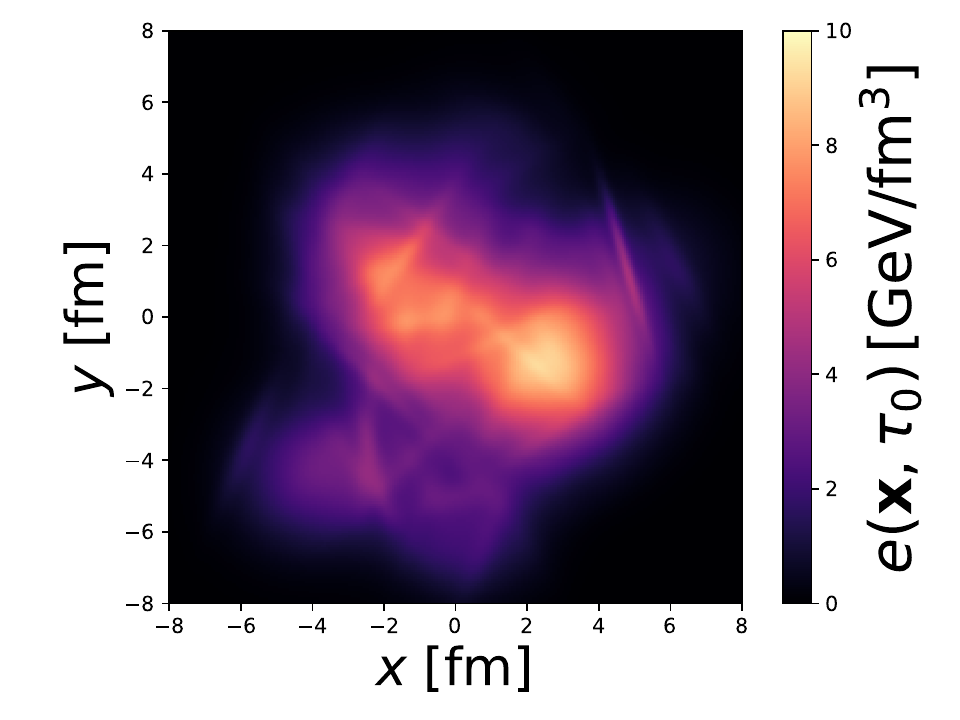}
            \caption{Energy density profile in the transverse plane of one event of a \ruru{} collision at $\sqrt{s} = 200$ GeV with an impact parameter of $b = 4.4$ fm.}
            \label{Fig:Edens}
        \end{figure}
    
        Figure~\ref{Fig:Edens} shows an exemplary event of the energy density profile of a Ruthenium-Ruthenium collision. The individual narrow lines can be attributed to hadrons with a large $p_T$. As a result of the Lorentz contraction (see Eq. (\ref{Eq:SmearingKernel})) the energy density of such a high $p_T$ particle then appears as a straight line.
    
        We now want to discuss our results for the isobar ratios. We simulate $\approx 10^7$ \ruru{} and \zrzr{} collisions at $\sqrt{s_{NN}} = 200$ GeV for each case from Tab. \ref{Tab:WSCases}. In addition, we run simulations were the nuclear configurations which incorporate NN SRC of the two full descriptions of $^{96}$Ru and $^{96}$Zr (case 1 + 6).
        Since the total entropy of the system is directly related with the number of produced (charged) particles $S\sim N_{ch}$ the quantities are presented as a function of $S$ to mimic experimental measurements as a function of $N_{ch}$ \cite{STAR:2021mii}. Since we are interested in the difference between observables in \ruru{} and \zrzr{} collisions, only ratios of Ru (case 1-5) / Zr (case 6) are shown.
        \begin{figure}[t]
            \centering
            \includegraphics[width=\linewidth]{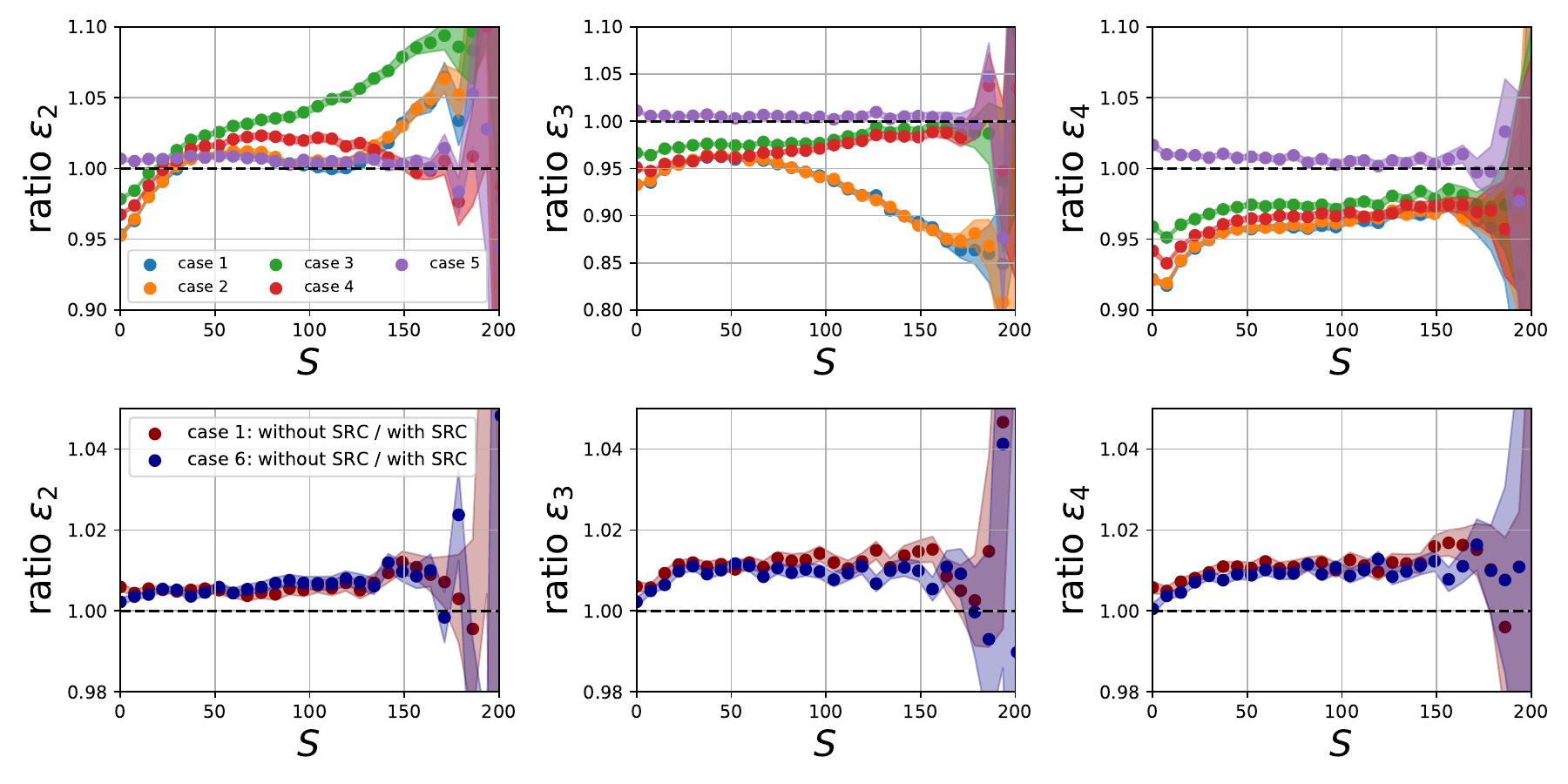}
            \caption{Ratio of $\varepsilon_2$ (left), $\varepsilon_3$ (center) and $\varepsilon_4$ (right) as a function of the total entropy $S$. In the upper row, the ratio is taken of the results of case 1-5 with respect to case 6 (see Tab. \ref{Tab:WSCases}). In the lower row the ratio is taken between the nuclear configurations without and with NN SRC of case 1 (red) and case 6 (blue).}
            \label{Fig:PannelEpsN}
        \end{figure}
    
        Figure~\ref{Fig:PannelEpsN} shows the five described ratios of the eccentricities $\varepsilon_n$ as a function of $S$.
        The ratio case 1 to case 6 is the baseline calculation. Overall, the trends are fully consistent with the previous results obtained from \trento{} and \trento{}+free-streaming initial conditions.
        
        Once more, comparing case 1 to case 2 we can see that the impact of a non-zero triaxiality, $\gamma$, has no influence on the eccentricities $\varepsilon_{2,3,4}$.
        Going to the ratio of case 3 which in comparison to case 2 highlights the different $\beta_3$ parameter. We find that the inclusion of $\beta_3$ increases both the eccentricity $\varepsilon_2$ and $\varepsilon_3$ of the initial state of \ruru{} collisions, respectively, in mid-central and central events.
        The influence of varying $\beta_2$ parameter is contained in the comparison between ratio 4 with 3. Here we see that mainly the second eccentricity is modified. If one goes to larger entropy bins, a larger $\beta_2$ parameter increases $\varepsilon_2$ meaning that the collision area appears to be less spherical. Both $\varepsilon_3$ and $\varepsilon_4$ are only slightly affected.
        Finally the diffuseness has a strong influence on the eccentricities of all orders that are presented here. With the matching of the diffuseness of $^{96}$Ru to the one of $^{96}$Zr (going from case 4 to case 5) we find that the main driver of the non-monotonic behavior of the ratios of $\varepsilon_n$ is originating from the difference in $a$. Finally, mild deviations from unity in case 5/case 6 are seen due to the different half-width radius.
    
        The impact of NN SRC can be seen in the lower row of Fig.~\ref{Fig:PannelEpsN}. Throughout the shown eccentricities of order $n=2,3,4$ we find that the inclusion of NN SRC reduces the eccentricity by about $\approx 1\%$. As a result of the improved description of the nuclei on the level of $\rho^{(2)}$, the initial state of the hydrodynamical description appears to be more \textit{round} in comparison to that obtained from an uncorrelated sample of the Woods-Saxon distribution. However, we find that the isobar ratios in the upper row of Fig.~\ref{Fig:PannelEpsN} are not affected. Note that this seems to challenge the finding of Fig.~\ref{fig:parameterScan_trento}. This could imply that including NN SRC by means of a $d_{\rm min}$ parameter is not realistic. These findings deserve further investigation.
        \begin{figure}[!ht]
            \centering
            \includegraphics[width=\linewidth]{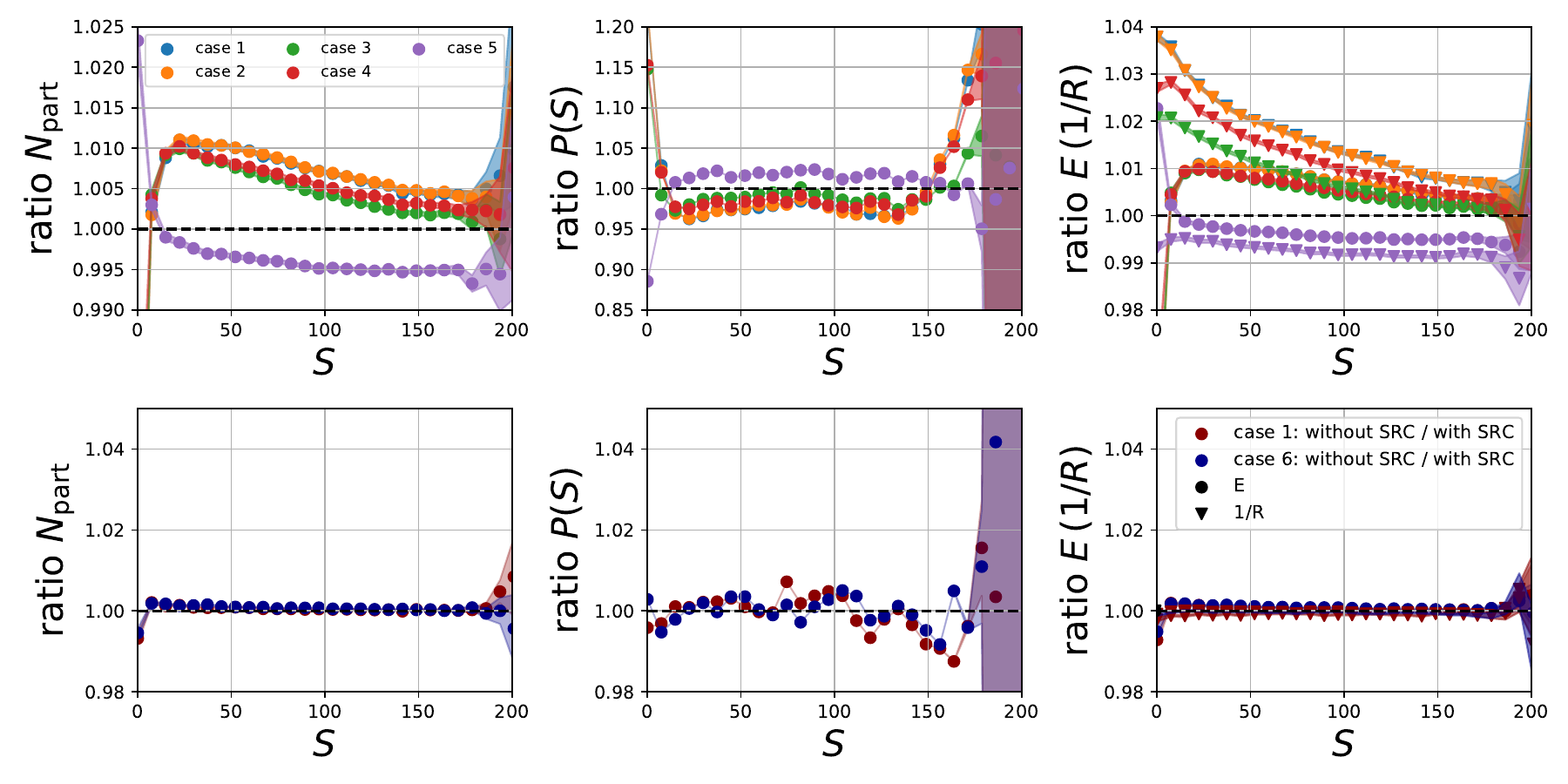}
            \caption{Ratio of number of paricipants $N_{\mathrm{part}}$ (left), probability distribution $P(S)$ (center), total energy $E$ and inverse radius $1/R$ (right). In the upper row ratios of case 1-5 with respect to case 6 are presented, whereas the lower row highlights the effect of the NN SRC.}
            \label{Fig:PannelObsN}
        \end{figure}
    
    For completeness, we also study the isobar ratios for the number of participants, $N_{\mathrm{part}}$, for the probability density of $S$, $P(S)$, for the total energy of the system, $E$, and for the inverse mean squared radius, $1/R$. Our results are shown in Fig. \ref{Fig:PannelObsN}.
    
        First, we observe that the NN SRC have no impact on any of the quantities shown. This is due to the fact that the observables presented in Fig. \ref{Fig:PannelObsN} are mainly sensitive to modifications of the one-body rather than the two-body density of the colliding ions. In contrast to the eccentricities $\varepsilon_n$, which probe to leading order the two-point function of the density field and two-body correlations, $\rho^{(2)}$, in the colliding isotopes [as discussed in Eq.~(\ref{eq:1p2p})], the effects of the two-body correlations on $N_{\mathrm{part}}$, $P(S)$ and $E$ or $1/R$ are subleading and  washed out in the isobar ratios. We see, then, that for all four of these observables we find that the main driver of the isobar ratio is the difference in the diffuseness of the nuclei. These results are fully consistent with those shown in the previous sections.

\subsection{Impact of small-\texorpdfstring{$x$}{x} evolution on isobar geometries}
The Color Glass Condensate framework uses Wilson lines $V(x_\perp)$ to depict the eikonal propagation of a high-energy probe. These Wilson lines serve as stochastic variables that capture the characteristics of the target hadron, and they are linked to the weight functional, which exhibits a dependence on rapidity $Y$ described by the JIMWLK renormalization group equation \cite{Jalilian-Marian:1997qno, Jalilian-Marian:1997jhx, Iancu:2000hn,Iancu:2001ad}. In numerical calculations, the JIMWLK hierarchy is formulated in terms of the Langevin equation \cite{Weigert:2000gi,Blaizot:2002np} governing the dynamics of the Wilson lines \cite{Lappi:2012vw} as
\begin{align}
  V_{\xT}(Y+dY)&{}=\exp \Big\{-i \frac{\sqrt{\alpha_s dY}}{\pi} \int_{\zT} K_{\xT-\zT}\cdot \left(V_{\zT}\boldsymbol{\xi}_{\zT}V^\dag_{\zT}\right) \Big\}\notag\\ &{}\times V_{\xT}(Y) \exp \Big\{ i \frac{\sqrt{\alpha_s dY}}{\pi} \int_{\zT} K_{\xT-\zT}\cdot \boldsymbol{\xi}_{\zT}\Big\}\,,
\end{align}
with Gaussian white noise $\boldsymbol{\xi}_{\zT} = (\xi_{\zT,1}^{a} t^a,\xi_{\zT,2}^{a} t^a)$ that is local in transverse coordinate, color, and rapidity, i.e., $\langle \xi_{\zT, i}^b(Y)\rangle = 0$ and
\begin{equation}\label{eq:noise}
  \langle \xi_{\xT, i}^a(Y) \xi_{\mathbf{y}_\perp, j}^b(Y')\rangle = \delta^{ab}\delta^{ij}\delta_{ \xT\mathbf{y}_\perp }^{(2)} \delta(Y-Y')\,.
\end{equation}
In accordance with \cite{Schlichting:2014ipa}, the undesired exponential growth of the cross-section, which would violate unitarity as we go to smaller and smaller $x$, is controlled by introducing a modification to the JIMWLK kernel $K_{x_\perp}$ 
\begin{equation}
  K_{\xT-\zT} = m|\xT-\zT|~K_{1}(m |\xT-\zT|)~\frac{\xT-\zT}{(\xT-\zT)^2}\,,
\end{equation}
where $K_1(x)$ is the modified Bessel function of the second kind which suppresses emission at large distance scales and
limits growth in impact parameter space.

The initial condition for the JIMWLK evolution, i.e., the Wilson line at the initial rapidity $Y$ is obtained using the IP-Glasma model \cite{Schenke:2012wb, Schenke:2012hg} where color charge densities $\rho^a(x_\perp)$ are sampled from a Gaussian distribution
\begin{align}
    \langle \rho_k^a(x_\perp)\rho_l^b(y_\perp)\rangle =\delta^{ab}\delta^{kl}\delta^2(x_\perp - y_\perp) \frac{g^2\mu^2(x_\perp)}{N_y}
\end{align}
where the indices $k,l=1,2,\dots, N_y$ represent a discretized longitudinal coordinate, and we set $N_y=50$. The color charge density distribution $g^2\mu(x_\perp, x)$ is proportional to the saturation scale $Q_s(T(x_\perp), x)$ and is obtained using the IP-Sat model \cite{Bartels:2002cj,Kowalski:2003hm} which provides $Q_s$ as a function of nuclear thickness $T(x_\perp)$ at a given Bjorken $x$, which corresponds to the initial rapidity at a given collision energy $\sqrt{s}$. The nuclear thickness $T(x_\perp)$ is obtained by sampling the position of each nucleon from a Wood-Saxon distribution 
\begin{align}
    \rho(r,\theta) = \frac{\rho_0}{1+\exp{(r-R'(\theta))/a}}
\end{align}
with $R'(\theta) = R[1+\beta_2Y_2^0(\theta)+\beta_4Y_4^0(\theta)]$ and $\rho_0$ is the density at the center of the nucleus. Subsequently, the positions of $N_q = 3$ hotspots per nucleon are sampled from a Gaussian distribution characterized by a width of $B_p = 4~\rm GeV^{-2}$. Each individual hotspot is then assigned a two-dimensional Gaussian thickness profile, employing a width of $B_q = 0.3~\rm GeV^{-2}$. With this, we determine $Q_s(x_\perp,x)$ in a self-consistent manner by iteratively solving $x = x(b_\perp) = Q_s(x,T(b_\perp))/\sqrt{s}$.

In practice, one then determines the
Wilson lines $V(x_\perp)$  at initial rapidity by approximating the path ordered exponential by the product \cite{Lappi:2007ku}
\begin{align}
    V_{x_\perp} = \prod_{k=1}^{N_y}\exp\Big(-ig \frac{\rho_k(x_\perp)}{\nabla_\perp^2+m^2}\Big)
\end{align}
where $m = 0.2$~GeV is the infrared regulator required to suppress the long Coulomb tails.  

By following the aforementioned initialization procedure for the IP-Glasma model, we generate a total of 100 configurations of the Wilson line $V_{x_\perp}^{\rm Ru/Zr}$ for $^{96}$Ru and $^{96}$Zr nuclei. These configurations are obtained at the largest value of $x$, corresponding to the lowest collision energy of $\sqrt{s_L} = 70$ GeV. Starting from the initial rapidity (collision energy) Wilson line configuration, we employ JIMWLK evolution to evolve towards the smallest value of $x$, which corresponds to the highest collision energy of interest, $\sqrt{s_H} = 7000$ GeV. The rapidity dependence in the JIMWLK evolution is proportional to $ln(\sqrt{s}/\sqrt{s_L})$.

\begin{figure}
    \centering
    \includegraphics[width=0.6\textwidth]{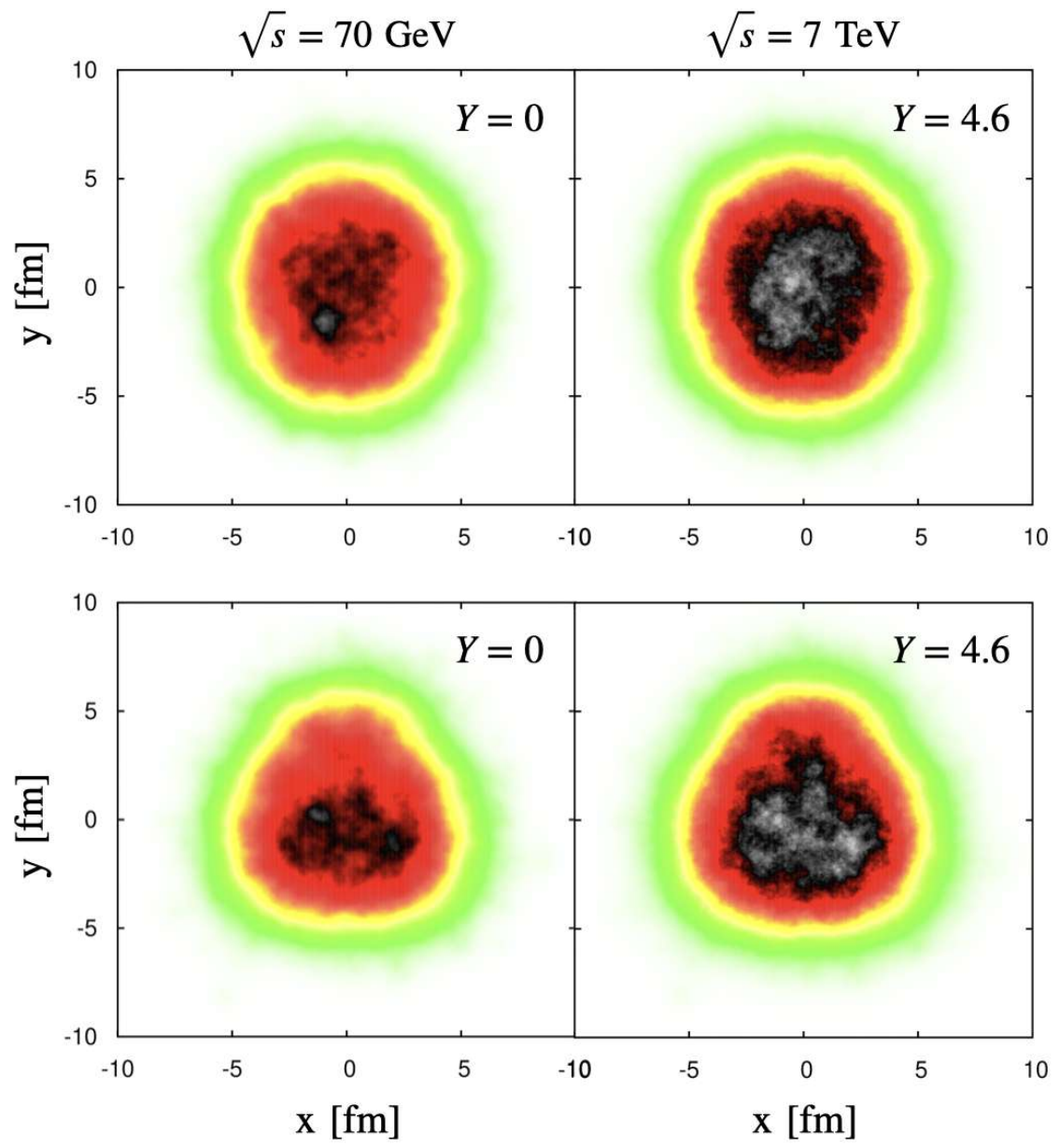}
    \caption{JIMWLK evolution of the trace of the Wilson line, $1-\mathrm{Re}[\mathrm{tr}(V_{x_\perp})]/N_c$, for $^{96}$Ru (top) and $^{96}$Zr (bottom) nuclei at two different values of $\sqrt{s}$.}
    \label{fig:JIMWLK_nuclei}
\end{figure}
\begin{figure}
    \centering
   \includegraphics[width=0.49\textwidth]{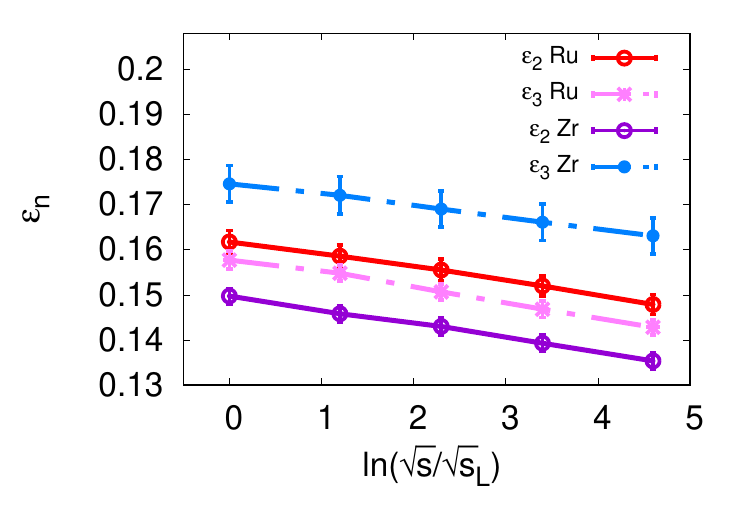}
  \includegraphics[width=0.49\textwidth]{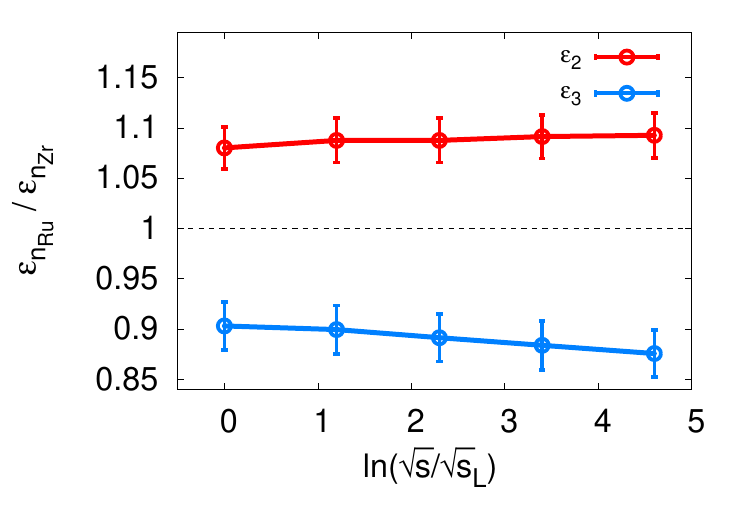}
    \caption{Geometric eccentricities $\varepsilon_n= \sqrt{\langle|\varepsilon_n(y)|^2\rangle}$ (left) and its ratio (right) for the isobars $^{96}$Ru and $^{96}$Zr as a function of collision energy}
    \label{fig:ecc_Ru_Zr}
\end{figure}

In Fig.~\ref{fig:JIMWLK_nuclei}, we illustrate the spatial distribution within $^{96}$Zr and $^{96}$Ru nuclei, with a fixed orientation of a nucleus ($\phi = 0$ and $\theta = \pi/2$ in Fig.~\ref{fig:rhoO}) by plotting the trace of the Wilson line $1-\mathrm{Re}[\mathrm{tr}(V_{x_\perp})]/N_c$ for two distinct values of $\sqrt{s}$. This representation provides a straightforward means of describing the distribution of the gluon field in transverse space. As previously observed in \cite{Schlichting:2014ipa,Schenke:2022mjv}, the increase in the saturation scale $Q_s$ due to an increase in collision energy $\sqrt{s}$ or a decrease in $x$ leads to a reduction in the characteristic transverse length scale, approximately $1/Q_s$. Consequently, this reduction facilitates the emergence of finer structures on the right panel. Remarkably, even with a 100-fold change in collision energy, the overall large-scale configuration of the nucleus exhibits only minor modifications. This implies that the small-$x$ evolution entails only a gradual smoothing of the geometric profile and growth in the impact parameter space (see the recent Ref.~\cite{Mantysaari:2024qmt} for a dedicated study on the subject). 

In the left panel of Fig.~\ref{fig:ecc_Ru_Zr}, we present the variation of eccentricity as a function of collision energy, represented by ${\rm ln}(\sqrt{s}/\sqrt{s_L})$, for $^{96}$Ru and $^{96}$Zr nuclei in the context of ultra central collisions ($b=0~\rm fm$). It is observed that $^{96}$Ru exhibits a higher ellipticity, while $^{96}$Zr possesses a larger triangularity due to the respective effects of quadrupole and octupole deformations. As we progress towards higher energies, the eccentricity reduces, primarily attributed to the gradual smoothening of the geometric profile resulting from the JIMWLK evolution. The right panel of Fig.~\ref{fig:ecc_Ru_Zr} displays the ratio of the n-th order eccentricity for $^{96}$Ru and $^{96}$Zr nuclei as a function of collision energy. Despite initiating with an equal number of large-$x$ partons as sources and subjecting both nuclei to the same amount of small-$x$ evolution, the observed deviation from unity in the ratio is attributed to variations in the nuclear structure. The energy dependence is very modest.

\subsection{New methods for efficient systematic study}

\subsubsection{Statistical limitations when studying nuclear structure effects in ultrarelativistic collisions}

Observables that are measured in high-energy nuclear collisions represent statistical properties of ensembles of collisions. Computing these in theoretical models, then, requires simulating many collision events.  Changes in the nuclear structure properties that are the focus of this task force typically are reflected in only small changes in measured observables.  As a result, a very large number of collision simulations must typically  be made in order to reduce the statistical uncertainty to be smaller than the effect to be studied.  The resulting computational cost significantly inhibits the opportunities for systematic study of nuclear structure in the context of high-energy collisions.   As seen in the various results contained in this report, one is typically limited to studying a small handful of discrete parameter values --- and even then, one is often restricted to computing initial-state estimators for observables due to the prohibitive cost of running full simulations.  

However, with a significant reduction in statistical demands, one can dramatically expand the possibilities for systematic study --- for example allowing for a comprehensive extraction of properties from a high-dimensional Bayesian analysis, as has recently become standard in the field of ultrarelativistic heavy-ion collisions.

Indeed, we developed methods that greatly reduce the number of necessary simulations for just such systematic study \cite{Luzum:2023gwy}.  The main idea is that, rather than always using completely independent statistics when comparing different sets of nuclear properties, one instead ensures that the statistical uncertainty is strongly correlated between the different scenarios.  In this way, one can very accurately probe the variation of observables with varying nuclear structure, using a surprisingly small number of simulation events.

\subsubsection{Changing nuclei by shifting nucleons}

In order to compare two nuclei with the same number of nucleons and slightly different structural properties, we implement these differences by shifting the positions of nucleons. 
The simplest case is a difference in the nuclear size, which is implemented by multiplying the coordinates of all nucleons by a scale factor. 

\begin{figure}[h]
\begin{center}
\includegraphics[width=0.5\linewidth]{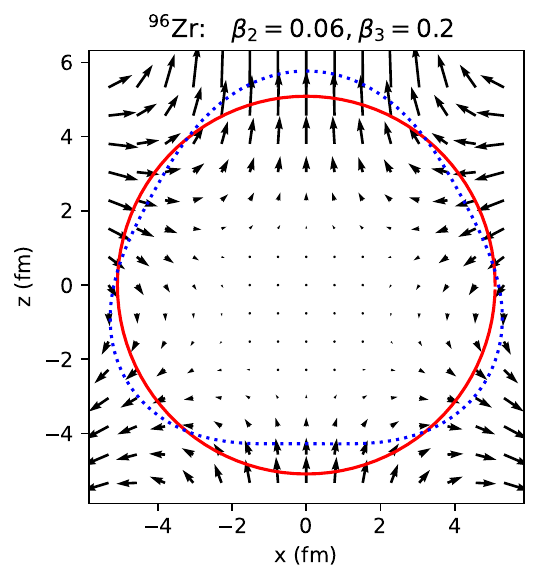}
\end{center}
\caption{
\label{fig:quiver}
Vector plot of shift $d\vec x$ in $x$-$z$ plane for case of axial quadrupole and octupole deformation, $\beta_{2,0} = 0.06, \beta_{3,0} = 0.2$.
The axis scales correspond to Woods-Saxon radius $R = 5.09$ fm.   
The curves represent the Woods-Saxon radius $R(\theta,\phi)$ for the starting spherical distribution (solid) and the final deformed distribution (dotted).}
\end{figure}

Differences in the nuclear skin and deformation can be implemented in a similar way. 
In practice, the deformation is implemented by starting from a spherical nucleus, and shifting the positions of nucleons, using a specific method which is detailed in Ref.~\cite{Luzum:2023gwy}. 
This is illustrated in Fig.~\ref{fig:quiver}, where we start from a nucleus made of $A=96$ nucleons sampled independently from a spherical Woods-Saxon distribution $\rho(r)=\left(1+\exp((r-R)/a)\right)$, and shift their positions in such a way that the resulting distribution is a deformed Woods-Saxon distribution $\tilde \rho(r, \theta, \phi)$  with axial quadrupole and octupole deformation: 
\begin{equation}
\label{eq:defWS}
\tilde \rho(r, \theta, \phi)  
\propto 
\rho\left(r - R\beta_{2,0} Y_{2,0}(\theta)-R\beta_{3,0} Y_{3,0}(\theta)\right),
\end{equation}
where $Y_{l,0}(\theta)$ are the Laplace spherical harmonics, and we have chosen the deformation $\beta_{2,0}$ and $\beta_{3,0}$ to be those expected for the $^{96}$Zr nucleus. 

Short-range two-body correlations, resulting from the nucleon-nucleon interaction, can be implemented in a similar way, by changing the position of every nucleon in a way that depends on its distance from all other nucleons~\cite{Luzum:2023gwy}. 
This implementation is superior to that of Refs.~\cite{Alver:2008aq,Moreland:2014oya} in that it approximately preserves the one-body distribution, and also applies to nuclear shapes without an axial symmetry. 

\subsubsection{Application to heavy-ion collisions}

Once the nucleon positions have been shifted so as to implement the desired structural change, one must make sure that parameters specific to the nucleus-nucleus collision are unchanged. 

Compared to the standard method where one compares two nuclear configurations through independent Monte Carlo simulations, the present method results in a significant efficiency gain. 
Smaller nucleon shifts result in larger efficiency gains, so that this method is superior to any other method for precision studies of the nuclear structure, in which one must study the effect of small changes on heavy-ion collision observables. 

\newpage

\section{Full dynamical simulations}
\label{sec:5}

\subsection{Mapping of initial-state geometry to final-state observables in isobar collisions}

\subsubsection{Linear and nonlinear hydrodynamic response}

As argued before, a key input for understanding nuclear structure in the context of relativistic heavy-ion collisions is the reliance on the strong (nearly linear) mapping between the geometrical shape of the initial state, $\mathcal{E}_n$, which is propagated into momentum space that can eventually be described by collective flow harmonics, $V_n$. 

We quantify the geometrical shape of the initial state through vectors that quantify both the magnitude of the initial geometry, $\varepsilon_n$, as well as the angle, $\Phi_n$ such that we can define an eccentricity that is a complex vector:
\begin{equation}
    \mathcal{E}_n\equiv \varepsilon_n e^{in\Phi_n}
\end{equation}
where $n$ indicates the harmonic i.e. $n=2$ describes elliptical shapes, $n=3$ describes triangular shapes, $n=4$ describe square or ``plus sign" shapes and so forth.  

In practice, these eccentricities are calculated over some sort of density  field $\rho$  (e.g. entropy density or energy density) such that
\begin{equation}
    \varepsilon_{m,n}=-\frac{\int r^m e^{in\phi}\rho(r,\phi)rdrd\phi}{\int r^m \rho(r,\phi)rdrd\phi}
\end{equation}
where this is often abbreviated as 
\begin{equation}
    \varepsilon_{m,n}=-\frac{\left\{r^m e^{in\phi}\right\}}{\left\{r^m \right\}}
\end{equation}
such that the curly brackets $\left\{\dots\right\}$ indicate the density weighted integral over the two-dimensional space. Often the most important eccentricities are the limit where $m=n$, i.e. the magnitude of elliptical eccentricity $\varepsilon_{2,2}$, but there are cases when $m\neq n$ is the leading order term i.e. for directed flow $v_1$ (rapidity even, we ignore rapidity odd directed flow since we consider only 2+1 hydrodynamic simulations here) where the leading order eccentricity is $\varepsilon_{1,3}$ since the $\varepsilon_{1,1}=0$ if the eccentricity is computed  with respect to the center of mass of the density.

Then for a given flow harmonic vector $V_n$, one can define it as 
\begin{equation}
    V_n\equiv v_n e^{in\Psi_n}
\end{equation}
where its respective magnitude is $v_n$ and angle is $\Psi_n$. Each flow harmonic of the $n^{th}$ order can be described then as a response to at least one $\mathcal{E}_n$ and sometimes a series of eccentricities (see \cite{Teaney:2010vd,Gardim:2011xv,Gardim:2014tya} for the generating function to derive this series expansion). Thus, one can define a $\mathcal{V}_n^{est}$ that contains at least one eccentricity or more, depending on both the properties of that flow harmonic and the precision with which one would like to predict that flow harmonic based on the eccentricities themselves.  

To determine the accuracy of an estimator one can calculate a Pearson coefficient
\begin{equation}\label{eqn:Pearson}
    Q_n\equiv \frac{\langle V_n (V_n^{est})^*\rangle }{\sqrt{\langle |V_n|^2\rangle\langle |V_n^{est}|^2\rangle}}
\end{equation}
wherein $Q_n\rightarrow \pm 1$ implies either a perfect mapping from $V_n^{est} \rightarrow V_n$ for a positive sign or perfect anti-correlation for a negative sign.  For $Q_n\rightarrow 0$ there is no correlation whatsoever between the estimator and the flow harmonic.  

If we consider the example of elliptical and triangular flows, the leading order terms are the linear responses, i.e., 
\begin{eqnarray}
 \nonumber   V_2&=&\kappa_2\mathcal{E}_2+\mathcal{O}(\mathcal{E}_2\mathcal{E}_2^*\mathcal{E}_2)+\mathcal{O}(\mathcal{E}_4\mathcal{E}_2^*)+\ldots \,,\\
     V_3&=&\kappa_3\mathcal{E}_3+\mathcal{O}(\mathcal{E}_3\mathcal{E}_3^*\mathcal{E}_3) + \mathcal{O}(\mathcal{E}_5\mathcal{E}_2^*)+\ldots\,,
\end{eqnarray}
where $\kappa_n$ are so-called linear response coefficients, and higher order corrections include either higher-order powers of the lowest-order anisotropies or any other terms allowed by symmetry. The (linear) leading-order terms provide excellent descriptors of the final flow coefficients in central collisions, while nonlinear terms become more important for peripheral collisions \cite{Noronha-Hostler:2015dbi} and small systems \cite{Sievert:2019zjr}.  For $n>3$ couplings of different harmonics must be considered \cite{Teaney:2012ke,Yan:2015jma}.  This can be intuitively understood for $n=4$, for instance, because both an overall elliptical shape and a square or plus sign shape will both contribute to the fourth harmonic. Directed flow ($n=1$) is also highly dependent on higher order terms such that the leading order term is insufficient to describe its behavior well \cite{Gardim:2014tya}.  Thus, in this work we will only focus on the linear mapping including $n=2$ and $n=3$. They provide an excellent approximation of the final flow fluctuations in central collisions, which are the most sensitive to nuclear structure and the $\beta_2$ and $\beta_3$ deformations. 
\begin{table}[t]
    \centering
    \begin{tabular}{|c|ccc|}
     \hline
    Ion &  $\beta_2$ & $R$ [fm] & a [fm] \\
    \hline
      $^{96}_{44}$Ru & 0.1 & 5.085 & 0.46 \\
     $^{96}_{44}$Ru   & 0.158  & 5.085 & 0.46 \\
       $^{96}_{44}$Ru  & 0.2  & 5.085 & 0.46 \\
       $^{96}_{44}$Zr   & 0 & 5.02 & 0.46 \\
        \hline
    \end{tabular}
    \caption{Summary of initial state parameters for the deformation and ion size within the \trento{}+v-USPhydro analysis of the mapping.}
    \label{tab:tntvUSP}
\end{table}

Ideally, for high-precision extractions of nuclear structure information from the initial state, one would like a nearly perfect linear mapping between $V_n$ and $\mathcal{E}_n$.  However, up until now this has not been tested extensively for deformed, medium sized nuclei such as the $A=96$ isobars under consideration here\footnote{A previous study did test the mapping against the effect of the quadrupole deformation, $\beta_2$, in a large deformed system, namely \uuuu{} \cite{Wertepny:2019yye}, and of the octupole deformation, $\beta_3$ in \pbpb{} collisions \cite{Carzon:2020xwp}. In both cases a strong, almost linear mapping for $n=2$ and $n=3$ was found.}. Here, we study this by means of the same setup employed for \auau{} collisions at top RHIC energies in Refs.~\cite{Alba:2017hhe,Rao:2019vgy}.

Our framework uses \trento{} initial conditions \cite{Moreland:2014oya} with the standard parameterization based on the Duke Bayesian analysis from Ref.~\cite{Bernhard:2016tnd} wherein we have reduced thickness parameter $p=0$, gamma fluctuation parameter $k=1.6$, and the nucleon width $\sigma =0.51$ fm. The normalization constant for the entropy density is chosen to accurately reproduce the measured \auau{} multiplicities. For the study of nuclear deformations we vary both the radius parameter $R$ and the $\beta_2$ deformations.  Future work is planned for studying the mapping of variations of $\beta_3$, $\beta_4$, and the diffusive parameter $a$ (as well as cross-correlations between them) but we will report on this separately. A summary table of the parameters used in \trento{} is shown in Tab.\ \ref{tab:tntvUSP}.

Approximately 6000 \trento{} initial conditions (with each configuration shown in Tab.\ \ref{tab:tntvUSP}) were ran by means of the v-USPhydro hydrodynamic code \cite{Noronha-Hostler:2013gga,Noronha-Hostler:2014dqa,Noronha-Hostler:2015coa} within the $0-60\%$ centrality range, where centrality bins are determined by binning in the initial total entropy of the events, which is a good proxy of the final multiplicity. We use the best fit parameters from Ref.~\cite{Alba:2017hhe} that include the initial time of $\tau_0=0.6$ fm, shear viscosity over entropy density ratio of $\eta/s=0.05$, bulk viscosity over entropy density ratio of  $\zeta/s=0$, freeze-out temperature $T_{FO}=150$ MeV, and the PDG16+/2+1[WB] equation of state from. Following freeze-out, direct decays with the full PDG16+ list \cite{Alba:2017mqu} are taken into account and the final-state flow harmonics calculated. Statistical error on the results are computed by means of jackknife resampling.

\begin{figure}[t]
    \centering
    \begin{tabular}{cc}
      \includegraphics[width=0.45\linewidth]{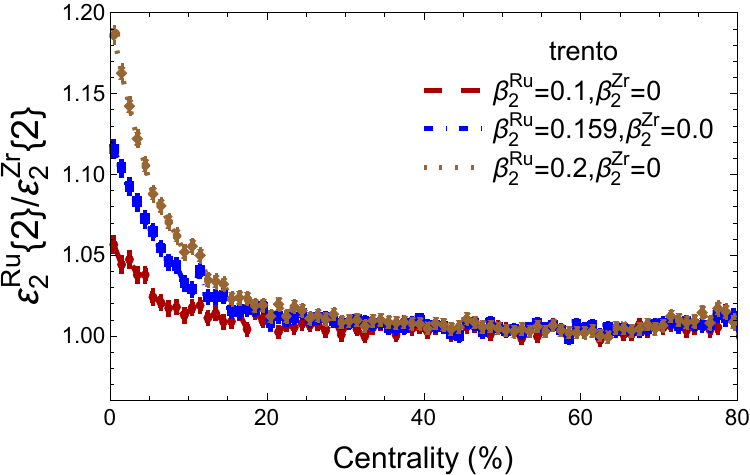}   &   \includegraphics[width=0.45\linewidth]{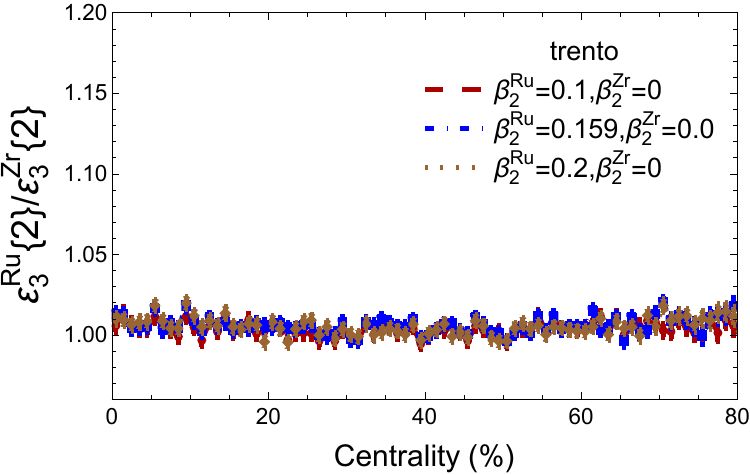}
    \end{tabular}
    \caption{Isobar ratios of rms eccentricities, $\varepsilon_n\left\{2\right\}$, for $n=2$ (left) and $n=3$ (right), computed from the initial conditions of the v-USPhydro simulations.}
    \label{fig:ecc_2part}
\end{figure}

Before discussing the full hydrodynamic results, it can be useful to doublecheck that we obtain results for the influence of the $\beta_2$ deformation on the eccentricities that are consistent with those shown in the previous sections.  Running million of initial-state simulations, for two-particle correlations we define 
\begin{equation}
    \varepsilon_n\{2\}=\sqrt{\langle \varepsilon_n^2\rangle}
\end{equation}
which we evaluate in centrality bins of 1\% width. In Fig.\ \ref{fig:ecc_2part} we show the isobar ratio for $\varepsilon_n\{2\}$ taken between \ruru{} over \zrzr{} collisions for elliptical flow ($n=2$, left) and triangular flow ($n=3$, right). Note that here we assumed a perfectly spherical $^{96}$Zr nucleus. As anticipated, the eccentricity in central collisions is strongly affected by the quadrupole deformation parameter. The isobar ratio is as high as 1.20 when the deformation of $^{96}$Ru is set to $\beta_2=0.20$. For the isobar ratio of triangular eccentricities, we do not find a significant impact of the quadrupole deformation of ruthenium, consistent with the \trento{} results shown in the previous sections.

Although we do not show it here, we also look at the isobar ratio of higher-order multi-particle cumulants as an alternative probe of the nuclear geometry. Specifically, we study $\varepsilon_n\left\{4\right\}/\varepsilon_n\left\{2\right\}$ where the four-particle correlator is defined as
\begin{equation}
    \varepsilon_n\left\{4\right\}=\left(2\langle v_n^2\rangle^2-\langle v_n^4\rangle\right)^{1/4},
\end{equation}
which is given in terms of the second and the fourth moments of the $v_n$ distribution in the considered multiplicity class.  Taking the ratio of $\varepsilon_n\left\{4\right\}/\varepsilon_n\left\{2\right\}$ cancels the response coefficients if one assumes linear response, such that
\begin{equation}
    \frac{v_n\left\{4\right\}}{v_n\left\{2\right\}}\approx \frac{\varepsilon_n\left\{4\right\}}{\varepsilon_n\left\{2\right\}}.
\end{equation}
This works very well in central collisions \cite{Yan:2013laa,Giacalone:2017uqx,Carzon:2020xwp}, and provides us with a convenient means to assess the impact of the nuclear structure on the final-state observables. Specifically, the fourth-order cumulant of the $v_n$ distribution in the limit of central collision quantifies the non-Gaussianity (kurtosis) of the two-dimensional flow vector ($V_n$) fluctuations \cite{Alqahtani:2024ejg}. 

We confirmed for the isobars that $\varepsilon_n\left\{4\right\}/\varepsilon_n\left\{2\right\}$ and $\varepsilon_3\left\{4\right\}/\varepsilon_3\left\{2\right\}$ are enhanced by the inclusion of the parameters $\beta_2$ and $\beta_3$ in the limit of central collisions. This is consistent with previous studies done with either \uuuu{} \cite{Giacalone:2018apa,Mehrabpour:2023ign} or \pbpb{} \cite{Carzon:2020xwp,Xu:2025cgx} collisions. Overall, these ratios involving higher-order cumulants in central collisions offer an additional experimental handle to pin down targeted features of the nuclear geometry.

\subsubsection{Collective flow and mapping}

\begin{figure}[t]
    \centering
    \begin{tabular}{cc}
      \includegraphics[width=0.45\linewidth]{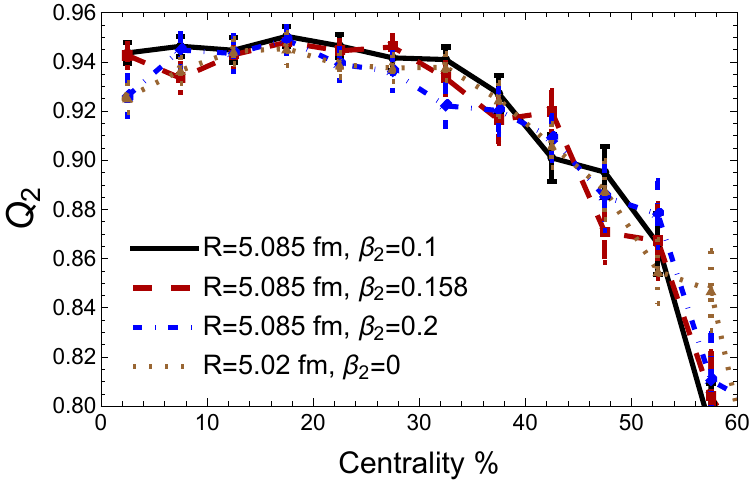}   &   \includegraphics[width=0.45\linewidth]{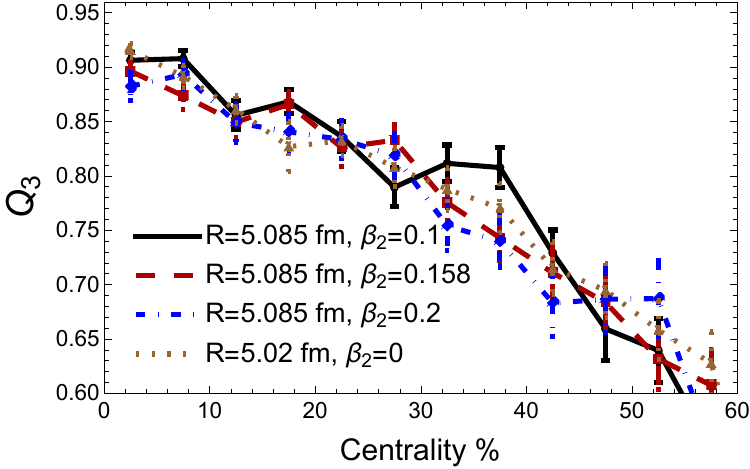}
    \end{tabular}
    \caption{Pearson coefficient $Q_n$, as defined by Eq.~(\ref{eqn:Pearson}), for the linear mapping of $\mathcal{E}_n$ to $V_n$ for $n=2$ (left) and $n=3$ (right) in \ruru{} collisions. Different line styles represent different choices of the deformation parameter $\beta_2$.}
    \label{fig:Qn}
\end{figure}

\begin{figure}[t]
    \centering
    \includegraphics[width=0.5\linewidth]{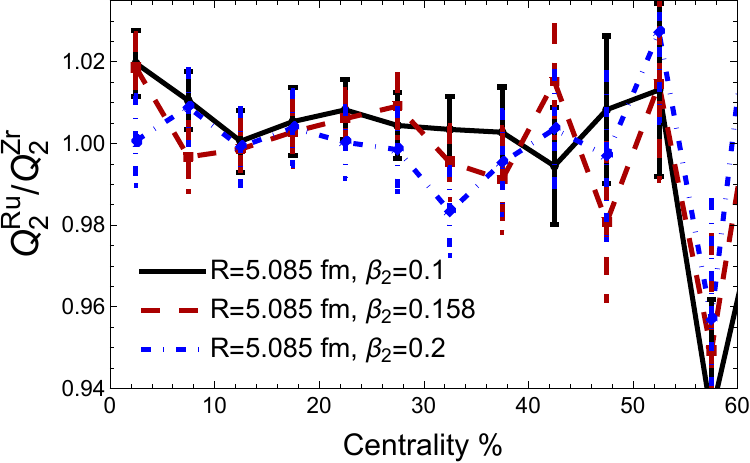}
    \caption{Isobar ratio of the Pearson coefficients $Q_2$  for the linear mapping of $\mathcal{E}_2$ to $V_2$ as a function of the collisions centrality. Different line styles correspond to different $\beta_2$ parameters in $^{96}$Ru.}
    \label{fig:Qrats}
\end{figure}

The results of the full hydrodynamic simulations are now discussed. We first focus on the hydrodynamic response of the \ruru{} collision systems. We study the Pearson coefficient, as defined in Eq.\ (\ref{eqn:Pearson}), that is shown in Fig.\ \ref{fig:Qn} for $n=2$ and $n=3$.  We find for both elliptical and triangular flows that the linear correlation is quite strong in central collisions. The quality of these correlations does not present a sensitivity to the value of the quadrupole deformation parameter. 

We now study the variation of the Pearson coefficient moving from \ruru{} to \zrzr{} collisions. The results are shown in Fig.~\ref{fig:Qrats}. For this ratio, the error bars are determined using error propagation. We find that all 3 configurations for Ru appear to have nearly identical ratios with Zr. Therefore, with a relative statistical accuracy of about 2\%, we conclude that the linear hydrodynamic response is equally strong in \ruru{} and \zrzr{} collisions, across a wide centrality interval, and that this holds irrespective of the choice of deformation of $^{96}$Ru. Our statistical precision does not allow for an in-depth study of the ratio of the $Q_3$ coefficient as well.

\subsection{Isobar ratios in 3+1D hydrodynamic simulations}

At the top RHIC energy, the Bjorken boost-invariant assumption is expected to work well. Therefore, it is a good approximation to simplify the dynamic description of the collision systems to 2+1 dimensions. Under such an assumption, the system's anisotropic flow coefficients strongly correlate with the system's transverse geometry. However, as the experimental measurements of the RHIC isobar collisions reach the sub-percent level, we need to understand the influence of non-trivial longitudinal fluctuations in heavy-ion collisions on imaging the nuclear shape established with the boost-invariant picture. Along the longitudinal direction, different incoming nucleons inside the projectile and target would lose different amounts of energy as they penetrate the other nucleus. The transverse shape of the energy density profile could vary as a function of the space-time rapidity $\eta_s$, known as the twisted/tilted effects. Microscopically, the longitudinal evolution of the energy density profile is controlled by the small-$x$ gluon radiation. It can be interpreted as a random walk process as the transverse shape decorrelates with each other with an increasing rapidity gap.

In experiment, the standard flow analysis requires us to correlate particles with a large enough rapidity gap to suppress non-flow correlation. As the precision of the measurement reaches the sub-percent level in the RHIC isobar collisions, we would need to quantify the effects of longitudinal fluctuations when interpreting the physics driving the measurements. 

In this section, we explore the impact of nuclear structure on the full three-dimensional geometry in the RHIC isobar collisions at the top RHIC energy using the iEBE-MUSIC framework with the 3D MC-Glauber + MUSIC + UrQMD model~\cite{Shen:2022oyg}. We calibrate all the model parameters with \auau{} collisions at 200 GeV \cite{Shen:2023awv} and study the effects of different nuclear shapes between $^{96}$Ru and $^{96}$Zr on the experimental observables.

The effect of fireball shape evolution along the longitudinal direction can be quantified by the following ratios,
\begin{equation}
    r_n(\eta) = \frac{\Re\{\langle V_n(-\eta) V^*_n(\eta^\mathrm{ref}) \rangle \}}{\Re\{\langle V_n(\eta) V^*_n(\eta^\mathrm{ref}) \rangle\} },
    \label{eq:rn}
\end{equation}
where $\langle \cdots \rangle$ averages over collision events within a given centrality bin and $V_n(\eta) = \sum_j e^{i n \phi_j}$ is the $n$-th order charged hadron anisotropic flow vector at a given pseudorapidity. The reference flow vector $V_n(\eta^\mathrm{ref})$ was chosen to be at forward rapidity, with $\eta^\mathrm{ref} \in [3.1, 5.1]$ as in the STAR measurements \cite{Nie:2020trj}. Because the rapidity gap in the numerator is larger than that in the denominator for positive $\eta$, we expect the ratio $r_n$ to be smaller than unity. Within the model simulations, we can build an initial-state estimator by using the initial-state eccentricity vectors $\mathcal{E}_n$,
\begin{equation}
    r^{\varepsilon}_n(\eta_s) = \frac{\Re\{\langle \mathcal{E}_n(-\eta_s) \mathcal{E}^*_n(\eta_s^\mathrm{ref}) \rangle \}}{\Re\{\langle \mathcal{E}_n(\eta_s) \mathcal{E}^*_n(\eta_s^\mathrm{ref}) \rangle\} }.
    \label{eq:rn_ecc}
\end{equation}
Here we use the same rapidity region $\eta^\mathrm{ref}_s \in [3.1, 5.1]$ as in the STAR measurements \cite{Nie:2020trj}.

\begin{figure}[t]
    \centering
    \includegraphics[width=0.49\textwidth]{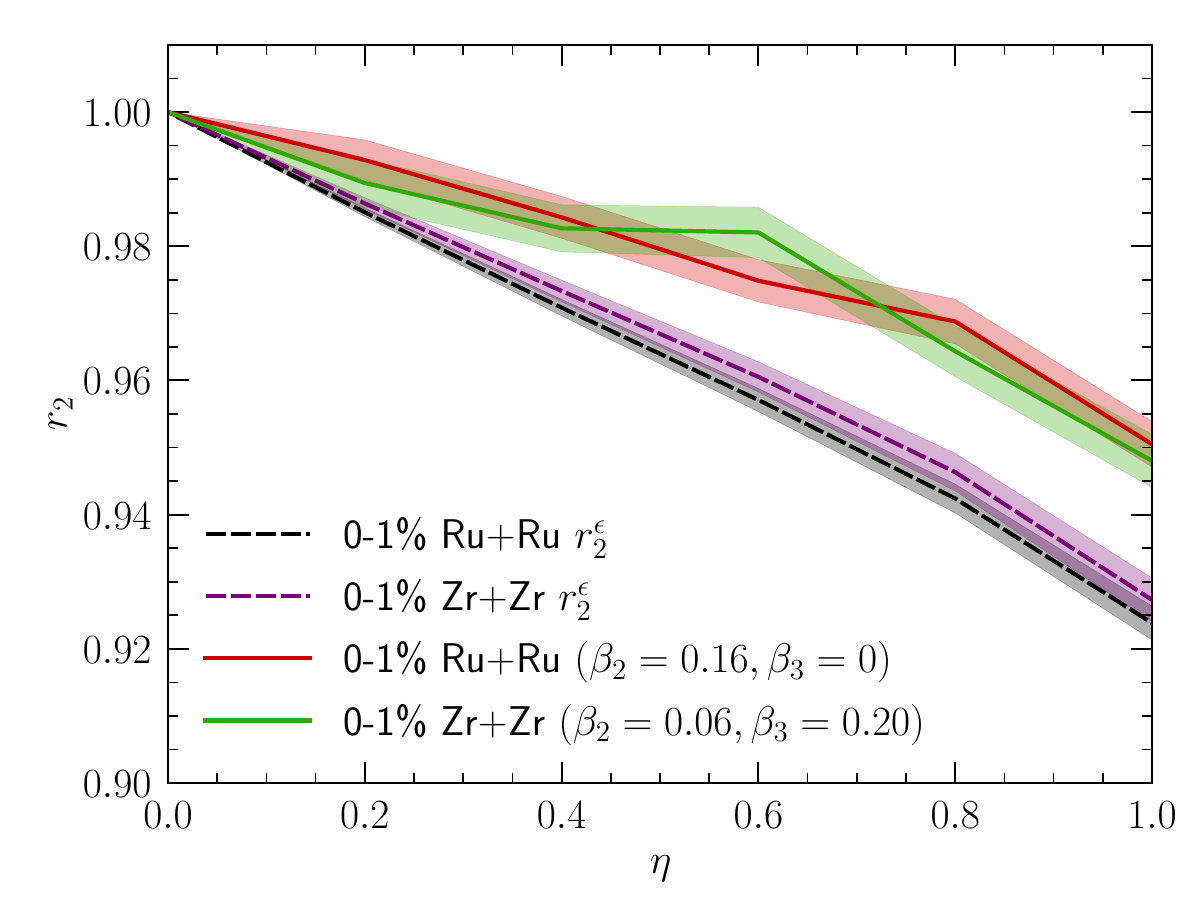}
    \includegraphics[width=0.49\textwidth]{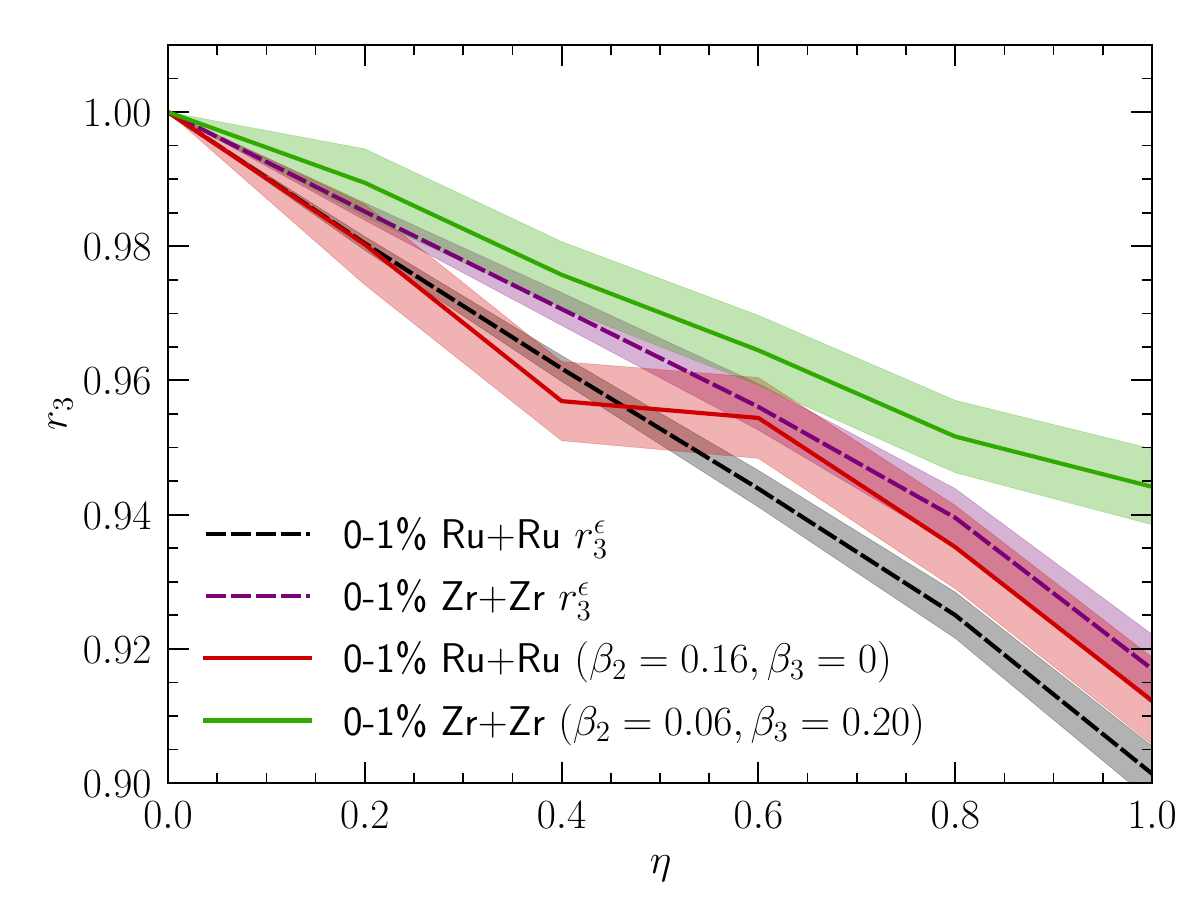}
    \caption{The longitudinal flow decorrelation coefficients for elliptic flow $v_2$ (left) and triangular flow $v_3$ (right) in 0-1\% central isobar collisions. Dashed lines represent the results obtained for eccentricity-based correlations as given by Eq.~(\ref{eq:rn_ecc}).}
    \label{Fig:3DIsobar_rn}
\end{figure}

Figure~\ref{Fig:3DIsobar_rn} shows the longitudinal flow decorrelation coefficients for 0-1\% isobar collisions. We select the ultra-central collision events to ensure that colliding nuclei almost fully overlap so that the resulting energy density profile imprints the nuclear shape. We assume the $^{96}$Ru nuclei have a relatively large $\beta_2$ deformation while the $^{96}$Zr nuclei have small $\beta_2$ but large $\beta_3$ deformation. 
First, we note that the flow decorrelation coefficients $r_n(\eta)$ from the full (3+1)D simulations deviate significantly from their initial-state estimators $r^\varepsilon_n(\eta)$, while the difference between Ru+Ru and Zr+Zr collisions is close. 
Therefore, one needs to perform full (3+1)D high-statistics model simulations to predict the absolute value of longitudinal decorrelation. At the same time, the initial-state estimators provide a reasonable expectation about the effects of nuclear structure on $r_n$ of the isobar collisions (see also Ref.~\cite{Nie:2022gbg} for a recent study with the AMPT model, and Ref.~\cite{Mehrabpour:2025rzt} for hydrodynamic calculations involving \oooo{} and \nene{} collisions).
For elliptic flow, we find that the two isobar collision system gives very close results. Our understanding is that this is due to the fact that the spatial eccentricity $\mathcal{E}_2$ receives contributions from $\beta_2$ and $\beta_3$ deformations,
\begin{equation}
    \langle |\mathcal{E}_2|^2 \rangle \propto a_2 + b_2 \beta_2^2 + b_{2, 3} \beta_3^2.
    \label{eq:e2response}
\end{equation}
Therefore, the large $\beta_2$ deformation in the Ru nuclei results in similar effects as those from the large $\beta_3$ deformation in the Zr nuclei in $r_2(\eta)$. 
Meanwhile, the $r_3(\eta)$ ratios show a significant system dependence. The $r_3(\eta)$ from \zrzr{} collisions is closer to unity than those in \ruru{} collisions. The ordering is because the large $\beta_3$ deformation in $^{96}$Zr results in large triangularity in the initial state of the central \zrzr{} collision,
\begin{equation}
    \langle |\mathcal{E}_3|^2 \rangle \propto a_3 + b_{3} \beta_3^2.
    \label{eq:e3response}
\end{equation}
The large initial $\mathcal{E}_3$ in the collision geometry reduces the longitudinal decorrelation. The full (3+1)D simulations show that the $r_3(\eta)$ difference between \ruru{} and \zrzr{} collisions is preserved throughout evolution and can be measured in the final-state observables. Figure~\ref{Fig:3DIsobar_rn} provides an additional theoretical prediction to study the connections between nuclear structure and high-energy heavy-ion collisions. 

\subsection{A Multi Phase Transport Model (AMPT)}

To set the stage, we stress that a main challenge in nuclear physics research is to characterize the shape and radial structure of atomic nuclei across the Segr\`e chart, and understand how these features emerge from the interactions among the constituent nucleons. In high-energy heavy-ion collisions, the initial condition of the QGP and its evolution are sensitive to the deformation and the radial profile of the colliding nuclei~\cite{Bally:2022vgo,STAR:2024wgy,Zhang:2024bcb,Zhang:2025hvi,STAR:2025vbp}. The nuclear density is often described by a deformed Woods-Saxon (WS) parametrization, 
\begin{align}
    \rho(r, \theta, \phi) \propto \frac{1}{1+e^{\left[r-R_0\left(1+\beta_2 Y_2^0(\theta, \phi)+\beta_3 Y_3^0(\theta, \phi)\right)\right] / a}},
\end{align}
with quadrupole deformation, $\beta_2$, octupole deformation, $\beta_3$, half-density radius, $R_0$, and surface diffuseness, $a$. Here, we study how these parameters manifest themselves in dynamical simulations of high-energy isobar collisions.

\subsubsection{The AMPT model}
We simulate the dynamics of the QGP using the multi-phase transport model (AMPT). The AMPT model~\cite{Lin:2004en} is a hybrid framework with four main phases: fluctuating initial conditions from the HIJING model, elastic parton cascade simulated by the ZPC model, quark coalescence for hadronization and hadronic re-scattering based on the ART model. The AMPT model was reasonably tuned to describe collective flow and charged particle spectra data in large systems at both RHIC and LHC energies. We use AMPT v2.26t5 in string-melting mode at $\sqrt{s_{NN}}$ = 200 GeV with a partonic cross section of 3.0 $m$b. We study the signatures of the nuclear structure parameters in isobar ratios compute from $^{96}$Ru+$^{96}$Ru and $^{96}$Zr+$^{96}$Zr collisions. 

\subsubsection{Taylor-expansion approach to isobar ratios}
To explore the scaling behavior of the ratios with respect to the WS parameters, we simulate collisions covering a wide range of $\beta_2$, $\beta_3$, $R_0$ and $a_0$. The default values assumed for these parameters in $^{96}$Ru and $^{96}$Zr are listed in Tab.~\ref{amptTab1}. For small variations of $\beta_n$, $R_0$, and $a_0$ from their default reference values, the observable $\mathcal{O}$ has the following leading-order form~\cite{Jia:2021oyt,Zhang:2021kxj}, 
\begin{align}
    \mathcal{O} \approx b_0+b_1 \beta_2^2+b_2 \beta_3^2+b_3\left(R_0-R_{0, \text { ref }}\right)+b_4\left(a-a_{\text {ref }}\right)
\end{align}
where $b_0$ is the value for spherical nuclei at some reference radius and diffuseness, and $b_1-b_4$ are centrality dependent response coefficients that encode the final-state dynamics. Most dependence on mass number is carried by $b_0$, while $b_1-b_4$ are expected to be weak functions of mass number. The ratio of $\mathcal{O}$ between $^{96}$Ru+$^{96}$Ru and $^{96}$Zr+$^{96}$Zr collisions then has a simple scaling relation:
\begin{align}
\label{eq:RO}
R_{\mathcal{O}} \equiv \frac{\mathcal{O}_{\mathrm{Ru}}}{\mathcal{O}_{\mathrm{Zr}}} \approx 1+c_1 \Delta \beta_2^2+c_2 \Delta \beta_3^2+c_3 \Delta R_0+c_4 \Delta a\,,
\end{align}
where $\Delta \beta_n^2=\beta_{n, \mathrm{Ru}}^2-\beta_{n, \mathrm{Zr}}^2, \Delta R_0=R_{0, \mathrm{Ru}}-R_{0, \mathrm{Zr}}, \Delta a=$ $a_{\mathrm{Ru}}-a_{\mathrm{Zr}}$ and $c_n=b_n / b_0$. 
As anticipated, this approach demonstrates that isobar collisions are a precision tool to probe the multiple features in the structure of the colliding isotopes. 
\begin{table}
\centering
\begin{tabular}{ccccc}
 & $R_0$ (fm) & $a_0$ (fm) & $\beta_2$ & $\beta_3$ \\ \hline
Case 1: $^{96}$Ru & 5.09 & 0.46 & 0.162 & 0 \\
Case 2 & 5.09 & 0.46 & 0.06 & 0 \\
Case 3 & 5.09 & 0.46 & 0.06 & 0.20 \\
Case 4 & 5.09 & 0.52 & 0.06 & 0.20 \\
Case 5: $^{96}$Zr & 5.02 & 0.52 & 0.06 & 0.20 \\ \hline \hline  
Ratios & $\frac{\text{Case 1}}{\text{Case 2}}$ & $\frac{\text{Case 1}}{\text{Case 3}}$ & $\frac{\text{Case 1}}{\text{Case 4}}$ & $\frac{\text{Case 1}}{\text{Case 5}}$ \\ \hline \hline 
\multicolumn{1}{c|}{Case 1 and 5 difference} & $\Delta R_0$ & $\Delta a$ & $\Delta \beta_2^2$ & $\Delta \beta_3^2$ \\ \cline{2-5}
 & 0.07 fm & -0.06 fm & 0.0226 & -0.04 \\ \hline
\end{tabular}
\caption{Nuclear structure parameters employed to parametrize $^{96}$Ru and $^{96}$Zr and and relevant parameter differences.}
\label{amptTab1}
\end{table}

\begin{figure}[t]
	\begin{centering}
		\includegraphics[width=\linewidth]{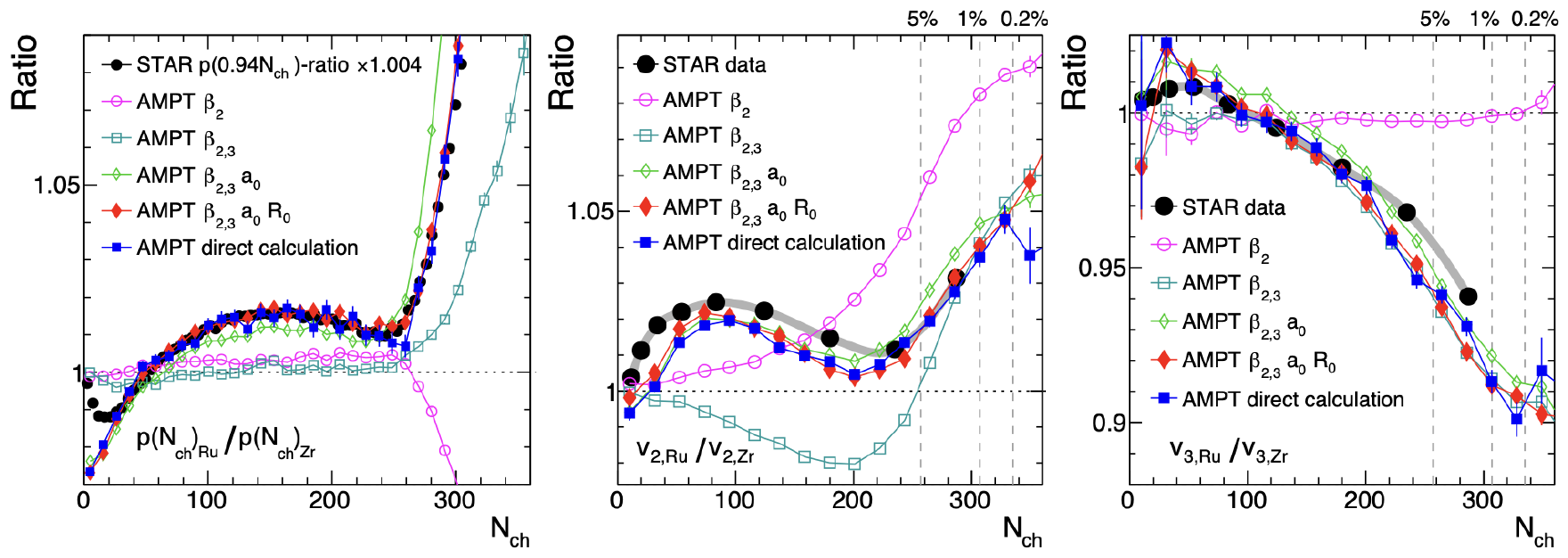}
	\par\end{centering}
	\caption{Isobar ratios of $p\left(N_{\mathrm{ch}}\right)$ (left panel), rms $v_2$ (middle panel), and rms $v_3$ (right panel) obtained from AMPT simulations with the nuclear structure parameters in Tab.~\ref{amptTab1} or calculated by means of Eq.~(\ref{eq:RO}) from the response coefficients~\cite{Jia:2021oyt} (labeled "direct calculation").}
    \label{figAMPT:vn}
\end{figure}

We now want to show that different observables probe different corners of the parameter space. Figure~\ref{figAMPT:vn} shows a step-by-step construction of the prediction for isobar ratios in AMPT, and how they compare to STAR data \cite{STAR:2021mii}. Each panel shows the isobar ratio obtained directly from simulations of $^{96}$Ru+$^{96}$Ru and $^{96}$Zr+$^{96}$Zr collisions where we vary the parameters one by one in agreement with Tab.~\ref{amptTab1}. In addition, we show a result labeled ``direct calculation'' which corresponds to the evaluation of Eq.~(\ref{eq:RO}) with coefficients $c_n$ computed in AMPT.

We start with the full simulations for different parameter sets. The left panel of Fig.~\ref{figAMPT:vn} shows that the characteristic broad peak and non-monotonic behavior of the isobar ratio of multiplicity distributions, $p\left(N_{\mathrm{ch}}\right)_{\mathrm{Ru}}/p\left(N_{\mathrm{ch}}\right)_{\mathrm{Zr}}$, is a clear signature of the skin thickness difference, $\Delta a$. In the most central collisions, the ratio is sensitive to all four parameters. The mid panel of Fig.~\ref{figAMPT:vn} shows instead the ratio $v_{2, \mathrm{Ru}}/ v_{2, \mathrm{Zr}}$. In central collisions, this is dominated by $\Delta \beta_2^2$, while in non-central collisions it receives a significant negative contribution from $\Delta \beta_3^2$. In midcentral and peripheral collisions, the impact of $\Delta a$ is pronounced, causing the characteristic non-monotonic behavior as a function of multiplicity. The influence of $\Delta R_0$ is negligible. Lastly, the right-most panel of Fig.~\ref{figAMPT:vn} shows the ratio $v_{3, \mathrm{Ru}} / v_{3, \mathrm{Zr}}$. This is mainly influenced by $\Delta \beta_3^2$, although $\Delta a$ and $\Delta R_0$ have some significant contributions over a broad $N_{\text {ch }}$ range.

Moving then to the result obtained from the application on Eq.~(\ref{eq:RO}) [``direct calculation'' in Fig.~\ref{figAMPT:vn}], near-perfect agreement with the previous results is obtained. We conclude that the Taylor expansion approximation is excellent, and that the isobar ratios enable indeed to resolve relative properties of neighboring ground states with unprecedented precision. 

\subsubsection{Additional probes of the nuclear geometry} 
Along these lines, let us stress that in Ref.~\cite{Jia:2022qgl} we demonstrate how selected isobar ratios enable one to isolate the impact of focused nuclear properties. For example, decomposing the elliptic flow into a contribution from fluctuations, $\sigma_{v_2}$, plus a contribution from the reaction plane flow, $v_{\rm RP}$, our analysis shows that the isobar ratio of $\sigma_{v_2}$ is only sensitive to the deformations of the isobars, while the isobar ratio of $v_{\rm RP}$ is exclusively sensitive to the difference in the skin thickness. As it is well-known that the value of the fourth-order cumulant, $v_2\{4\}$, is very close to that of $v_{\rm RP}$ in off-central heavy-ion collisions, we conclude that the isobar ratio of $v_2\{4\}$ observables provides the strongest sensitivity to the nuclear skin (this was also noted in the analysis of \pbpb{} data in Ref.~\cite{Giacalone:2023cet}). Similarly, we expect the study of the isobar ratio of the spectator-based elliptic flow \cite{STAR:2021mii,ALICE:2022xhd} to provide unique insights into the neutron skin of the collided species. Techniques to study as well ultra-central collisions to probe the skin thickness via free spectator neutron have also been proposed~\cite{Liu:2022kvz,Liu:2022xlm}.

In Ref.~\cite{Jia:2022qrq,Wang:2024mce}, within the same AMPT setup we study in addition the isobar ratios of the nonlinear response coefficients characterizing the coupling between elliptic and triangular flows to quadrangular and pentagonal flows, which can also serve as an independent means to constrain nuclear structure information.

\subsection{Hydrodynamic calculations with iEBE-VISNHU}

\newcommand {\snn}	{\sqrt{s_{_{\rm NN}}}}
\newcommand{\ac}{{\rm ac}_{2}\{3\}}
\newcommand {\Zr}	{$^{96}$Zr}
\newcommand {\Ru}	{$^{96}$Ru}
\newcommand {\RuRu}	{$^{96}_{44}$Ru+$^{96}_{44}$Ru}
\newcommand {\ZrZr}	{$^{96}_{40}$Zr+$^{96}_{40}$Zr}
\newcommand{\betaZr} {\beta_{\rm 3,Zr}}
\newcommand{\betaRu} {\beta_{\rm 2,Ru}}
\newcommand {\mean}[1]	{\langle #1\rangle}
\newcommand {\vtt}	{v_{2}\{2\}}
\newcommand {\vtf}	{v_{2}\{4\}}
\newcommand {\vft}	{v_{4}\{2\}}

Ultra-relativistic heavy-ion collisions at RHIC and the LHC are aimed at creating the QGP and studying the properties of the hot QCD matter. Relativistic hydrodynamics is a successful tool for describing and predicting various flow data measured in experiments, showing that the created QGP droplet behaves like perfect liquids with very small viscosity~\cite{Heinz:2013th, Gale:2013da, Shuryak:2014zxa, Song:2013gia, Song:2017wtw}. 

The RHIC isobar runs, originally designed to search for the chiral magnetic effect (CME), open new opportunities to utilize heavy-ion collisions as a tool to probe nuclear structure.  As mentioned in the previous sections, the nuclear deformations are, in particular, imprinted in the initial energy density distribution of the created QGP fireball, and their impacts are transferred to the final-state multi-particle correlations following the evolution of the bulk matter. Using the iEBE-VISHNU hybrid model~\cite{Shen:2014vra, Song:2010aq}, we study flow observables in \ruru\ and \zrzr collisions at $\snn=200$ GeV and analyze their sensitivity to the quadrupole and octupole deformations of \Ru\ and \Zr~\cite{Zhao:2022uhl}.

In the iEBE-VISHNU hybrid model, the initial entropy density distribution for the hydrodynamic evolution is generated from the \trento{} model~\cite{Moreland:2014oya, Bernhard:2016tnd}. The colliding nucleons for each nucleus are sampled via a deformed Woods-Saxon distribution:
\begin{align}
    \rho(r,\theta,\phi)&=\frac{\rho_0}{1+\exp[(r-R(\theta,\phi))/a]}, \notag
\end{align}
where $R(\theta,\phi)$ parametrizes the nuclear surface, while $a$ is the diffuseness parameter. As done previously, for a deformed nucleus the radius function can be expanded using the spherical harmonics $R=R_0(1+\beta_2Y_2^0(\theta,\phi)+\beta_3Y_3^0(\theta,\phi)+\cdots)$, where $R_{0}$ is the half-width radius parameter, while $\beta_{2}$ and $\beta_{3}$ are, respectively, the quadrupole and octupole deformation parameters.

After initialization with the nucleon coordinates, 2+1D viscous hydrodynamics with longitudinal boost-invariance~\cite{Heinz:2005bw, Song:2007ux, Song:2007fn} is implemented to simulate the evolution of QGP, which numerically solves the transport equations of the energy-momentum tensor $T^{\mu\nu}$ and the second-order evolution equations of the shear stress tensor $\pi^{\mu\nu}$ and the bulk viscous pressure $\Pi$, together with the input of the equation of state s95-PCE~\cite{Huovinen:2009yb, Shen:2010uy}. The QGP fireball hadronizes near $T_c$ to produce thermal hadrons. This is performed by the iSS event generator according to the Cooper-Frye formula~\cite{Song:2010aq}. These thermal hadrons are then fed to the Ultra-relativistic Quantum Molecular Dynamics (UrQMD) code for subsequent hadronic evolution with scatterings and resonance decays~\cite{Bass:1998ca, Bleicher:1999xi}.  

The parameters for the iEBE-VISHNU simulations, including the \trento{} initial condition parameters, shear and bulk viscosity, the switching temperature, etc., are fixed by fitting the multiplicity, mean $p_T$, flow harmonics of all charged and identified hadrons in other collision systems~\cite{Moreland:2018gsh, Zhao:2017yhj, Xu:2016hmp}.  For \zrzr\  and \ruru\ collisions  at $\snn=200$ GeV discussed here,  these tuned parameters can be found in~\cite{Zhao:2022uhl}. The Woods-Saxon parameters relevant for the present study are listed in Tab.~\ref{tab:WSDFT}, which assumes that \Ru\ and \Zr\ have only the quadrupole and octupole deformations. The larger skin of \Zr\ is also included, using the EDF results that predict a slope parameter of the symmetry energy of $L(\rho_c) = 47.3$ MeV ~\cite{Xu:2021uar}. Early nuclear structure experiments suggest that $\betaRu =0.16$ \cite{Pritychenko:2013gwa} with negligible octupole deformation and $\betaZr=0.20$ with very small quadrupole deformation \cite{kibedi_2002}. These are similar, albeit significantly different from the best fit parameters that we employ here. More details are provided in Ref.~\cite{Zhao:2022uhl}.
\begin{table}[t]
	\centering{}%
    \begin{tabular}
    {p{2.0cm}p{1.4cm}p{1.4cm}p{1.4cm}p{1.4cm}}
    \hline
	       &  $\beta_{2}$ &   $\beta_{3}$  &$R_{0}$    & $a$  \\
    \hline
	    Ru-para-I  & 0.12 & 0.00 & 5.093 & 0.487   \\
	    Zr-para-I   & 0.00 & 0.16 & 5.021 & 0.524   \\
	\hline
	\end{tabular}
	\caption{WS parameterizations of \Ru\ and \Zr\  in our tuned model~\cite{Zhao:2022uhl}. The quoted values for $R_0$ and $a$ are in fm. \label{tab:WSDFT}}
\end{table}

\begin{figure}[t]
    \centering
    \begin{tabular}{cc}
      \includegraphics[width=0.32\linewidth]{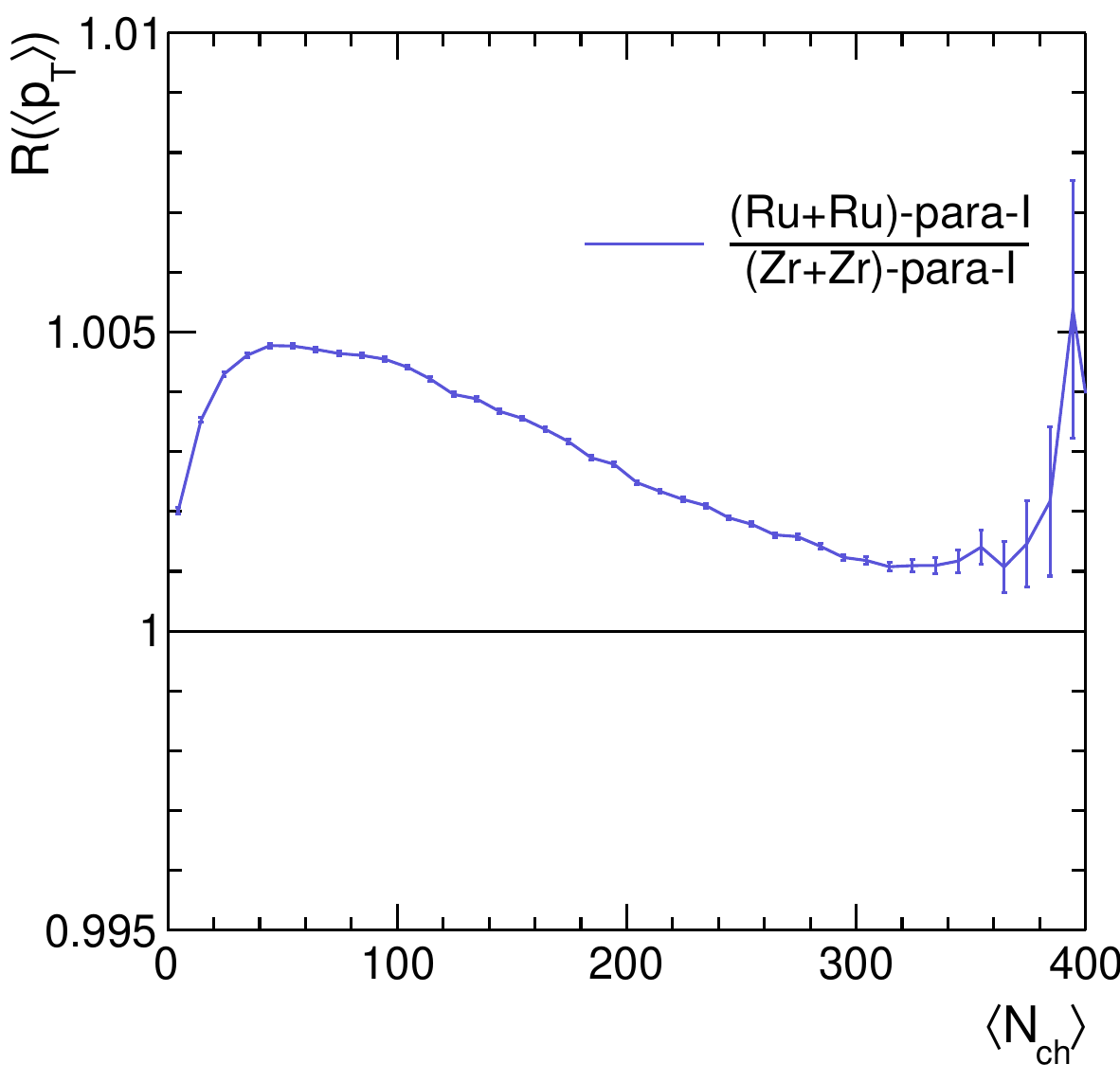}
      \includegraphics[width=0.32\linewidth]{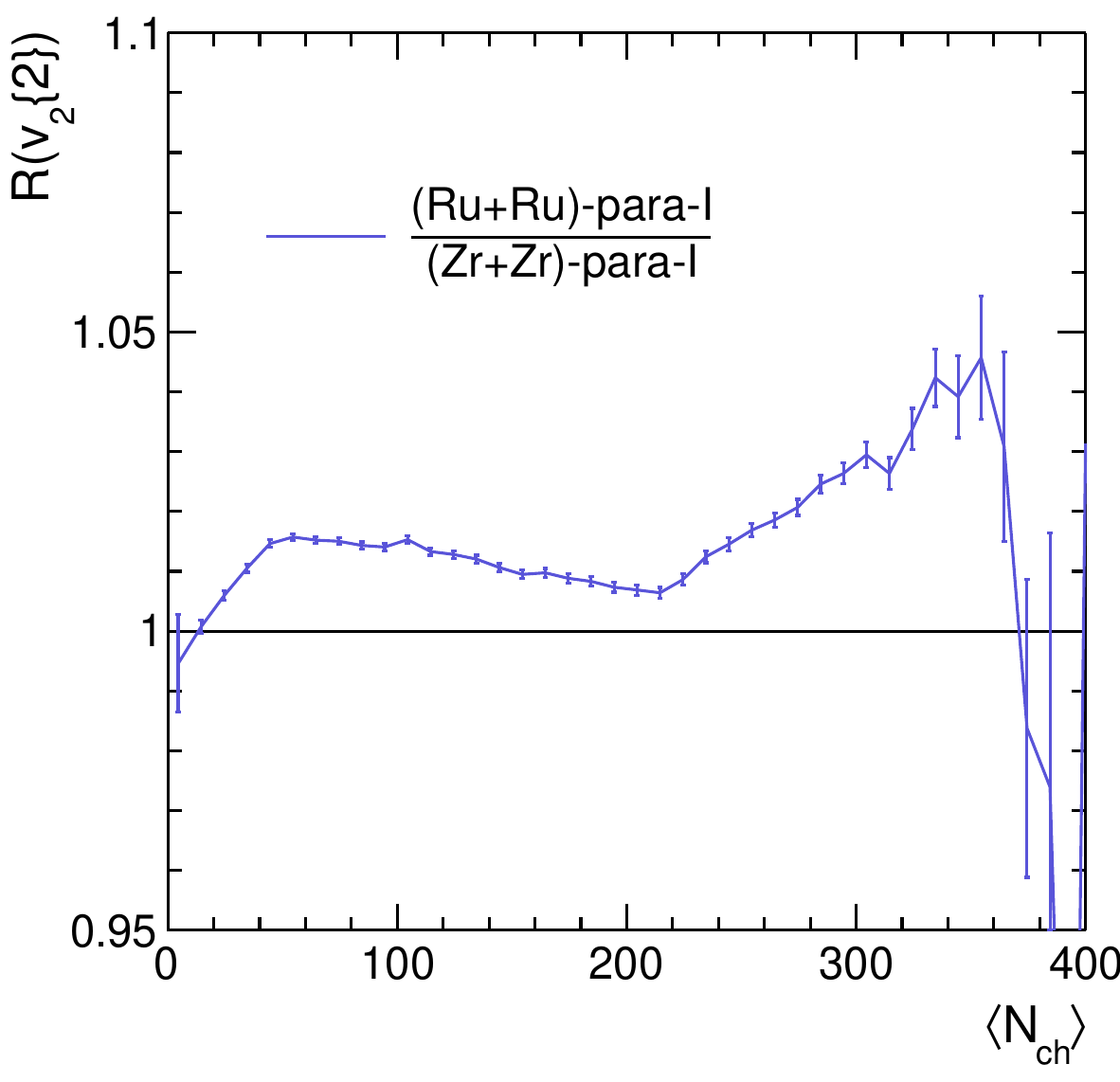}   
      \includegraphics[width=0.32\linewidth]{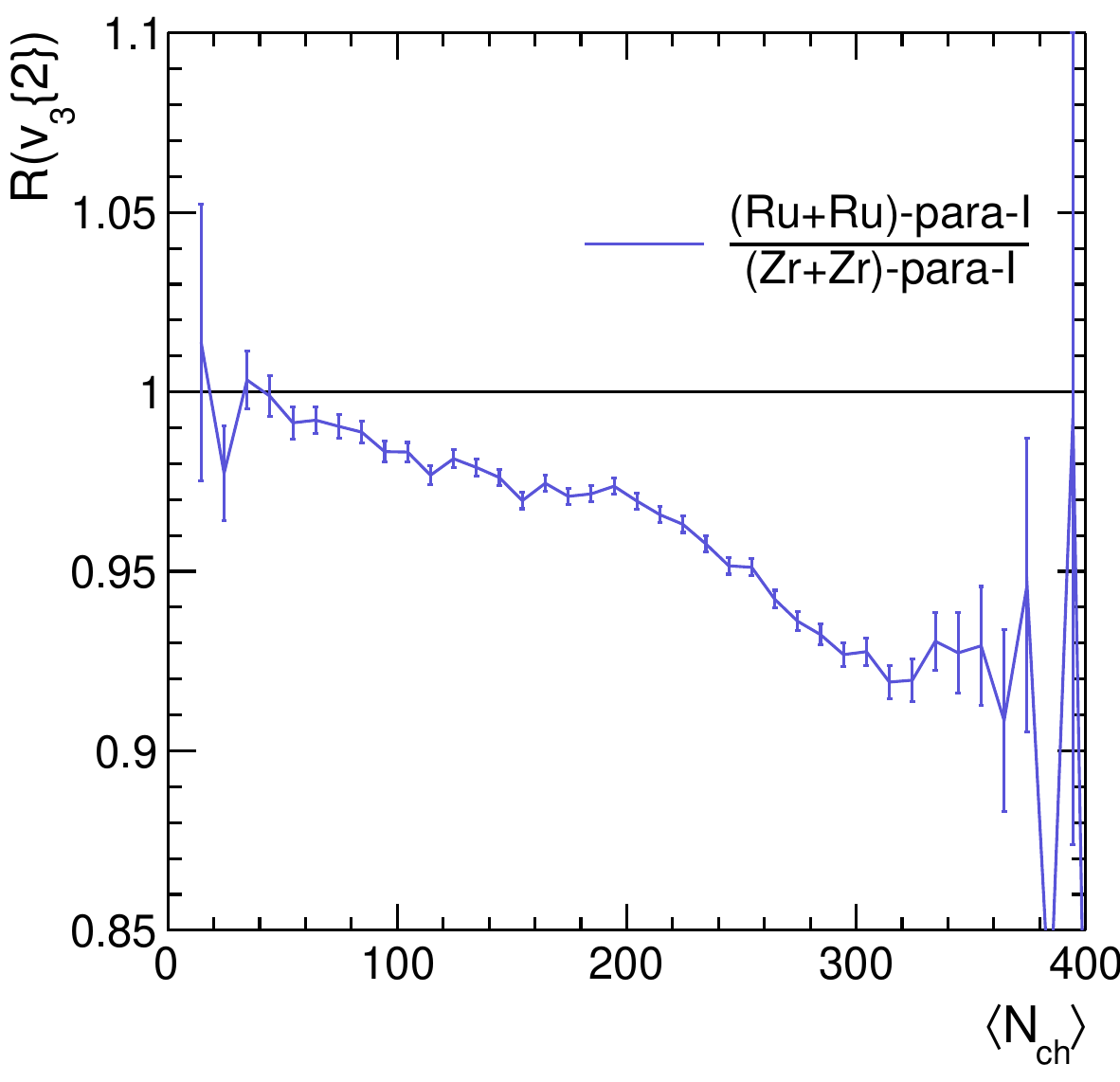}      
    \end{tabular}
    \caption{Multiplicity dependence of  $R(\langle p_T\rangle)$, $R(v_2\{2\})$ and $R(v_3\{2\})$ computed from all charged hadrons in \ruru\ and \zrzr collisions at $\snn=200$ GeV, calculated by the iEBE-VISHNU model with the deformation parameters-I for \Ru\ and \Zr.}
    \label{fig:flow}
\end{figure}

Using the iEBE-VISHNU model with the deformation parameter-I for \Ru\ and \Zr, we calculate the isobar ratio of average transverse momenta, $R(\langle p_T \rangle)$, of rms elliptic flow coefficients, $R(v_2\{2\})$, and of rms triangular flow coefficients, $R(v_3\{2\})$, for all charged hadrons in \RuRu\ and \ZrZr\ collisions at $\snn=200$ GeV.  As usual, the ratio is defined as $R(X) \equiv \frac{X_{\rm RuRu}}{X_{\rm ZrZr}}$. The trends we find are reported in Fig.~\ref{fig:flow}, and provide a picture that is fully consistent with that provide by either the initial-state \trento{} simulations of the AMPT calculations of the previous section. In particular, the quadrupole deformation of \Ru\ leads to an enhancement of $R(v_2\{2\})$ in the most central collisions, while the octupole deformation of \Zr\ leads to a suppression of $R(v_3\{2\})$. Meanwhile, the thicker halo-type neutron skin in \Zr\
leads to the non-monotonic trend of  $R(v_2\{2\})$ from central to peripheral collisions~\cite{Xu:2021vpn, Zhang:2021kxj}. These results are overall in excellent agreement with preliminary STAR data binned as a function of the collision multiplicity. We conclude that Bayesian fit of the isobar collision data based on the hydrodynamic model will be able to pin down with precision the parameters of the nuclear shapes.

Besides the single flow harmonics, additional multi-particle azimuthal correlations, routinely measured at colliders, are also expected to be sensitive to the initial state of the QGP and to the deformation of the colliding nuclei. In~\cite{Zhao:2022uhl}, we focus in particular on the effects of the deformations of \Ru\ and \Zr\ on the three-particle asymmetric cumulant $\ac = \mean{\mean{e^{i(2\varphi_{1}+2\varphi_{2}-4\varphi_{3}})}}$ \cite{ATLAS:2019peb}, where $\mean{\mean{...}}$ means that all particles of interest are summed in a single event and then averaged over an ensemble of all events. In the absence of non-flow effects,  $\ac$ can be written as  $\ac = \mean{v_{2}^{2}v_{4}\cos{4(\Phi_2-\Phi_4)}}$, where $\Phi_{2,4}$ is the event-plane of the associated flow harmonics. In~\cite{Zhao:2022uhl} we show that the ratio $R(\ac)$ is actually extremely sensitive to the deformations of the isobars, providing, thus, a means to place additional independent constraints on these parameters.

\subsection{Hydrodynamic simulations with the \textit{Trajectum} framework}

\subsubsection{Model framework and nuclear structure inputs}

The \textit{Trajectum} framework employed in this work consists of three main stages: an initial condition setup, a hydrodynamic evolution phase, and a final freeze-out into hadrons, subsequently propagated via the UrQMD to model the hadronic gas stage \cite{Bass:1998ca,Bleicher:1999xi}. This study uses version 1.2 of the code \cite{Nijs:2020roc,Nijs:2020ors,Nijs:2021clz}\footnote{Publicly available at \url{https://sites.google.com/view/govertnijs/trajectum}.}. Our calculations are based on those of Ref.~\cite{Nijs:2021kvn}, where we use a maximum likelihood setup following Ref.~\cite{Nijs:2021clz}. 

Nuclear structure takes part in shaping the initial conditions of the hydrodynamic code, which are here generated by means of the \trento{} model \cite{Bernhard:2016tnd}, in which the positions of the nucleons within each colliding nucleus are sampled from the Woods-Saxon (WS) distribution:
\begin{equation}
\label{eq:roftheta}
P(r,\theta) \propto \left(1 + \exp\left(\frac{r - R(\theta)}{\sigma}\right)\right)^{-1},
\end{equation}
where we include quadrupole and octupole deformations, $R(\theta) = R  \left(1 + \beta_2Y_2^0(\theta) + \beta_3Y_3^0(\theta)\right)$. The other parameters include the half-width nuclear radius $R$, and the surface diffuseness $\sigma$. Separate WS profiles are employed for protons and neutrons, which allows us to include the presence of a neutron skin. We implement, in addition, a short-range repulsion of nucleons in the \trento{} computation through a minimum inter-nucleon spacing parameter, $d_\text{min}$.

As done for the previous AMPT study, we examine the impact of different features of the nuclear geometry by carying among five distinct WS parameter sets for both $^{96}$Zr and $^{96}$Ru. Our choices of parameters are shown in Tab.\ref{tab:WSparameters}, and very close to those employed in the other parts of this document. Case 1 is derived from $e$–A scattering measurements \cite{Pritychenko:2013gwa,Nayak:2014zba}, while case 2 comes from finite-range liquid-drop model calculations \cite{nucl-th/9308022}. Cases 3 and 4 are motivated by density functional calculations of these nuclei (although assuming spherical nuclei) and correspond to the upper and lower rows of Tab.~2 in Ref.~\cite{Xu:2021vpn}. In case 4, a quadrupole deformation parameter of $\beta_2=0.16$ in $^{96}$Ru is added. Case 5 mirrors case 4 but incorporates $\beta_2$ and $\beta_3$ values in $^{96}$Zr in accordance to Ref.~\cite{Zhang:2021kxj}. For reference, based on these parameters we expect the largest effect of the difference in skin thickness to manifest in the collisions moving from case 2 to case 3.

\begin{table}[t]
\centering
\begin{tabular}{cccccccc}
\hline
\hline
nucleus & $R_p\,$[fm] & $\sigma_p\,$[fm] & $R_n\,$[fm] & $\sigma_n\,$[fm] & $\beta_2$ & $\beta_3$ & $\sigma_\text{AA}\,$[b] \\
\hline
$_{44}^{96}$Ru(1) & 5.085 & 0.46 & 5.085 & 0.46 & 0.158 & 0 & 4.628 \\
$_{40}^{96}$Zr(1) & 5.02 & 0.46 & 5.02 & 0.46 & 0.08 & 0 & 4.540 \\
\hline
$_{44}^{96}$Ru(2) & 5.085 & 0.46 & 5.085 & 0.46 & 0.053 & 0 & 4.605 \\
$_{40}^{96}$Zr(2) & 5.02 & 0.46 & 5.02 & 0.46 & 0.217 & 0 & 4.579 \\
\hline
$_{44}^{96}$Ru(3) & 5.06 & 0.493 & 5.075 & 0.505 & 0 & 0 & 4.734 \\
$_{40}^{96}$Zr(3) & 4.915 & 0.521 & 5.015 & 0.574 & 0 & 0 & 4.860 \\
\hline
$_{44}^{96}$Ru(4) & 5.053 & 0.48 & 5.073 & 0.49 & 0.16 & 0 & 4.701 \\
$_{40}^{96}$Zr(4) & 4.912 & 0.508 & 5.007 & 0.564 & 0.16 & 0 & 4.829 \\
\hline
$_{44}^{96}$Ru(5) & 5.053 & 0.48 & 5.073 & 0.49 & 0.154 & 0 & 4.699 \\
$_{40}^{96}$Zr(5) & 4.912 & 0.508 & 5.007 & 0.564 & 0.062 & 0.202 & 4.871 \\
\hline
\hline
\end{tabular}
\caption{\label{tab:WSparameters} Woods-Saxon parameters employed to describe ${96}$Ru and ${96}$Zr, and total inelastic nucleus-nucleus cross sections ($\sigma_{\rm AA}$) resulting from their respective collisions. Subscripts $p$ and $n$ denote proton and neutron parameters respectively. The $\beta_n$ coefficients are identical for both protons and neutrons.}
\end{table}

\subsubsection{Isobar ratios of multiplicity and collective flow}

\begin{figure}[t]
\centering
\includegraphics[width=0.45\linewidth]{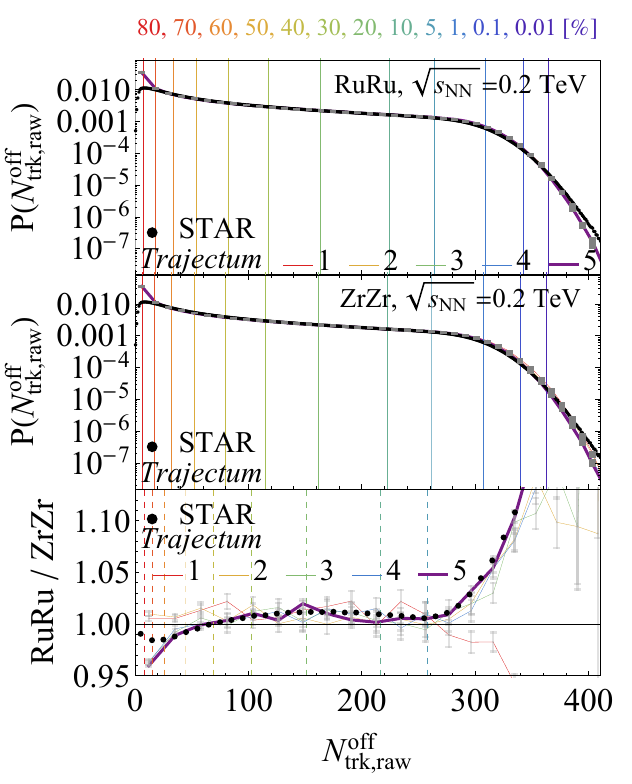}
\caption{We show the comparison between (inversely-corrected) multiplicity distributions obtained in \ruru{} and \zrzr{} collisions within the \textit{Trajectum} framework (lines) to STAR data \cite{STAR:2021mii} on $N_{\rm trk,raw}^{\rm off}$ distributions (symbols). Centrality classes are displayed, for our case 5 (colored lines) and for STAR data (dashed lines).}
\label{fig:Traj1}
\end{figure}

We first study the probability distributions of multiplicities in \ruru{} and \zrzr{} collisions. The STAR track distributions used to draw the histograms are raw and must be corrected for the detector response before they can be compared to the (true) charge multiplicities predicted by the \textit{Trajectum} calculations. Here we apply the inverse correction. Following the interpolation method discussed in Refs.~\cite{STAR:2008med,STAR:2021mii}, and considering $|\eta| < 0.5$ and $p_T > 0.2$ GeV acceptance cuts, we correct the \textit{Trajectum} predictions to be able to match them to STAR data. We correct, in addition, the normalization of our multiplicities to ensure that the area under the histograms for central to mid-central events matches that of the measurement \cite{Nijs:2021kvn}. Our final results are shown in Fig.~\ref{fig:Traj1}. They are in excellent agreement with the STAR curves, except for the region with less than 20 detected tracks which is plagued by detection inefficiencies. Based on the comparison with the different WS parametrizations, especially on the isobar ratio in the bottom panel, we conclude that the data strongly favors a higher quadrupole deformation in $^{96}$Ru, as well as a larger skin thickness for $^{96}$Zr. This in full agreement with the results shown so far in this document. 

\begin{figure}[t]
\centering
\includegraphics[width=\linewidth]{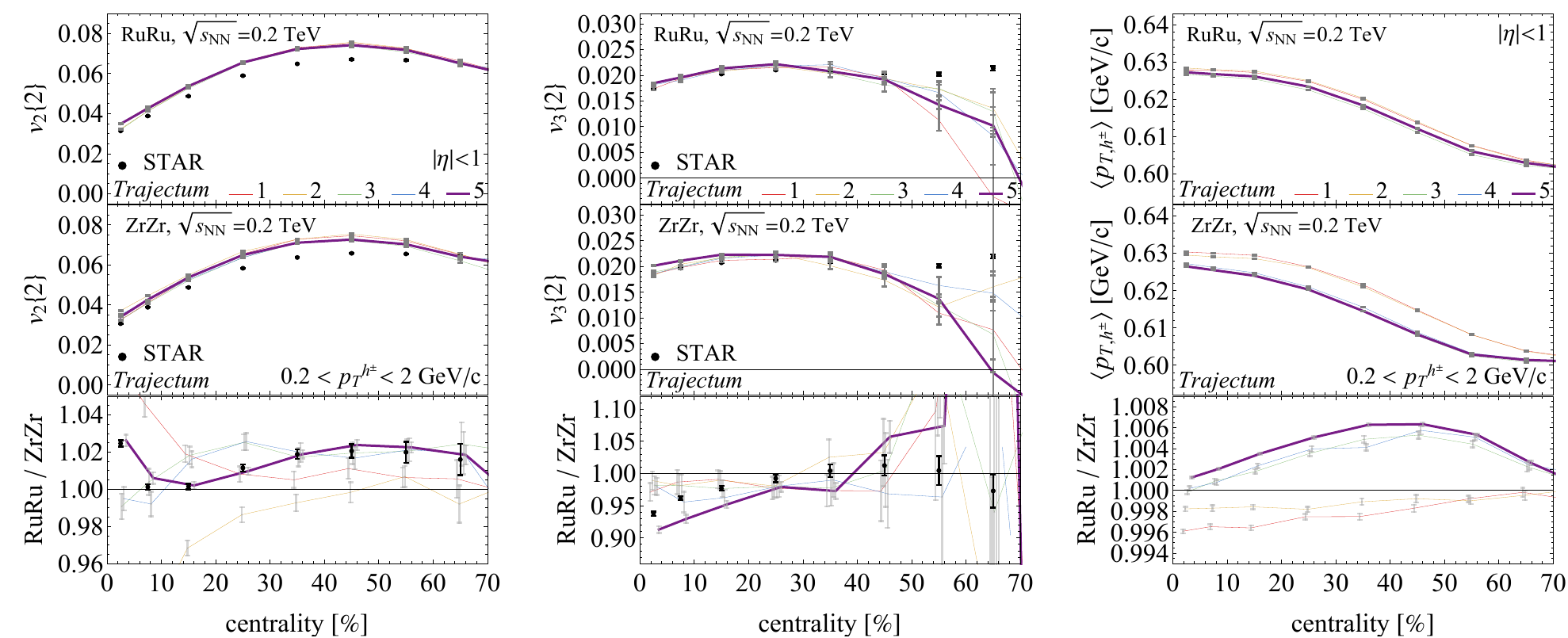}
\caption{We show hydrodynamic results for $v_2\{2\}$ (left), $v_3\{2\}$ (center), and $\langle p_T\rangle $ (right) in \ruru{} collisions (top) and \zrzr{} collisions (middle) at RHIC energy for all WS parameter sets considered in Tab.~\ref{tab:WSparameters}. The isobar ratios of these observables are shown in the bottom panels. The STAR data is from Ref.~\cite{STAR:2021mii}.}
\label{fig:Traj2}
\end{figure}

We now move on to discuss the collective flow of isobar collisions. Our results are compared to STAR data in Fig.~\ref{fig:Traj2}.

The first two panels display the rms elliptic flow $v_2{2}$, and the rms triangular flow, $v_3{2}$. For both these quantities, our calculations tend to over-predict their absolute values. This is not surprising, as the model parameters in our framework were tuned to LHC data rather than RHIC data. However, the isobar ratios of these quantities (lower panels) eliminate this error are are nicely consistent with measurements. The isobar ratio of $v_2\{2\}$ confirms that the combinations of effects coming from a larger $\beta_2$ in $^{96}$Ru, a very large $\beta_3$ as well as significantly larger skin thickness in $^{96}$Zr, all included in our case 5, permit the hydrodynamic simulations to yield an excellent description of the STAR results. Note that this is the only case among those considered that enables us to do so. For the isobar ratio of $v_3\{2\}$, we see that the deviation from unity in case 5 is too large compared to the one experimentally observed. This suggests that $\beta_3=0.20$ may be an overestimate. This is consistent with the iEBE-VISNHU analysis of the previous subsection, where a tune to RHIC data prefers $\beta_2=0.12$ in $^{96}$Ru and $\beta_3=0.16$ in $^{96}$Ru. A global Bayesian analysis of STAR data will be required to resolve this issue.

The right-most panel of  Fig.~\ref{fig:Traj2} shows instead a prediction fo the mean transverse momenta in isobar collisions. Of particular interest is the isobar ratio in the lower panel. Focusing on the case 5 results, we predict that the charged hadron $\langle p_T \rangle$ is higher in \ruru{} collisions, and that the deviation from \zrzr{} collisions is at maximum 0.5\% around 40\% centrality. This is consistent with other hydrodynamic simulations (e.g. in Fig.~\ref{fig:flow}) and with preliminary STAR measurements. The crucial observation to make is that, as we move from case 2 to case 3, including a difference in skin thickness between isobars, the isobar ratio of $\langle p_T \rangle$ goes from being lower than unity to above unity, confirming the trends observed, e.g., in Sec.~\ref{sec:trento}. This showcases once more the unprecedented capability of isobar ratios in discerning fine relative differences among the structure of the isotopes under study.

\subsubsection{Viscous effects on isobar ratios}

\begin{figure}[t]
\centering
\includegraphics[width=\linewidth]{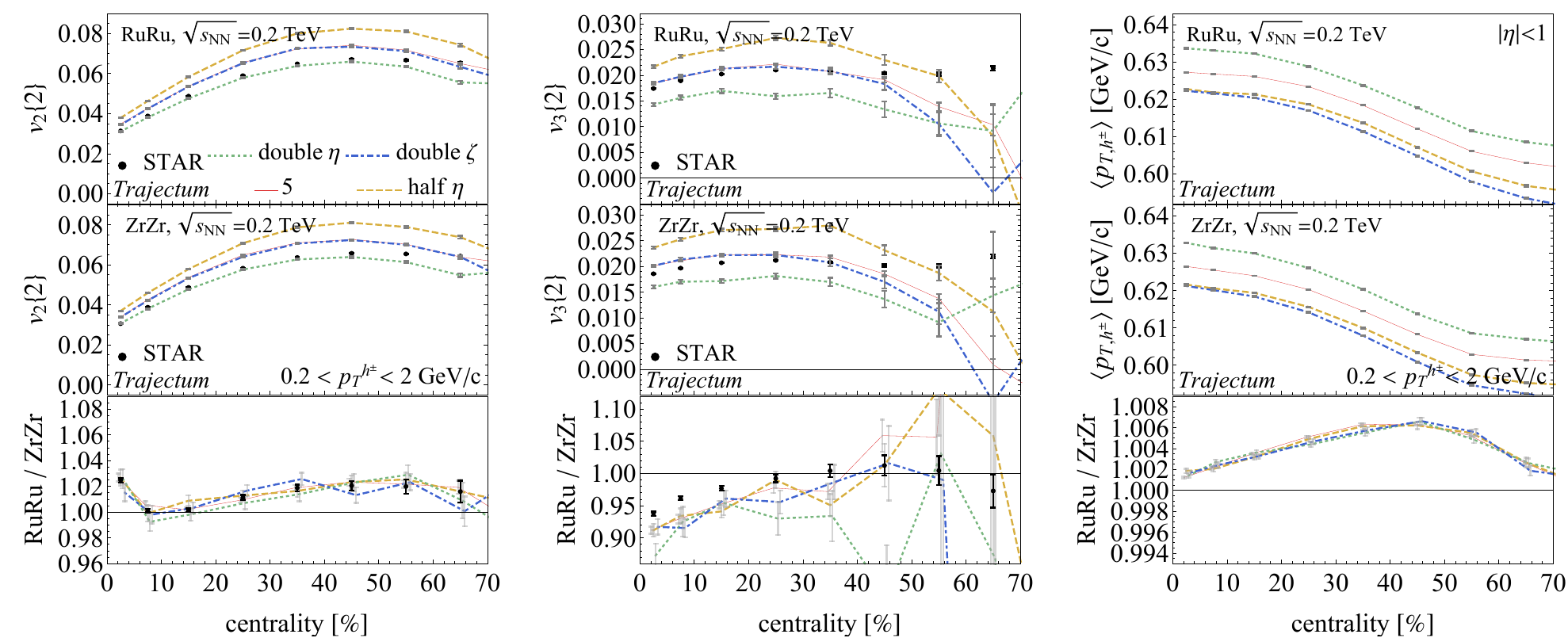}
\caption{Same as Fig.~\ref{fig:Traj2}, but for nuclear structure case 5 only, and different variations of the specific shear ($\eta/s$) and specific bulk ($\zeta/s$) viscosities of the QGP.}
\label{fig:Traj3}
\end{figure}

Here we want to give an explicit illustration of the fact that uncertain QGP properties, in particular viscosities, do not play a role in shaping the isobar ratios, which can be used, hence, as robust probes of the initial states of the collisions and of the nuclear geometry. To this end, Fig.~\ref{fig:Traj3} illustrates that while variations in shear and bulk viscosities significantly impact the absolute values of $v_2\{2\}$, $v_3\{2\}$, and $\langle p_T \rangle$, they do not significantly alter the isobar ratios, consistent with similar findings from AMPT-based studies \cite{Jia:2021oyt,Zhang:2022fou}. This indicates in particular that although the linear response coefficients, $\kappa_n$, to the initial eccentricities  depend much on viscosities, such dependence cancels out in the isobar ratio.

\subsection{Remarks}

In conclusion to this section, we would like to emphasize the non-trivial result that the three independent frameworks that we have considered (AMPT, the Peking University tune of the iEBE-VISHNU code, and \textit{Trajectum}) yield results for the isobar ratios that are fully consistent with each other\footnote{To this, one can also add the results of the additional iEBE+VISHNU hydrodynamic simulations discussed in Ref.~\cite{Xu:2021uar}, which are also consistent with those presented in this section.}. Although in some sense this amounts to mere code validation, it is an unprecedented result, as the effects we are after are sometimes only sub-percent deviations from unity. The \textit{Trajectum} and iEBE-VISHNU results for the ratio of mean transverse momentum are, for instance, in full agreement even if the departure from unity in the isobar ratio is only at the level of 0.1\%. This means that potential numerical artifacts in the numerical solvers are fully under control, and that even a variation of 0.06 fm in the skin thickness of the colliding isotopes leads to significant effects that can be reliably predicted.

\newpage

\section{Discussion and recommendations}
\label{sec:6}

\subsection{Summary of the results}

In summary, isobar collisions at RHIC \cite{STAR:2021mii} have established a new method for precision studies of the many-body structure of nuclear ground states. By studying how observables vary across collision systems that are similar in mass, one can isolate signatures of features such as nuclear deformations and skins in unprecedented detail.  We emphasize that this is one of the most important legacies of the high-energy nuclear collision program at RHIC.

Our first conclusion is that $^{96}$Ru and $^{96}$Zr exhibit fundamentally different structural properties in a region of the nuclear chart, $A\approx100$, characterized by complex isotopic behavior that is still poorly understood. STAR data clearly points to a higher value of $\beta_2$ for $^{96}$Ru, and to an effect akin to a large octupole deformation in the ground state of $^{96}$Zr. Both features are in qualitative agreement with low-energy spectroscopic information, but a quantitative understanding of the associated intrinsic nuclear properties appears to be currently out of reach for beyond-mean-field calculations based on state-of-the-art energy density functional models. Hopefully, upcoming \textit{ab initio} calculations that do not rely on the notion of an intrinsic nuclear shape, such as NLEFT simulations coupled to high-fidelity chiral EFT interactions, will be able to elucidate these issues further in a not too distant future. However, this is a significant endeavor and will likely require a benchmark of results from a variety of different \textit{ab initio} frameworks. 

However, with appropriate choices of deformation and skin thickness parameters, we find that hydrodynamic and initial-state calculations can reproduce \textit{quantitatively} the observations made by the STAR collaboration. We do not find any way of describing the peculiar dissimilarities reported between \ruru{} and \zrzr{} collisions other than assigning such effects to dissimilarities in the structure of these isotopes. The effects of QGP-related properties do not play a significant role in the measured isobar ratios.

Ultimately, drawing quantitative conclusions about the role of the strong magnetic field in shaping collective flow measurements in isobar collisions remains premature without a rigorous uncertainty quantification of the nuclear geometries.

\subsection{Physics opportunities with isobar collisions}

 This unexpected spin-off of the RHIC program opens the way for a number of potentially groundbreaking studies at the intersection of heavy-ion collisions, nuclear structure, and other disciplines. Here we isolate five such research opportunities based on isobar collisions that we deem especially important and compelling. The following is a refined version of our report from June 2022 \cite{GSI_Discussion_2023}.

\subsubsection{Search for the Chiral Magnetic Effect}

In regard to the search for the CME, the isobars with $A=96$ appear to be too complex and poorly understood for drawing firm conclusions from STAR data. As discussed above, doing so will require a much more refined knowledge of the structure of these isotopes from cutting-edge \textit{ab initio} computations of nuclear structure, which may take a long time to achieve. If there is interest in repeating an isobar run to further optimize the CME search, the situation could dramatically improve if the selected species were $^{136}$Xe and $^{136}$Ce. Some study would be required on the structure of $^{136}$Ce, but one could take advantage of the fact that $^{136}$Xe is essentially a spherical nucleus \cite{nudat3}. In addition, $^{136}$Xe has been studied extensively in low-energy nuclear research due to its relevance for dark matter and other BSM-related searches \cite{Belley:2023btr}. Furthermore, $^{136}$Xe can be used in the SMOG2 system of the LHCb detector \cite{BoenteGarcia:2024kba}, so that one should be able to get beforehand a rather clear picture of the structure of this isotope (as probed in a collider) by means of $^{208}$Pb+$^{136}$Xe collisions.

\subsubsection{Light-ion collisions for precision small-system studies}

We can exploit the highly deformed structure of light nuclei to shed new light on the long-standing issue of the origin of the collective behavior observed in small collisions systems at RHIC and LHC \cite{Grosse-Oetringhaus:2024bwr}. Experiments with $pp$ and $pA$ collisions (including notably $d$+Au and He+Au collisions) have provided solid \textit{qualitative} evidence of a collective response of the system to its initial-state geometry \cite{CMS:2010ifv,ALICE:2012eyl,ATLAS:2012cix,PHENIX:2018lia,STAR:2022pfn}. For example, $v_2$ in $d$+Au collisions is larger than that of $p$+Au collisions at the same multiplicities, indicating the impact of the deuteron geometry \cite{PHENIX:2018lia,STAR:2022pfn}. However, turning the small system question into a \textit{quantitative} problem is difficult because of intrinsic theoretical uncertainties related to the modeling of the initial-state geometry formed in collisions that involve individual nucleons \cite{Schenke:2021mxx}. To make a substantial leap toward a discussion that is more quantitative, one needs isobar collisions with light nuclei. The discussions of the RRTF have in particular pointed out that the nucleus $^{20}$Ne, due to its extreme shape and closeness in mass to $^{16}$O \cite{nudat3}, would be ideal to complement the planned efforts with \oooo{} collisions \cite{Brewer:2021kiv,ALICE:2021wim} and pave to way to a quantitative analysis of the collective flow. This has led to a detailed proposal, where \textit{ab initio} evaluations of the ground states of $^{16}$O and $^{20}$Ne have been coupled to hydrodynamic simulations to highlight the possibility of having quantitative predictions for such systems, with theoretical uncertainties under control \cite{Giacalone:2024luz}. Remarkably, following this proposal, one day of $^{20}$Ne+$^{20}$Ne collisions was recently added to the schedule of LHC Run 3 \cite{lhc}. This is in part an outcome of the EMMI RRTF discussions.

\subsubsection{Measuring the neutron skin of \texorpdfstring{$^{48}$Ca}{48Ca}}

Theoretical studies show that the neutron skin of atomic nuclei is tightly linked to the so-called symmetry energy of the equation of state of nuclear matter, a crucial parameter in the determination of the stability and properties of neutron stars \cite{Vinas:2013hua,Thiel:2019tkm}. As such, the neutron skin of nuclei (especially spherical ones) has been the subject of very active investigations, both in theory \cite{Hu:2021trw} and in experiment \cite{Lattimer:2023rpe}. The neutron skin is formally defined as the difference between the root-mean-square (rms) radii of the neutron and proton distributions:
    \begin{equation}
    \delta_{np} = \sqrt{\langle r^2 \rangle_n} - \sqrt{\langle r^2 \rangle_p} .    
    \end{equation}
    One nucleus of particular interest in this context is $^{48}$Ca, a doubly magic nucleus with a pronounced neutron excess. The dedicated CREX experiment at Jefferson Lab has reported a neutron skin for $^{48}$Ca of $\delta_{np}= 0.121 \, \pm \, 0.026 ({\rm exp}) \, \pm \, 0.024 \, ({\rm model})$ \cite{CREX:2022kgg}. This should be compared to the PREX-II value for the neutron skin of $^{208}$Pb, which is $\delta_{np}=0.278 \, \pm \, 0.078 \, ({\rm exp}) \,  \pm \, 0.012 \, ({\rm model})$ \cite{PREX:2021umo}. Remarkably, the large difference between these two values appears to be inconsistent with the predictions of nuclear models \cite{Reed:2023cap}, which has sparked a substantial debate in the community \cite{Mammei:2024qow,Sammarruca:2023mxp}. Interestingly enough, estimating the neutron skin of $^{48}$Ca through high-energy nuclear collisions is relatively straightforward. The calcium isotopic chain includes two doubly magic nuclei: $^{48}$Ca and $^{40}$Ca. The latter is symmetric, with equal numbers of protons and neutrons ($Z = 20$), and thus a near-vanishing neutron skin. Notably, both isotopes have the same measured charge radius, implying that the difference in overall size arises solely from their neutron content. As discussed in this report, for isobars one can robustly determine the differences in skin thickness. For example, analyses of RHIC data have revealed \cite{Xu:2022ikx}
    \begin{equation}
    \delta_{\rm np} (^{96}{\rm Zr}) - \delta_{\rm np} (^{96}{\rm Ru}) \approx 0.06 \pm 0.02~\mathrm{fm}.  
    \end{equation}
    Given that $\delta_{\rm np}(^{48}{\rm Ca}) \gg \delta_{\rm np}(^{40}{\rm Ca}) \approx 0$, it follows that high-energy $^{40}$Ca+$^{40}$Ca and $^{48}$Ca+$^{48}$Ca collisions have the potential to isolate:
    \begin{equation}
    \delta_{\rm np} (^{48}{\rm Ca}) - \delta_{\rm np} (^{40}{\rm Ca}) \simeq \delta_{\rm np} (^{48}{\rm Ca}),    
    \end{equation}
    with a precision of approximately 0.02 fm. As the skin of $^{208}$Pb has also been determined in a Bayesian analysis of LHC data \cite{Giacalone:2023cet}, a combined analysis of $^{48}$Ca and $^{208}$Pb nuclei via LHC data would help us shed new light on the apparent CREX-PREX tension, while informing physical parameters relevant for neutron star research.

    \subsubsection{Triaxiality of rare earth nuclei}

One could even use the collider to detect unresolved properties of the nuclear ground states. Although the nuclear triaxiality (as quantified by the parameter $\gamma$ in the Woods-Saxon density) plays essentially no role in the observables discussed in this manuscript~\cite{ALICE:2024nqd,Lu:2023fqd}, it is now established that three-particle correlation measurements such as $v_n^2$-$[p_T]$ correlations or the skewness of the mean transverse momentum fluctuations \cite{Bally:2021qys,Jia:2021qyu,Nielsen:2023znu}, are strongly sensitive to its value. Experimentally, this has been demonstrated thanks to data on \xexe{} collisions from LHC \cite{ALICE:2021gxt,ATLAS:2022dov}, as $^{129}$Xe is a fully triaxial nucleus \cite{Bally:2021qys}. Some evidence of a small triaxiality in $^{238}$U has also been reported by the STAR collaboration \cite{STAR:2024wgy}.  Recent studies have proposed further investigating the nuclear shape phase transition in the $\gamma$-soft structures of $^{129}$Xe via six-particle correlations involving $v_{2}^{4}$ and $[p_{\rm T}]^2$~\cite{Zhao:2024lpc}. A {\it unified algorithm} for multi-particle correlations has also been developed for the implementation of such observables in experimental analyses~\cite{Nielsen:2025pkz}. With that in mind, a completely unique possibility in heavy-ion collisions is that of revealing the triaxial structure of highly-prolate isotopes in the region of the rare earths. Whether or not such isotopes present a mild triaxiality in their ground states (of order $\gamma\approx 5^\circ$) is an important open question, as this property is sensitive to detailed dynamical processes inherent to the underlying nuclear force \cite{Otsuka:2023yts}. However, quantifying the triaxiality of such isotopes with precision is not feasible on the basis of conventional low-energy spectroscopic techniques. The unique feature of heavy-ion collisions is that the associated probes of triaxiality depend on this parameter with a factor $\beta_2^3$ in front, that is, they are proportional to $\pm\beta_2^3 \cos(3\gamma)$ \cite{Jia:2021qyu}. Because of this, even the effects of a very small value of $\gamma$ would be strongly amplified to a visible signal whenever one deals with a large value of $\beta_2$.  This is precisely the situation for rare earth nuclei, which have intrinsic surface deformation parameters that can be as high as $\beta_2 = 0.35$. The candidates for these searches are $^{166}$Er and $^{154}$Sm, which are predicted to exhibit a significant difference in $\gamma$ \cite{Otsuka:2023yts}. Some preliminary calculations of the proposed effect can be found in Fig.~65 of the latest STAR Beam User Request \cite{STAR_BUR_2024}. If one prefers to make use of exactly isobaric species, one could choose optimal species (also based on machine limitations) across the gadolinium (Gd), erbium (Er), dysprosium (Dy), or ytterbium (Yr) isotopic chains.

\subsubsection{Nuclear matrix elements of neutrinoless double beta decay}

This point was not thoroughly discussed during the EMMI RRTF, but has emerged in recent times as a highly promising direction of investigation. In fact, a seemingly obvious application of isobar collisions is in relation to the issue of the determination of the nuclear matrix elements of the neutrinoless double beta ($0\nu\beta\beta$) decay transition \cite{Agostini:2022zub}.  The $0\nu\beta\beta$ decay is conjectured to occur between two stable isobaric states. This transition involves the expectation value of a two-body operator that is naturally sensitive to two-body correlations of nucleons across a range of length scales \cite{Yao:2021wst}. Thanks to the development of emulators and Bayesian analysis in low-energy nuclear theory, it is today possible to study in detail the correlation between the NME and other fundamental nuclear properties such as radii and deformations \cite{Horoi:2022ley,Zhang:2024tzr,Zhang:2024nqr,Belley:2024zvt}. For well-deformed candidates, the NME is strongly correlated with the quadrupole deformation of the parent and daughter isotopes. As shown in a recent publication, this implies that multi-particle correlation observables in the final states of heavy-ion collisions are also in a tight correlation with the NME values \cite{Li:2025vdp}. Therefore, one should investigate how such correlations may ultimately help reduce the theoretical uncertainties on the computed values of the matrix elements. Indeed, it seems highly probable that a collider run with, in particular, the isobars $^{76}$Ge and $^{76}$Se, relevant for next-generation $0\nu\beta\beta$ decay searches \cite{Brugnera:2023zgw}, could have a significant impact on this research.

\subsection{Final recommendation}

The main recommendation of the EMMI RRTF is that any future proposals for precision high-energy physics studies with nuclei at present and future collider facilities should be accompanied by a detailed assessment of the uncertainties associated with the structure of the species involved. This should be done in a close collaboration between the low- and high-energy nuclear physics communities. One of the most important points highlighted in this document is that there is no clear pathway for nuclear structure calculations based on energy density functional theory to arrive at a reliable picture of the ground state of $^{96}$Zr. Therefore, we urge the community of theorists working on \textit{ab initio} calculations of nuclear structure to further develop their models to obtain an accurate description of deformed transitional nuclei in the $A\sim100$ region. This is a major challenge, but may be achieved before the next decade. It will ultimately allow for a reanalysis of STAR data, potentially leading to a sounder statement about the presence or absence of CME signals in such data.

\bigskip
\noindent With that said, let us stress once more that the successful operation of isobar collisions at RHIC represents a major milestone in nuclear physics research. It opens unprecedented opportunities to reach into new corners of the landscape of phenomena accessible via the study of atomic nuclei and QCD matter in general. \textbf{The EMMI RRTF strongly recommends pursuing these novel scientific cases to potentially capitalize on the opportunities presented by high-energy experiments with isobars in future collider runs.} We emphasize that the study of isobar collisions at LHC would greatly benefit from the proposed installation of a second ion injector \cite{AlemanyFernandez:2025ixd} in the accelerator complex.

\newpage

\section{Acknowledgments}
The members of the RRTF are grateful to the ExtreMe Matter Institute EMMI at GSI for the generous financial support, as well as to the Institute for Theoretical Physics at Heidelberg University for hosting the events. The members of the RRTF thank Xavier Roca-Maza for his contribution as an external speaker during the first meeting of the Task Force, as well as Wolfram Korten and Magda Zieli\'nska for useful input and discussions. 

\bigskip
\noindent F. Capellino, G. Giacalone, and A. Kirchner are funded by the Deutsche Forschungsgemeinschaft (DFG, German Research Foundation) – Project-ID 273811115 – SFB 1225 ISOQUANT, and under Germany's Excellence Strategy EXC2181/1-390900948 (the Heidelberg STRUCTURES Excellence Cluster).
J.Jia is supported by the U.S. Department of Energy, Award number No. DE-SC0024602.
The work of T. R. Rodr\'iguez and L. M. Robledo is supported by the Spanish MICIN under PID2021-127890NB-I00. T. R. Rodr\'iguez and L. M. Robledo gratefully thank the support from GSI-Darmstadt computing facility.
Y. Zhou and E. G. Nielsen acknowledge the support from the European Research Council (ERC) under the European Union’s Horizon Europe research and innovation programme (Grant Agreement No. 101077147), VILLUM FONDEN (grant number 00025462), Danmarks Frie Forskningsfond (Independent Research Fund Denmark).
F. G. Gardim was supported by Conselho Nacional de Desenvolvimento Cient\'{\i}fico  e  Tecnol\'ogico  (CNPq grant 306762/2021-8).
K. P. Pala and W. M. Serenone acknowledge  support  from Funda\c{c}\~ao de Amparo \`a Pesquisa do Estado de S\~ao Paulo (respectively via grants 2020/15893-4 and 2021/01670-6,2022/11842-1 ).  
A. V. Giannini has been partially supported by CNPq. 
F. G. Gardim, A. V. Giannini, F. Grassi, K. P. Pala and W. M. Serenone acknowledge  support  from Funda\c{c}\~ao de Amparo \`a Pesquisa do Estado de S\~ao Paulo (FAPESP  grant  2018/24720-6) and project INCT-FNA Proc.~No. 464898/2014-5 as well as for computing resources,  the  Superintend\^encia de Tecnologia da Informa\c{c}\~ao at Universidade de S\~ao Paulo (Aguia Clauster)  and the National Laboratory for Scientific Computing (LNCC/MCTI, Brazil), through project COL-ISO as well as the ambassador program (UFGD), subproject FCNAE, which providing HPC resources from the SDumont supercomputer.
J. Noronha-Hostler acknowledges the support from the US-DOE Nuclear Science Grant No. DE-SC0023861  and the support from the Illinois Campus Cluster, a computing resource that is operated by the Illinois Campus Cluster Program (ICCP) in conjunction with the National Center for Supercomputing Applications (NCSA), and which is supported by funds from the University of Illinois at Urbana-Champaign.
K. Wimmer acknowledges support from the European Research Council (ERC) under the European Union’s Horizon 2020 research and innovation programme (grant agreement No 101001561).
B. Schenke is supported by the U.S. Department of Energy, Office of Science, Office of Nuclear Physics under DOE Contract No.~DE-SC0012704 and within the framework of the Saturated Glue (SURGE) Topical Theory Collaboration.
C. Shen was supported by the US Department of Energy under Award No. DE-SC0021969. C.S. acknowledges support from a DOE Office of Science Early Career Award.
W. B. Zhao is supported by the National Science Foundation (NSF) under grant number ACI-2004571 within the framework of the XSCAPE project of the JETSCAPE collaboration, and by the U.S. Department of Energy, Office of Science, Office of Nuclear Physics, within the framework of the Saturated Glue (SURGE) Topical Theory Collaboration.
The work of A. V. Afanasjev is supported by the U.S. Department of Energy, Office of Science, Office of Nuclear Physics under Award No. DE-SC0013037.
W. Ryssens is a research associate of the F.R.S.-FNRS (Belgium). The present research benefited from computational resources made available on the Tier-1 supercomputers Zenobe and Lucia of the Fédération Wallonie-Bruxelles, infrastructure funded by the Walloon Region under the grant agreement nr 1117545. Additional computational resources have been provided by the Consortium des Équipements de Calcul Intensif (CÉCI), funded by the Fonds de la Recherche  Scientifique de Belgique (F.R.S.-FNRS) under Grant No. 2.5020.11 and by the Walloon Region. 
S. F. Taghavi is supported by the Deutsche Forschungsgemeinschaft (DFG) through the grant number 517518417.
H. Song is supported in part by the National Natural Science Foundation of China under Grant Nos. 12247107, 12075007.
M. Luzum~was supported by São Paulo Research Foundation (FAPESP) under grants 2017/05685-2, 2018/24720-6, and 2023/13749-1, by project INCT-FNA Proc.~No.~464898/2014-5, and by CAPES - Finance Code 001.
D. Lee was partially supported by DOE grant DE-SC0013365, DE-SC0024586, and DE-SC0023175, with computational resources provided by the Gauss Centre for Supercomputing e.V. (www.gauss-centre.eu) for computing time on the GCS Supercomputer JUWELS at J{\"u}lich Supercomputing Centre (JSC) and special GPU time allocated on JURECA-DC as well as the Oak Ridge Leadership Computing Facility through the INCITE award ``Ab-initio nuclear structure and nuclear reactions.''.
C. Zhang is supported in part by the National Key Research and Development Program of China under Contract Nos. 2024YFA1612600 and 2022YFA1604900, the National Natural Science Foundation of China under Grant Nos. 12025501 and 12147101, and Shanghai Pujiang Talents Program under Contract No. 24PJA009.
H. Mehrabpour is partly supported by the National Natural Science Foundation of China under Grant No. 12247107.

\newpage

\phantomsection
\addcontentsline{toc}{section}{References}
\printbibliography

\end{document}